\documentclass[twoside,11pt]{article}

\usepackage[preprint]{jmlr2e}
\usepackage{amsmath}
\usepackage{float}
\usepackage{booktabs}
\usepackage{arydshln}
\usepackage[]{algorithm}
\usepackage{algpseudocode}
\usepackage{xcolor}
\usepackage[normalem]{ulem} 
\usepackage{subfig}
\usepackage{mathtools}
\mathtoolsset{showonlyrefs}
\usepackage{jmlr2e}
\usepackage{tikz}

\usepackage{forloop}
\newcounter{ct}
\newcommand{\indt}[1][1]{\forloop{ct}{0}{\value{ct} < #1}{\hspace{\algorithmicindent}}}

\jmlrheading{1}{2024}{1-xx}{8/24}{xx/2x}{meila00a}{Teemu H\"ark\"onen and Simo S\"arkk\"a}

\ShortHeadings{Log-Gaussian gamma process estimation with iterated posterior linearization}{H\"ark\"onen and S\"arkk\"a}
\firstpageno{1}

\begin{document}
\title{Estimation of log-Gaussian gamma processes with iterated posterior linearization and Hamiltonian Monte Carlo}

\jmlrheading{1}{2024}{1-xx}{8/24}{xx/2x}{meila00a}{Teemu H\"ark\"onen and Simo S\"arkk\"a}


\ShortHeadings{Log-Gaussian gamma processes with iterated posterior linearization and HMC}{H\"ark\"onen and S\"arkk\"a}
\firstpageno{1}

\title{Estimation of log-Gaussian gamma processes with iterated posterior linearization and Hamiltonian Monte Carlo}

\author{\name Teemu H\"ark\"onen \email teemu.h.harkonen@aalto.fi \\
       \addr Department of Electrical Engineering and Automation\\ Aalto University\\
       Espoo, FI-02150, Finland
       \AND
       \name Simo Särkkä \email simo.sarkka@aalto.fi \\
       \addr Department of Electrical Engineering and Automation\\ Aalto University\\
       Espoo, FI-02150, Finland}

\editor{N.N.}

\maketitle
\begin{abstract}
Stochastic processes are a flexible and widely used family of models for statistical modeling.
While stochastic processes offer attractive properties such as inclusion of uncertainty properties, their inference is typically intractable, with the notable exception of Gaussian processes.
Inference of models with non-Gaussian errors typically involves estimation of a high-dimensional latent variable.
We propose two methods that use iterated posterior linearization followed by Hamiltonian Monte Carlo to sample the posterior distributions of such latent models with a particular focus on log-Gaussian gamma processes.
The proposed methods are validated with two synthetic datasets generated from the log-Gaussian gamma process and a multiscale biocomposite stiffness model.
In addition, we apply the methodology to an experimental Raman spectrum of argentopyrite.
\end{abstract}

\section{Introduction}
\label{sec:introduction}
Modeling of measurement data is a fundamental aspect of science and engineering.
The typical approach is to explicitly parameterize a model, optimize the model parameters with respect to the measurement data, and make predictions according to the optimized parameters.
This paradigm includes linear regression, possibly at its simplest, and, in contrast, the wide variety of modern neural network architectures  \citep{Abiodun:2018,Cong:2023, Xu:2023}.

Alternatively, it is possible to use a non-parametric model.
In such an approach, one can encode information, such as smoothness, on the function underlying the data generation to avoid explicitly defining a parametric model.
The typical example of such construction, a stochastic process, is the Gaussian process, which encodes this information via its covariance kernel function  \citep{WilliamsRasmussen:2006}.
While an attractive option for many applications such as magnetic field estimation, material science, remote sensing, spectroscopy  \citep{Solin:2018, Volker:2021, Harkonen:2023:emission, Kuitunen:2025} with interesting connections to neural networks  \citep{Neal:1995, Garriga:2018, Greg:2019}, Gaussian processes suffer from a cubic computational complexity with respect to the number of measurement data points.

This computational complexity has sired a wide range of approaches to circumvent or reduce the complexity through inducing points  \citep{Hensman:2013}, reduced-rank representations  \citep{Solin:2020}, sparse covariance structures  \citep{Susiluoto:2020}, and variational methods  \citep{Titsias:2009}, together with other approaches  \citep{Liu:2020:GP}.
The computational burden is further exacerbated with non-Gaussian measurement error models.

An example of a commonly used stochastic process with non-Gaussian errors is the log-Gaussian Cox process with geostatistical, ecological, seismic, and spectroscopic applications among many others  \citep{Diggle:2013, Waagepetersen:2015, Liu:2020, DAngelo:2022, Harkonen:2023}.
With log-Gaussian Cox processes, measurements are modeled as Poisson-distributed random variables with the latent intensity function of the process modeled as a Gaussian process  \citep{Moller:1998}.
This hierarchical structure results in intractable likelihood computation, which has bred numerous approaches for estimating the process \citep{Vanhatalo:2013, Teng:2017}, such as maximum a \textit{posteriori} estimates, Laplace approximation, variational methods, and Markov chain Monte Carlo with Hamiltonian Monte Carlo in particular \citep{Wainwright:2008, Betancourt:2017, Robert:2021}.
However, more sophisticated linearization-based approaches, such as posterior linearization \citep{Garcia-Fernandez:2015, Garcia:2017, Tronarp+Fernandez+Sarkka:2018}, have seen little application in estimating such models with latent stochasticity.

Motivated by the log-Gaussian Cox processes, log-Gaussian gamma processes were proposed to generate synthetic Raman spectra for training partially-Bayesian neural networks  \citep{Harkonen:2024}.
In the original study, the log-Gaussian gamma process parameter posterior distributions were estimated using Markov chain Monte Carlo.
The estimated posterior distributions were then used to generate statistically realistic Raman spectra, which were then used to generate data pairs for the particular application.
However, the log-Gaussian gamma process model was simplified to avoid Markov chain Monte Carlo sampling of high-dimensional posterior distributions of the underlying shape and rate processes.
Smoothed measurements were used as a proxy for the expectation of the process, and the rate of the gamma process was modeled as a scalar random variable.

\begin{figure}
    \centering
    \includegraphics[width = \textwidth]{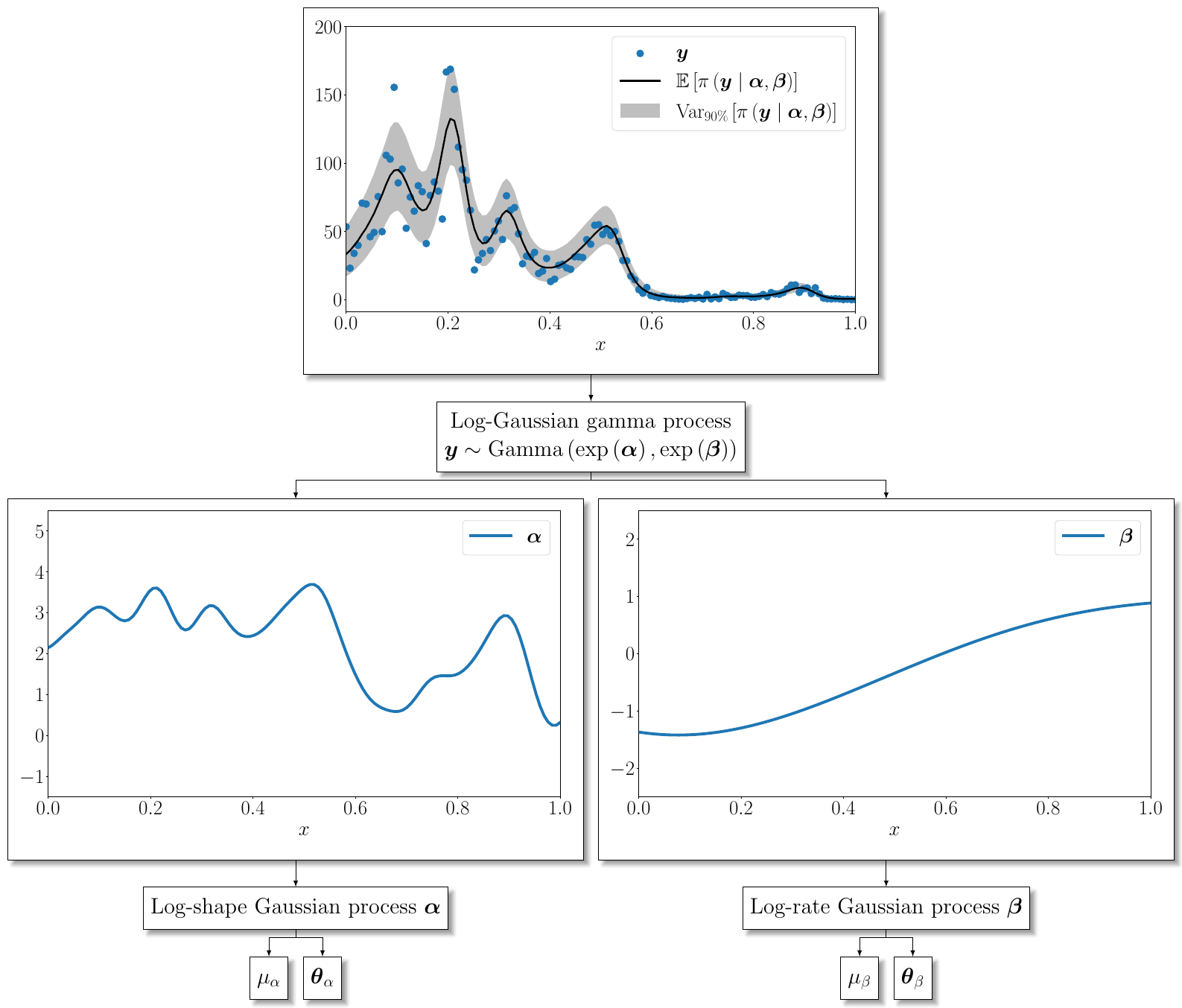}
    \caption{The log-Gaussian gamma process. Measurement data $\boldsymbol{y}$ is modeled as gamma-distributed random variables. The log-shape and log-rate parameters $\boldsymbol{\alpha}$ and $\boldsymbol{\beta}$ of the gamma distribution are modeled as Gaussian processes with parameters $\mu_\alpha$, $\boldsymbol{\theta}_\alpha$, $\mu_\beta$, and $\boldsymbol{\theta}_\beta$. $\mathbb{E}\left[ \pi( \boldsymbol{y} \mid \boldsymbol{\alpha}, \boldsymbol{\beta} ) \right]$ and $\textrm{Var}_{90\%}\left[ \pi( \boldsymbol{y} \mid \boldsymbol{\alpha}, \boldsymbol{\beta} ) \right]$ are the mean and 90\% confidence interval of the log-Gaussian gamma process.}
    \label{im:lggpDiagram}
\end{figure}

In this work, we consider posterior distribution sampling of the unsimplified log-Gaussian gamma process, where both the shape and rate processes are modeled as log-Gaussian processes.
An illustration of the log-Gaussian gamma process structure is shown in Figure \ref{im:lggpDiagram}.
Our main contribution is the introduction of iterated posterior linearization as a tool for sampling the posterior distributions of the shape and rate processes and their corresponding hyperparameters.
We use the iterated posterior linearization in two ways.
In both of our approaches, we perform the posterior distribution estimation by initially marginalizing over the Gaussian process parameters and performing the iterated posterior linearization for the log-shape and log-rate processes.
In our first, approximate approach, we follow the iterated posterior linearization with Hamiltonian Monte Carlo sampling using the estimated log-shape and log-rate posterior means and covariances to obtain posterior distributions for the Gaussian process hyperparameters.
Second, we construct a sequence of posterior distributions using the approximate posterior distribution of our first approach as a starting point of our sequence, which we ultimately temper to coincide with the true posterior distribution of the log-Gaussian gamma process.
The results of both methods are comparable to estimates yielded by direct Hamiltonian Monte Carlo sampling of the full log-Gaussian gamma process posterior distribution, and at a lower computational cost.
The methods are readily applicable to the log-Gaussian Cox process and similar models with latent stochastic structures.
Software and the datasets are publicly available on the corresponding author's GitHub repository.

We apply Hamiltonian Monte Carlo and our proposed methods to two synthetic datasets and an experimental Raman spectrum measurement.
The first synthetic dataset is generated from the log-Gaussian gamma process.
The second synthetic dataset is generated using a biocomposite microscale mechanics  \citep{Konigsberger:2024}, with the deterministic output of the biocomposite model used as the mean of the generated log-Gaussian gamma process data.
Third, we apply the method to an experimental Raman spectrum of argentopyrite from the RRUFF database  \citep{RRUFF:2016}.

The rest of the manuscript is structured as follows.
We detail the structure of the log-Gaussian gamma process in Section \ref{sec:lggp}.
In Section \ref{sec:hmc}, we discuss the main properties of Hamiltonian Monte Carlo sampling, which we follow with a discussion on our iterated posterior linearization approaches in Section \ref{sec:iteratedPosteriorLinearization}.
We present numerical results for our proposed methods in comparison to HMC sampling in Section \ref{sec:results}.
We conclude the study with a discussion and remarks on future work in Section \ref{sec:conclusions}.
\section{Log-Gaussian gamma processes}
\label{sec:lggp}
We model strictly positive real-valued measurements $\boldsymbol{y} = ( y_1, \dots, y_K)^\intercal \in \mathbb{R}^{K \times 1}$ with $y_k \in \mathbb{R}_+$ at $D$-dimensional measurement locations $\boldsymbol{x} = ( x_1, \dots, x_K)^\intercal \in \mathbb{R}^{K \times D}$ with $x_k = ( x_{k,1}, \dots, x_{k,D} )^\intercal \in \mathbb{R}^{D \times 1}$ as conditionally-independent gamma-distributed random variables
\begin{equation}
    y_k := y( x_k ) \sim \pi\big( y_k \mid \alpha_k, \beta_k \big)  = {\rm{Gamma}} \big( \exp\left\{ \alpha_k \right\}, \exp\left\{ \beta_k \right\} \big),
    \label{eq:dataDistribution}
\end{equation}
where $\alpha_k := \alpha( x_k ) \in \mathbb{R}$ denotes the logarithm of the shape parameter and $\beta_k := \beta( x_k ) \in \mathbb{R}$ is the logarithm of the scale parameter of the gamma distribution.
Logarithms of the shape and scale parameters are modelled as independent Gaussian processes (GP).
The GP for the shape parameter logarithm is modeled as
\begin{equation}
    \alpha( \boldsymbol{x} ) \sim {\rm{GP}} \big( \mu_{\alpha}, \Sigma_\alpha( \boldsymbol{x}, \boldsymbol{x}; \theta_\alpha) + \sigma_{ \alpha, \textrm{e} }^2 \boldsymbol{I} \big),
    \label{eq:alphaPriorGP}
\end{equation}
where $\alpha( \boldsymbol{x} ) = \left( \alpha( x_1 ), \dots, \alpha( x_K ) \right)^{\intercal} $, $\mu_{\alpha} \in \mathbb{R}$ is a constant mean, $\Sigma_\alpha( \boldsymbol{x}, \boldsymbol{x}; \theta_\alpha) \in \mathbb{R}^{K \times K}$ denotes the GP covariance matrix parameterized according to $\theta_\alpha$, $\sigma_{ \alpha, \textrm{e} }^2$ is an independent error variance with $\boldsymbol{I} \in \mathbb{R}^{K \times K}$ being the identity matrix.
Each $ij$th element $\left[ \Sigma_\alpha( \boldsymbol{x}, \boldsymbol{x}; \theta_\alpha) \right]_{ij}$ of the covariance matrix $\Sigma_\alpha( \boldsymbol{x}, \boldsymbol{x}; \theta_\alpha)$ is given by the squared exponential covariance function
\begin{equation}
    \left[ \Sigma_\alpha( \boldsymbol{x}, \boldsymbol{x}; \theta_\alpha) \right]_{ij} = \sigma_{ \alpha, \textrm{s} }^2 \prod_{d = 1}^D \exp\left\{ -\frac{1}{2} \frac{\left( x_{i,d} - x_{j,d} \right)^2 }{ l_{ \alpha, d}^2} \right\},
    \label{eq:alphaCovarianceFunctionSqrExp}
\end{equation}
with $\theta_\alpha = ( \sigma_{ \alpha, \textrm{e} }, \sigma_{ \alpha, \textrm{s} }, l_{ \alpha, 1}, \dots, l_{ \alpha, D} )^\intercal$ and where $\sigma_{ \alpha, \textrm{s} }^2$ denotes the signal variance and $l_{ \alpha,d}$ with $d \in \{ 1, \dots, D\}$ are length scale parameters along each measurement location dimension.
Note that, we absorb $\sigma_{ \alpha, \textrm{e} }$ into $\theta_\alpha$ for notational convenience.

Similarly, the logarithm of the scale parameter is given as
\begin{equation}
    \beta( \boldsymbol{x} ) \sim {\rm{GP}} \big( \mu_{\beta}, \Sigma_\beta( \boldsymbol{x}, \boldsymbol{x}; \theta_\beta) + \sigma_{ \beta, \textrm{e} }^2 \boldsymbol{I} \big),
    \label{eq:betaPriorGP}
\end{equation}
where $\beta( \boldsymbol{x} ) = \left( \beta( x_1 ), \dots, \beta( x_K ) \right)^\intercal $, $\mu_{\beta} \in \mathbb{R}$ denotes a constant mean of the GP, $\Sigma_\beta( \boldsymbol{x}, \boldsymbol{x}; \theta_\beta)$ is the GP covariance matrix parameterized according to $\theta_\beta$, and $\sigma_{ \beta, \textrm{e} }^2$ is an independent error variance.
As with the log-scale $ \alpha( \boldsymbol{x} )$, elements $\left[ \Sigma_\beta( \boldsymbol{x}, \boldsymbol{x}; \theta_\beta) \right]_{ij}$ of the covariance matrix $\Sigma_\beta( \boldsymbol{x}, \boldsymbol{x}; \theta_\beta)$ are given by
\begin{equation}
    \left[ \Sigma_\beta( \boldsymbol{x}, \boldsymbol{x}; \theta_\beta) \right]_{ij} = \sigma_{ \beta, \textrm{s} }^2 \prod_{d = 1}^D \exp\left\{ -\frac{1}{2} \frac{\left( x_{i,d} - x_{j,d} \right)^2 }{ l_{ \beta, d}^2} \right\},
    \label{eq:betaCovarianceFunctionSqrExp}
\end{equation}
where $\theta_\beta = ( \sigma_{\beta, e}, \sigma_{ \beta, \textrm{s} }, l_{ \beta, 1}, \dots, l_{ \beta, D} )^\intercal$, $\sigma_{ \beta, \textrm{s} }^2$ denotes the signal variance, and $l_{ \beta, d}$ with $d \in \{ 1, \dots, D\}$ are length scale parameters along each measurement location dimension of log-shape process, $ \beta( \boldsymbol{x} )$.

In addition, we assign prior distributions for the GP mean and covariance parameters, $\mu_\alpha, \theta_\alpha \sim \pi_0( \mu_\alpha, \theta_\alpha)$ and $\mu_\beta, \theta_\beta \sim \pi_0( \mu_\beta, \theta_\beta)$.
We follow the advice given in existing literature \citep{Gelman:2006}, and use independent half-normal distributions as the prior distributions for the GP variance parameters.
Independent truncated normal distributions are used for the GP length scale parameters.
The truncation bounds the length scale parameters from below.
For the GP means, we use normal distributions.
We further detail the prior distributions and their parameterizations in Table \ref{table:hyperparameterPriorDistributions}.

The above can be compiled into the following hierarchical model
\begin{equation}
    \begin{split}
        y(\boldsymbol{x}) \mid \boldsymbol{\alpha}, \boldsymbol{\beta}, \mu_\alpha, \theta_\alpha, \mu_\beta, \theta_\beta &\sim {\rm{Gamma}}   \big( \exp\left( \boldsymbol{\alpha} \right), \exp\left( \boldsymbol{\beta} \right) \big),\\
        \boldsymbol{\alpha} \mid \mu_\alpha, \theta_\alpha &\sim {\rm{GP}} \big( \mu_{\alpha}, \Sigma_\alpha( \boldsymbol{x}, \boldsymbol{x}; \theta_\alpha) + \sigma_{ \alpha, \textrm{e} }^2 \boldsymbol{I} \big),\\
        \boldsymbol{\beta} \mid \mu_\beta, \theta_\beta &\sim {\rm{GP}} \big( \mu_{\beta}, \Sigma_\beta( \boldsymbol{x}, \boldsymbol{x}; \theta_\beta) + \sigma_{ \beta, \textrm{e} }^2 \boldsymbol{I} \big),\\
        \mu_\alpha, \theta_\alpha &\sim \pi_0( \mu_\alpha, \theta_\alpha),\\
        \mu_\beta, \theta_\beta &\sim \pi_0( \mu_\beta, \theta_\beta),
    \end{split}
    \label{eq:lggpModel}
\end{equation}
with $\boldsymbol{\alpha} := \alpha( \boldsymbol{x} )$ and $\boldsymbol{\beta} := \beta( \boldsymbol{x} )$ used for brevity above and in the following exposition.a
We refer to the above model as the log-Gaussian gamma process.
The log-Gaussian gamma process model yields a posterior distribution for the log-Gaussian shape and scale processes together with the GP mean and covariance parameters as
\begin{equation}
\begin{split}
    \pi( \boldsymbol{\alpha}, \boldsymbol{\beta}, \mu_\alpha, \theta_\alpha, \mu_\beta, \theta_\beta \mid \boldsymbol{y} ) \propto \mathcal{L}( \boldsymbol{y} \mid \boldsymbol{\alpha}, \boldsymbol{\beta}, \mu_\alpha, \theta_\alpha, \mu_\beta, \theta_\beta) &\pi_0( \boldsymbol{\alpha} \mid \mu_\alpha, \theta_\alpha ) \pi_0( \boldsymbol{\beta} \mid \mu_\beta, \theta_\beta )\\ &\times \pi_0( \mu_\alpha, \theta_\alpha) \pi_0( \mu_\beta, \theta_\beta),
\end{split}
    \label{eq:lggpPosteriorDistribution}
\end{equation}
where $\mathcal{L}( \boldsymbol{y} \mid \boldsymbol{\alpha}, \boldsymbol{\beta}, \mu_\alpha, \theta_\alpha, \mu_\beta, \theta_\beta)$ denotes the likelihood function with $\pi_0( \boldsymbol{\alpha} \mid \mu_\alpha, \theta_\alpha )$ and $\pi_0( \boldsymbol{\beta} \mid \mu_\beta, \theta_\beta )$ being the log-GP prior distributions for the shape and scale processes defined in Equations~\eqref{eq:alphaPriorGP} and \eqref{eq:betaPriorGP}.
According to the assumption of gamma-distributed measurements $\boldsymbol{y}$ in Equation~\eqref{eq:dataDistribution}, the likelihood function $\mathcal{L}( \boldsymbol{y} \mid \boldsymbol{\alpha}, \boldsymbol{\beta}, \mu_\alpha, \theta_\alpha, \mu_\beta, \theta_\beta)$ can be given as
\begin{equation}
    \mathcal{L}( \boldsymbol{y} \mid \boldsymbol{\alpha}, \boldsymbol{\beta}, \mu_\alpha, \theta_\alpha, \mu_\beta, \theta_\beta) \propto \prod\limits_{k = 1}^K \frac{ b_k^{ a_k} 
  }{ \Gamma( a_k ) } y_k^{ a_k - 1}\exp( -b_k y_k ),
    \label{eq:gammaLikelihoodFunction}
\end{equation}
where $ a_k := \exp\left\{ \alpha_k \right\}$, $ b_k := \exp\left\{ \beta_k \right\}$, and $\Gamma( \cdot )$ is the gamma function.
The hierarchical log-GP prior distribution $ \pi_0( \boldsymbol{\alpha} \mid \mu_\alpha, \theta_\alpha) $ can be explicitly stated as
\begin{equation}
\begin{split}
     \pi_0( \boldsymbol{\alpha} \mid \mu_\alpha, \theta_\alpha) = &\frac{1}{\sqrt{(2\pi)^K}} \left\vert \Sigma_\alpha( \boldsymbol{x}, \boldsymbol{x}; \theta_\alpha) + \sigma_{ \alpha, \textrm{e} }^2 \boldsymbol{I} \right\vert^{-1/2}\\ & \times \exp\left\{
     -\frac{1}{2} \left( \boldsymbol{\alpha} - \mu_\alpha \right)^\intercal \left( \Sigma_\alpha( \boldsymbol{x}, \boldsymbol{x}; \theta_\alpha) + \sigma_{ \alpha, \textrm{e} }^2 \boldsymbol{I} \right)^{-1} \left( \boldsymbol{\alpha} - \mu_\alpha \right) \right\},
\end{split}
    \label{eq:alphaPriorGpDensity}
\end{equation}
and respectively for $\pi_0( \boldsymbol{\beta} \mid \mu_\beta, \theta_\beta)$ as
\begin{equation}
\begin{split}
     \pi_0( \boldsymbol{\beta} \mid \mu_\beta, \theta_\beta) = &\frac{1}{\sqrt{(2\pi)^K}} \left\vert \Sigma_\beta( \boldsymbol{x}, \boldsymbol{x}; \theta_\beta) + \sigma_{ \beta, \textrm{e} }^2 \boldsymbol{I} \right\vert^{-1/2}\\ & \times \exp\left\{
     -\frac{1}{2} \left( \boldsymbol{\beta} - \mu_\beta \right)^\intercal \left( \Sigma_\beta( \boldsymbol{x}, \boldsymbol{x}; \theta_\beta) + \sigma_{ \beta, \textrm{e} }^2 \boldsymbol{I} \right)^{-1} \left( \boldsymbol{\beta} - \mu_\beta \right) \right\},
\end{split}
    \label{eq:betaPriorGpDensity}
\end{equation}
where $\left\vert \,\cdot\, \right\vert$ denotes the matrix determinant.
For further details on GPs, see for example  \citep{WilliamsRasmussen:2006}.
The total number of parameters in the posterior distribution $\pi( \boldsymbol{\alpha}, \boldsymbol{\beta}, \mu_\alpha, \theta_\alpha, \mu_\beta, \theta_\beta \mid \boldsymbol{y} )$ is $2K + 2D + 6$.
\begin{table}[!t]
\centering
\begin{tabular}{c|c}
\toprule
Parameter & Prior distribution \\
\midrule
$\mu_{\alpha}$ & $ \mathcal{N} \left( \gamma_{ \mu, \alpha}, \rho_{ \mu, \alpha} \right) $ \\
$\mu_{\beta}$ & $ \mathcal{N} \left( \gamma_{ \mu, \beta}, \rho_{ \mu, \beta} \right) $ \\
\hline
$\sigma_{\alpha, \textrm{e} }$ & $ \mathcal{N}_{1/2} \left( \rho_{ \sigma, \textrm{{e}}, \alpha} \right) $ \\
$\sigma_{ \beta, \textrm{e} }$ & $ \mathcal{N}_{1/2} \left( \rho_{ \sigma, \textrm{{e}}, \beta} \right) $ \\
$\sigma_{ \alpha, \textrm{s} }$ & $ \mathcal{N}_{1/2} \left( \rho_{ \sigma, \textrm{{s}}, \alpha} \right) $ \\
$\sigma_{ \beta, \textrm{s} }$ & $ \mathcal{N}_{1/2} \left( \rho_{ \sigma, \textrm{{s}}, \beta} \right) $ \\
\hline
$ l_{ \alpha, d} $ & $ \mathcal{N}_\textrm{TR} \left( \gamma_{ l, \alpha}, \rho_{ l, \alpha}, B_{\alpha} \right)$ \\
$ l_{ \beta, d} $ & $ \mathcal{N}_\textrm{TR} \left( \gamma_{ l, \alpha}, \rho_{ l, \beta}, B_\beta \right) $
\\
\bottomrule
\end{tabular}
\caption{Gaussian process hyperparameter prior distribution specifications and their parameterizations. The constant means are modeled with normal distributions $\mathcal{N} \left( \gamma_{ \mu, \bullet}, \rho_{ \mu, \bullet} \right)$ parameterized according to means $\gamma_{ \mu, \bullet}$ and standard deviations $\rho_{ \mu, \bullet}$. The error and signal variances are modelled as half-normal distributions $\mathcal{N}_{1/2} \left( \rho_{ \sigma, \bullet, \bullet} \right)$ with $\rho_{\sigma, \bullet, \bullet}$ denoting their respective standard deviations. The length scales are assigned truncated normal distributions $ \mathcal{N}_\textrm{TR} \left( \gamma_{ l, \bullet}, \rho_{ l, \bullet}, B_\bullet \right)$ with means $\gamma_{ l, \bullet}$, standard deviations $\rho_{l,\bullet}$, and lower bounds $B_\bullet$.}
\label{table:hyperparameterPriorDistributions}
\end{table}
For predictions $\boldsymbol{y}^* = ( y_1^*, \dots, y_{K^*}^* )^\intercal$, $y_{k^*} \in \mathbb{R}_+$, at $K^*$ unobserved locations $\boldsymbol{x}^* = ( x_1^*, \dots, x_{K^*}^* )^\intercal$, $x_{k^*} \in \mathbb{R}^D$, we first construct the posterior predictive distribution for the log-shape and log-rate processes at $\boldsymbol{x}^*$.
We denote the predictive log-shape and log-rate processes at $\boldsymbol{x}^*$ as $\boldsymbol{\alpha}^* := \alpha( \boldsymbol{x}^* ) = \left( \alpha( x_1^* ), \dots, \alpha( x_{K^*}^* ) \right)^\intercal $ and $\boldsymbol{\beta}^* := \boldsymbol{\beta}( \boldsymbol{x}^* ) = \left( \beta( x_1^*), \dots, \beta( x_{K^*} ) \right)^\intercal$.
In general, the posterior predictive distribution $\pi( \boldsymbol{\alpha}^*, \boldsymbol{\beta}^* \mid \boldsymbol{y} )$ can be obtained via marginalization with respect to the posterior distribution $\pi( \boldsymbol{\alpha}, \boldsymbol{\beta}, \mu_\alpha, \theta_\alpha, \mu_\beta, \theta_\beta \mid \boldsymbol{y} )$ in Equation~\eqref{eq:lggpPosteriorDistribution}
\begin{equation}
    \pi( \boldsymbol{\alpha}^*, \boldsymbol{\beta}^* \mid \boldsymbol{y} ) = \int \pi( \boldsymbol{\alpha}^*, \boldsymbol{\beta}^* \mid \boldsymbol{\alpha}, \boldsymbol{\beta}, \mu_\alpha, \theta_\alpha, \mu_\beta, \theta_\beta ) \pi( \boldsymbol{\alpha}, \boldsymbol{\beta}, \mu_\alpha, \theta_\alpha, \mu_\beta, \theta_\beta \mid \boldsymbol{y} ) \textrm{d}\omega,
    \label{eq:alphaBetaPosteriorPredictiveDistribution}
\end{equation}
where $\textrm{d}\omega = \textrm{d}\boldsymbol{\alpha} \textrm{d}\boldsymbol{\beta} \textrm{d}\mu_\alpha \textrm{d}\mu_\beta \textrm{d}\theta_\alpha \textrm{d}\theta_\beta$.
Due to the GP assumption on $\boldsymbol{\alpha}$ and $\boldsymbol{\beta}$ and with fixed $\boldsymbol{\alpha}, \boldsymbol{\beta}, \mu_\alpha, \theta_\alpha, \mu_\beta$, and $\theta_\beta$, we have a closed-form solution for the predictive distributions of the log-shape $\boldsymbol{\alpha}^*$ and the log-rate $\boldsymbol{\beta}^*$
\begin{equation}
\begin{split}
    \pi( \boldsymbol{\alpha}^* \mid \boldsymbol{\alpha}, \mu_\alpha, \theta_\alpha ) &= \mathcal{N} \big( \mu_\alpha^*( \boldsymbol{x}^*; \boldsymbol{\alpha}, \mu_\alpha, \theta_\alpha), \Sigma_\alpha^*( \boldsymbol{x}^*, \boldsymbol{x}^*; \theta_\alpha) \big),\\
    \pi( \boldsymbol{\beta}^* \mid \boldsymbol{\beta}, \mu_\beta, \theta_\beta ) &= \mathcal{N} \big( \mu_\beta^*( \boldsymbol{x}^*; \boldsymbol{\beta}, \mu_\beta, \theta_\beta), \Sigma_\beta^*( \boldsymbol{x}^*, \boldsymbol{x}^*; \theta_\beta) \big),
\end{split}
\label{eq:alphaBetaPredictiveGP}
\end{equation}
with predictive means $\mu_\alpha^*( \boldsymbol{x}^*; \boldsymbol{\alpha}, \mu_\alpha, \theta_\alpha)$ and $\mu_\beta^*( \boldsymbol{x}^*; \boldsymbol{\beta}, \mu_\beta, \theta_\beta)$ given by
\begin{equation}
\begin{split}
    \mu_\alpha^*( \boldsymbol{x}^*; \boldsymbol{\alpha}, \mu_\alpha, \theta_\alpha) &= \Sigma_\alpha\left( \boldsymbol{x}^*, \boldsymbol{x}; \theta_\alpha \right) \left( \Sigma_\alpha( \boldsymbol{x}, \boldsymbol{x}; \theta_\alpha) + \sigma_{ \alpha, \textrm{e} }^2 \boldsymbol{I} \right)^{-1} \left( \boldsymbol{\alpha} - \mu_\alpha \right) + \mu_\alpha, \\
    \mu_\beta^*( \boldsymbol{x}^*; \boldsymbol{\beta}, \mu_\beta, \theta_\beta) &= \Sigma_\beta\left( \boldsymbol{x}^*, \boldsymbol{x}; \theta_\beta \right) \left( \Sigma_\beta( \boldsymbol{x}, \boldsymbol{x}; \theta_\beta) + \sigma_{ \beta, \textrm{e} }^2 \boldsymbol{I} \right)^{-1} \left( \boldsymbol{\beta} - \mu_\beta \right) + \mu_\beta,
\end{split}
\label{eq:predictiveMeansGP}
\end{equation}
with their respective predictive covariances $\Sigma_\alpha^*( \boldsymbol{x}^*, \boldsymbol{x}^*; \theta_\alpha)$ and $\Sigma_\beta^*( \boldsymbol{x}^*, \boldsymbol{x}^*; \theta_\beta)$ as
\begin{equation}
\begin{split}
    \Sigma_\alpha^*( \boldsymbol{x}^*, \boldsymbol{x}^*; \theta_\alpha) &= \Sigma_\alpha\left( \boldsymbol{x}^*, \boldsymbol{x}^*; \theta_\alpha \right) - \Sigma_\alpha\left( \boldsymbol{x}^*, \boldsymbol{x}; \theta_\alpha \right) \left( \Sigma_\alpha( \boldsymbol{x}, \boldsymbol{x}; \theta_\alpha) + \sigma_{ \alpha, \textrm{e} }^2 \boldsymbol{I} \right)^{-1} \\
    &\hspace{6cm}\times \Sigma_\alpha\left( \boldsymbol{x}^*, \boldsymbol{x}; \theta_\alpha \right)^\intercal, \\
    \Sigma_\beta^*( \boldsymbol{x}^*, \boldsymbol{x}^*; \theta_\beta) &= \Sigma_\beta\left( \boldsymbol{x}^*, \boldsymbol{x}^*; \theta_\beta \right) - \Sigma_\beta\left( \boldsymbol{x}^*, \boldsymbol{x}; \theta_\beta \right) \left( \Sigma_\beta( \boldsymbol{x}, \boldsymbol{x}; \theta_\beta) + \sigma_{ \beta, \textrm{e} }^2 \boldsymbol{I} \right)^{-1} \\
    &\hspace{6cm}\times \Sigma_\beta\left( \boldsymbol{x}^*, \boldsymbol{x}; \theta_\beta \right)^\intercal,
\end{split}
\label{eq:predictiveMeansCovariancesGP}
\end{equation}
which constitute $\pi( \boldsymbol{\alpha}^*, \boldsymbol{\beta}^* \mid \boldsymbol{\alpha}, \boldsymbol{\beta}, \mu_\alpha, \theta_\alpha, \mu_\beta, \theta_\beta ) = \pi( \boldsymbol{\alpha}^* \mid \boldsymbol{\alpha}, \mu_\alpha, \theta_\alpha ) \pi( \boldsymbol{\beta}^* \mid \boldsymbol{\beta}, \mu_\beta, \theta_\beta )$.

Given the predictive distribution $\pi( \boldsymbol{\alpha}^*, \boldsymbol{\beta}^* \mid \boldsymbol{y} )$ in Equation~\eqref{eq:alphaBetaPosteriorPredictiveDistribution}, we can further construct a predictive distribution $\pi( \boldsymbol{y}^* \mid \boldsymbol{y} )$ for new measurements $\boldsymbol{y}^*$ by marginalizing the observation model in Equation~\eqref{eq:dataDistribution} over the predictive distribution
\begin{equation}
    \pi( \boldsymbol{y}^* \mid \boldsymbol{y} ) = \int \pi\big( \boldsymbol{y}^* \mid \boldsymbol{\alpha}^*, \boldsymbol{\beta}^* \big) \pi( \boldsymbol{\alpha}^*, \boldsymbol{\beta}^* \mid \boldsymbol{y} ) \textrm{d} \boldsymbol{\alpha}^* \textrm{d} \boldsymbol{\beta}^*.
    \label{eq:dataPosteriorPredictiveDistribution}
\end{equation}
The posterior and predictive distributions in Equations~\eqref{eq:lggpPosteriorDistribution}, \eqref{eq:alphaBetaPosteriorPredictiveDistribution}, and \eqref{eq:dataPosteriorPredictiveDistribution} are analytically intractable.
Thus, we must resort to numerical sampling.
We discuss the chosen and proposed sampling methods in the following Sections.
\section{Hamiltonian Monte Carlo}
\label{sec:hmc}
Markov chain Monte Carlo (MCMC) methods \citep{Brooks:2011} are the gold standard for estimating posterior distributions.
While powerful and proven to converge to the target distribution given unlimited computational resources, standard MCMC implementations are known to suffer with high-dimensional posterior distributions  \citep{Cui:2016, Morzfeld:2019} which are typical for hierarchical Bayesian models  \citep{Turek:2016}.

Hamiltonian Monte Carlo (HMC) \citep{Brooks:2011}, or hybrid Monte Carlo \citep{Duane:1987}, circumvents the problems of standard MCMC methodology by combining the posterior distribution of interest, here given in Equation~\eqref{eq:lggpPosteriorDistribution}, with artificial momentum variables for each variable of interest in the posterior distribution  \citep{Brooks:2011}.
Denoting the vector of the artificial momentum variables as $\boldsymbol{p} \in \mathbb{R}^{ \left( 2K + 2D + 6 \right) \times 1}$, we can formulate an augmented joint posterior distribution
\begin{equation}
\begin{split}
    \pi( \boldsymbol{\alpha}, \boldsymbol{\beta}, \mu_\alpha, \theta_\alpha, \mu_\beta, \theta_\beta, \boldsymbol{p} \mid \boldsymbol{y} ) \propto \exp\left\{ U( \boldsymbol{\alpha}, \boldsymbol{\beta}, \mu_\alpha, \theta_\alpha, \mu_\beta, \theta_\beta \mid \boldsymbol{y} ) \right\} \exp\left\{ Q\left( \boldsymbol{p} \right) \right\},
\end{split}
\label{eq:hmcAugmentedPosteriorDistribution}
\end{equation}
where $U( \cdot )$ is the log-posterior distribution
\begin{equation}
    U( \boldsymbol{\alpha}, \boldsymbol{\beta}, \mu_\alpha, \theta_\alpha, \mu_\beta, \theta_\beta \mid \boldsymbol{y} ) = -\log \pi( \boldsymbol{\alpha}, \boldsymbol{\beta}, \mu_\alpha, \theta_\alpha, \mu_\beta, \theta_\beta \mid \boldsymbol{y} ),
    \label{eq:logPosteriorLGGP}
\end{equation}
and $ Q( \boldsymbol{p} ) $ denotes kinetic energy in Hamiltonian mechanics.
The typical HMC implementation sets $Q( \boldsymbol{p} )$ to correspond to the form of a zero-mean Gaussian log-probability 
\begin{equation}
    Q( \boldsymbol{p} ) = -\frac{1}{2}\boldsymbol{p}^\intercal \boldsymbol{M}^{-1} \boldsymbol{p},
    \label{eq:hmcMomentumDistribution}
\end{equation}
with covariance $\boldsymbol{M} \in \mathbb{R}^{ (2K + 2D + 6) \times (2K + 2D + 6) }$.
Proposals for the augmented posterior distribution in Equation~\eqref{eq:hmcAugmentedPosteriorDistribution} are generated by solving a system of differential equations describing Hamiltonian dynamics in time $\tau$
\begin{equation}
\begin{split}
    \frac{\textrm{d} \boldsymbol{z} }{ \textrm{d} \tau} = -\Upsilon \nabla H( \boldsymbol{z} ) = -\Upsilon \nabla U( \boldsymbol{\alpha}, \boldsymbol{\beta}, \mu_\alpha, \theta_\alpha, \mu_\beta, \theta_\beta \mid \boldsymbol{y} ) Q( \boldsymbol{p} ),
\end{split}
\label{eq:hamiltonianDynamics}
\end{equation}
where $\nabla$ denotes the gradient with respect to $\boldsymbol{\alpha}, \boldsymbol{\beta}, \mu_\alpha, \theta_\alpha, \mu_\beta, \theta_\beta, $ and $\boldsymbol{p}$ and
\begin{equation}
    \Upsilon = \begin{pmatrix}
        \boldsymbol{0} & \widetilde{\boldsymbol{I}}\\
        -\widetilde{\boldsymbol{I}} & \boldsymbol{0}
    \end{pmatrix},
    \label{eq:dynamicsMapping}
\end{equation}
where $ \widetilde{\boldsymbol{I}} $ is an appropriately sized identity matrix, $ \widetilde{\boldsymbol{I}} \in \mathbb{R}^{ (2K + 2D + 6) \times (2K + 2D + 6) }$.

Numerically, Equation~\eqref{eq:hamiltonianDynamics} can be solved with any suitable integrator.
In practice, the simple leapfrog integration scheme is used. 
The leapfrog integrator involves setting a step size and the number of performed integration steps.
These two parameters are noted to be of vital importance for efficient application of the HMC sampler  \citep{Brooks:2011}.
The No-U-Turn Sampler (NUTS) enables automatic specification of the number of integration steps and combines this with the dual averaging algorithm  \citep{Nesterov:2009} for setting the step size.
For a more detailed description of HMC and the NUTS sampler, please refer to the book  \citep{Brooks:2011} and paper \citep{Hoffman:2014}, respectively.
For our numerical examples, we use the PyMC implementation of the NUTS sampler  \citep{Abril-Pla:2023}.

While inarguably a powerful and effective sampling tool, NUTS is still computationally demanding due to the repeated need to evaluate the log-posterior distribution gradient in Equation~\eqref{eq:hamiltonianDynamics} over the leapfrog steps.
While this can be performed via automatic differentiation \citep{Margossian:2019}, the computational overhead nevertheless exists.
In the following Section, we propose two sampling schemes, an approximate and an exact scheme, for estimating the posterior distribution $ \pi( \boldsymbol{\alpha}, \boldsymbol{\beta}, \mu_\alpha, \theta_\alpha, \mu_\beta, \theta_\beta \mid \boldsymbol{y} ) $  of the log-Gaussian gamma process using iterated posterior linearization.
\section{Iterated posterior linearization}
\label{sec:iteratedPosteriorLinearization}
Iterated posterior linearization \citep{Garcia-Fernandez:2015, Garcia:2017, Tronarp+Fernandez+Sarkka:2018} was developed as a generalization to statistical linear regression and statistical linear regression  \citep{Gelb:1974, Lefebvre+Bruyninckx+DeSchutter:2002}. The core idea is to find a linearization that is statistically optimal with respect to a given parameter distribution, which is taken to be the approximate posterior distribution. 

In this study, we perform iterated posterior linearization to approximate the marginal posterior distribution of the log-shape and log-rate processes
\begin{equation}
    \pi( \boldsymbol{\alpha}, \boldsymbol{\beta} \mid \boldsymbol{y} ) = \int \pi( \boldsymbol{\alpha}, \boldsymbol{\beta}, \mu_\alpha, \theta_\alpha, \mu_\beta, \theta_\beta \mid \boldsymbol{y} ) \textrm{d}\mu_\alpha \textrm{d}\mu_\beta \textrm{d}\theta_\alpha \textrm{d}\theta_\beta,
    \label{eq:processMarginalLggpPosteriorDistribution}
\end{equation}
which we model as an affine relationship between the processes and the data, $\boldsymbol{y}$.
This affine relationship is modeled as
\begin{equation}
    \boldsymbol{y} \simeq \boldsymbol{A}^{(T)} \begin{bmatrix}
        \boldsymbol{\alpha}^{(T)} \\ \boldsymbol{\beta}^{(T)}
    \end{bmatrix} + \boldsymbol{b}^{(T)} + \boldsymbol{e},
    \label{eq:affineApproximation}
\end{equation}
where $\boldsymbol{A}^{(T)}  \in \mathbb{R}^{K \times 2K}$ denote linear coefficients, $\boldsymbol{b}^{(T)}  \in \mathbb{R}^{K \times 1} $ the bias, and $\boldsymbol{e} \sim \mathcal{N}\left( 0, \boldsymbol{\Lambda}^{(T)} \right)  \in \mathbb{R}^{K \times 1} $ is an error vector distributed according to covariance $\boldsymbol{\Lambda}^{(T)}  \in \mathbb{R}^{K \times K}$ at iteration $T$.
The affine relationship results in an approximate marginal posterior distribution $\pi^{(T)}( \boldsymbol{\alpha}, \boldsymbol{\beta} \mid \boldsymbol{y} )$
\begin{equation}
    \pi( \boldsymbol{\alpha}, \boldsymbol{\beta} \mid \boldsymbol{y} ) \approx \pi^{(T)}( \boldsymbol{\alpha}, \boldsymbol{\beta} \mid \boldsymbol{y} ) = \mathcal{N}\left( \boldsymbol{m}^{(T)}, \boldsymbol{P}^{(T)} \right),
    \label{eq:alphaBetaApproximateMarginalPosteriorFinal}
\end{equation}
where the posterior mean $\boldsymbol{m}^{(T)} \in \mathbb{R}^{2K \times 1}$ and covariance $\boldsymbol{P}^{(T)} \in \mathbb{R}^{2K \times 2K}$, jointly for $\boldsymbol{\alpha}$ and $\boldsymbol{\beta}$, are constructed iteratively as follows.

The initial mean and covariance at $t = 0$ are obtained as the expected mean and covariance of $\boldsymbol{\alpha}$ and $\boldsymbol{\beta}$ with respect to the prior distribution $ \pi_0( \boldsymbol{\alpha} \mid \mu_\alpha, \theta_\alpha ) \pi_0( \boldsymbol{\beta} \mid \mu_\beta, \theta_\beta ) \pi_0( \mu_\alpha, \theta_\alpha) \pi_0( \mu_\beta, \theta_\beta) $
\begin{equation}
\begin{split}
    \boldsymbol{m}^{(0)} &= \int \begin{bmatrix}
        \boldsymbol{\alpha} \\ \boldsymbol{\beta}
    \end{bmatrix} \pi_0( \boldsymbol{\alpha} \mid \mu_\alpha, \theta_\alpha ) \pi_0( \boldsymbol{\beta} \mid \mu_\beta, \theta_\beta ) \pi_0( \mu_\alpha, \theta_\alpha) \pi_0( \mu_\beta, \theta_\beta) \textrm{d}\mu_\alpha \textrm{d}\mu_\beta \textrm{d}\theta_\alpha \textrm{d}\theta_\beta,\\
    \boldsymbol{P}^{(0)} &= \int \left( \begin{bmatrix}
        \boldsymbol{\alpha} \\ \boldsymbol{\beta}
    \end{bmatrix} - \boldsymbol{m}^{(0)} \right) \left( \begin{bmatrix}
        \boldsymbol{\alpha} \\ \boldsymbol{\beta}
    \end{bmatrix} - \boldsymbol{m}^{(0)} \right)^\intercal \pi_0( \boldsymbol{\alpha} \mid \mu_\alpha, \theta_\alpha ) \pi_0( \boldsymbol{\beta} \mid \mu_\beta, \theta_\beta ) \\ 
    &\hspace{6.5cm} \times \pi_0( \mu_\alpha, \theta_\alpha) \pi_0( \mu_\beta, \theta_\beta) \textrm{d}\mu_\alpha \textrm{d}\mu_\beta \textrm{d}\theta_\alpha \textrm{d}\theta_\beta.
\end{split}
\label{eq:posteriorLinearizationInitialization}
\end{equation}
For subsequent iterations $0 < t \leq T$, the approximate marginal posterior distribution is
\begin{equation}
    \pi^{(t)}( \boldsymbol{\alpha}, \boldsymbol{\beta} \mid \boldsymbol{y} ) = \mathcal{N}\left( \begin{bmatrix}
        \boldsymbol{\alpha} \\ \boldsymbol{\beta}
    \end{bmatrix};  \boldsymbol{m}^{(t)}, \boldsymbol{P}^{(t)} \right),
    \label{eq:alphaBetaApproximateMarginalPosterior}
\end{equation}
where the mean $\boldsymbol{m}^{(t)}$ and covariance $\boldsymbol{P}^{(t)}$ are given by 
\begin{equation}
\begin{split}
    \boldsymbol{m}^{(t)} &= \boldsymbol{m}^{(0)} + \boldsymbol{K}^{(t)} \left( \boldsymbol{y} - \boldsymbol{\mu}^{(t)} \right),\\
    \boldsymbol{P}^{(t)} &= \boldsymbol{P}^{(0)} - \boldsymbol{K}^{(t)} \boldsymbol{S}^{(t)} \left[ \boldsymbol{K}^{(t)} \right]^\intercal,
\end{split}
\label{eq:plMomentUpdate}
\end{equation}
where $ \boldsymbol{\mu}^{(t)} \in \mathbb{R}^{K \times 1} $, $\boldsymbol{S}^{(t)} \in \mathbb{R}^{K \times K}$, and $\boldsymbol{K}^{(t)} \in \mathbb{R}^{2K \times K} $ are defined by
\begin{equation}
\begin{split}
    \boldsymbol{\mu}^{(t)} &= \boldsymbol{A}^{(t)} \boldsymbol{m}^{(0)} + \boldsymbol{b}^{(t)},\\
    \boldsymbol{S}^{(t)} &= \boldsymbol{A}^{(t)} \boldsymbol{P}^{(0)} \left[ \boldsymbol{A}^{(t)} \right]^\intercal + \boldsymbol{\Lambda}^{(t)},\\
    \boldsymbol{K}^{(t)} &= \boldsymbol{P}^{(0)} \left[ \boldsymbol{A}^{(t)} \right]^\intercal \left[ \boldsymbol{S}^{(t)} \right]^{-1},
\end{split}
\label{eq:plMomentUpdateParameters}
\end{equation}
with the statistical linear regression parameters $ \boldsymbol{A}^{(t)}$, $ \boldsymbol{b}^{(t)}$, and $ \boldsymbol{\Lambda}^{(t)}$ as
\begin{equation}
\begin{split}
    \boldsymbol{A}^{(t)} &= \left[\boldsymbol{P}^{ \alpha\beta y, (t - 1)} \right]^\intercal \left[ \boldsymbol{P}^{(t-1)} \right]^{-1}, \\
    \boldsymbol{b}^{(t)} &= \boldsymbol{\mu}^{+,(t - 1)} - \boldsymbol{A}^{(t)} \boldsymbol{m}^{(t - 1)}, \\
    \boldsymbol{\Lambda}^{(t)} &= \boldsymbol{P}^{yy,(t - 1)} - \boldsymbol{A}^{(t)} \boldsymbol{P}^{(t - 1)} \left[ \boldsymbol{A}^{(t)} \right]^\intercal, \\
\end{split}
\label{eq:slrParameters}
\end{equation}
where $ \boldsymbol{\mu}^{+,(t - 1)} \in \mathbb{R}_+^{K \times 1}$, $ \boldsymbol{P}^{yy,(t - 1)} \in \mathbb{R}^{K \times K}$, and $ \boldsymbol{P}^{ \alpha\beta y, (t - 1)} \in \mathbb{R}^{2K \times K}$ denote the predicted mean, covariance, and cross-covariance.
The predicted mean, covariance, and cross-covariance are in turn obtained through linearizing the model with respect to the previous best approximate marginal posterior distribution $\pi( \boldsymbol{\alpha}, \boldsymbol{\beta} \mid \boldsymbol{y} ) \approx \mathcal{N}\left( \begin{bmatrix}
        \boldsymbol{\alpha} \\ \boldsymbol{\beta}
    \end{bmatrix}; \boldsymbol{m}^{(t - 1)}, \boldsymbol{P}^{(t - 1)} \right) $
\begin{equation}
\begin{split}
    \boldsymbol{\mu}^{+,(t - 1)} &= \int \boldsymbol{y} \pi\left( \boldsymbol{y} \mid \boldsymbol{\alpha}^{(t - 1)}, \boldsymbol{\beta}^{(t - 1)} \right) \mathcal{N}\left( \begin{bmatrix}
        \boldsymbol{\alpha} \\ \boldsymbol{\beta}
    \end{bmatrix}; \boldsymbol{m}^{(t - 1)}, \boldsymbol{P}^{(t - 1)} \right) \textrm{d}\boldsymbol{y} \textrm{d} \boldsymbol{\alpha}^{(t - 1)} \textrm{d} \boldsymbol{\beta}^{(t - 1)},\\
    \boldsymbol{P}^{yy,(t - 1)} &= \int \left( \boldsymbol{y} - \boldsymbol{\mu}^{+,(t - 1)} \right) \Big( \boldsymbol{y} - \boldsymbol{\mu}^{+,(t - 1)} \Big)^\intercal \pi\left( \boldsymbol{y} \mid \boldsymbol{\alpha}^{(t - 1)}, \boldsymbol{\beta}^{(t - 1)} \right)\\
    &\hspace{5cm}\times \mathcal{N}\left( \begin{bmatrix}
        \boldsymbol{\alpha} \\ \boldsymbol{\beta}
    \end{bmatrix}; \boldsymbol{m}^{(t - 1)}, \boldsymbol{P}^{(t - 1)} \right) \textrm{d}\boldsymbol{y} \textrm{d} \boldsymbol{\alpha}^{(t - 1)} \textrm{d} \boldsymbol{\beta}^{(t - 1)},\\
    \boldsymbol{P}^{ \alpha\beta y, (t - 1)} &= \int \left(  \begin{bmatrix}
        \boldsymbol{\alpha}^{(t - 1)} \\\boldsymbol{\beta}^{(t - 1)}
    \end{bmatrix} - \boldsymbol{m}^{(t - 1)} \right) \Big( \boldsymbol{y} - \boldsymbol{\mu}^{+,(t - 1)} \Big)^\intercal \pi\left( \boldsymbol{y} \mid \boldsymbol{\alpha}^{(t - 1)}, \boldsymbol{\beta}^{(t - 1)} \right)\\ &\hspace{5cm}\times \mathcal{N}\left( \begin{bmatrix}
        \boldsymbol{\alpha} \\ \boldsymbol{\beta}
    \end{bmatrix};  \boldsymbol{m}^{(t - 1)}, \boldsymbol{P}^{(t - 1)} \right) \textrm{d}\boldsymbol{y} \textrm{d} \boldsymbol{\alpha}^{(t - 1)} \textrm{d} \boldsymbol{\beta}^{(t - 1)}.
\end{split}
\label{eq:plLinearization}
\end{equation}
In our numerical examples, we approximate the prior covariance $\boldsymbol{P}^{(0)}$ and the integrals in Equation~\eqref{eq:plLinearization} with Monte Carlo integration.
Note that the initial mean $\boldsymbol{m}^{(0)}$ is completely defined by the means of the prior distributions $\pi_0( \mu_\alpha )$ and $\pi_0( \mu_\beta )$.
Extended discussion and further details on iterated posterior linearization can be found in existing literature \citep{Garcia-Fernandez:2015, Garcia:2017, Tronarp+Fernandez+Sarkka:2018, BayesianFilteringSmoothing:2023}.

\begin{figure}
    \centering
    \includegraphics[width = 0.49\textwidth]{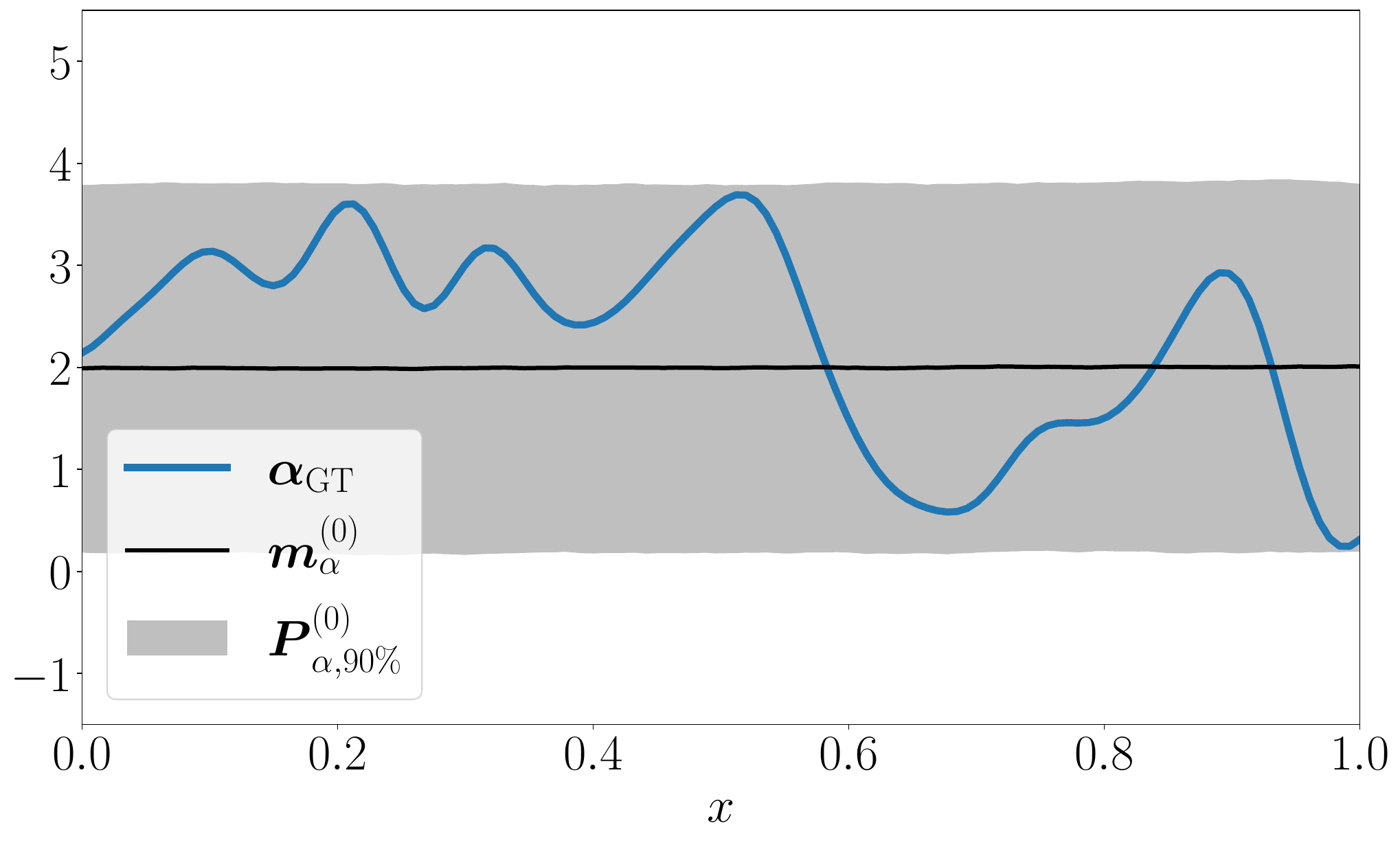}
    \includegraphics[width = 0.49\textwidth]{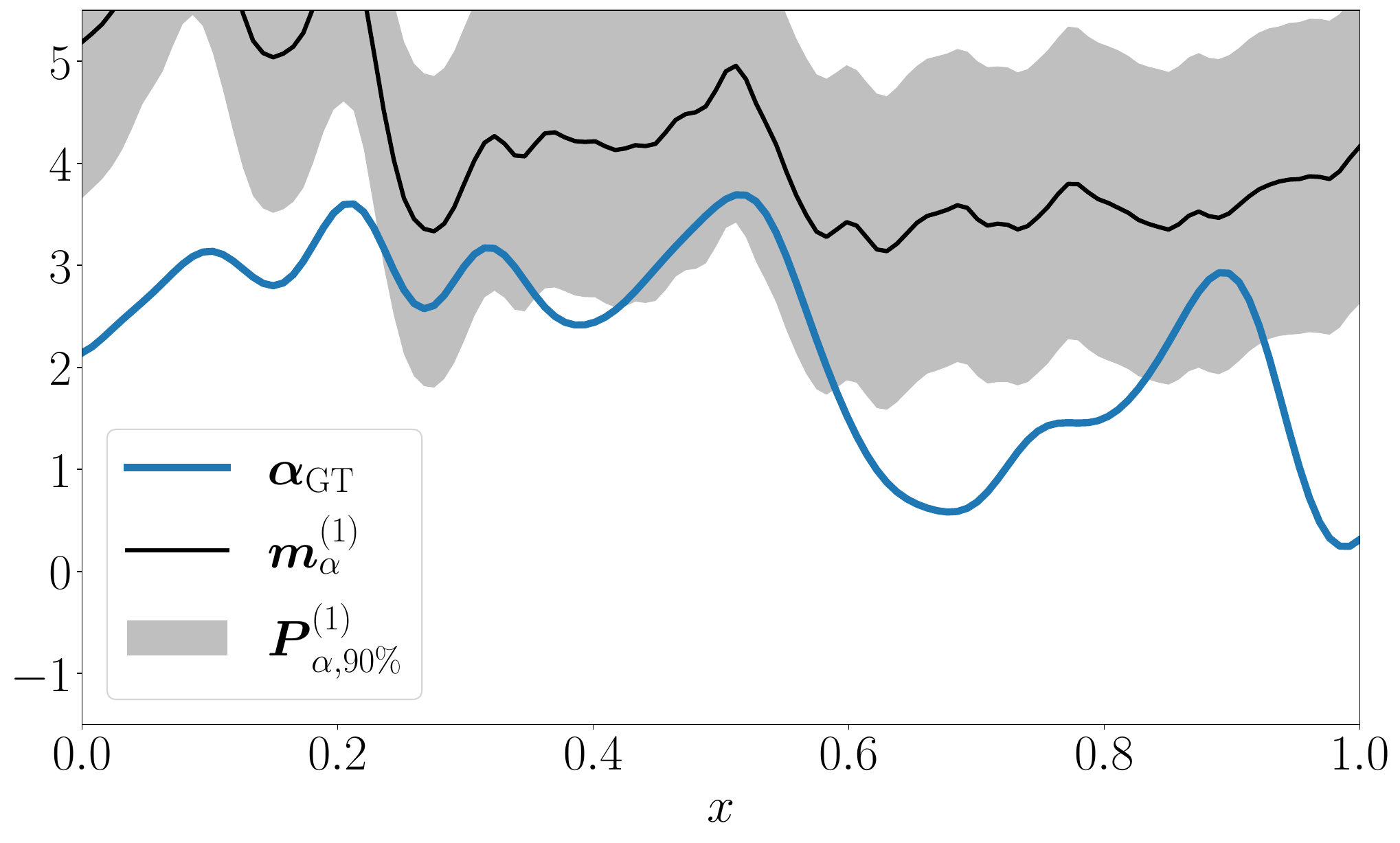}
    \includegraphics[width = 0.49\textwidth]{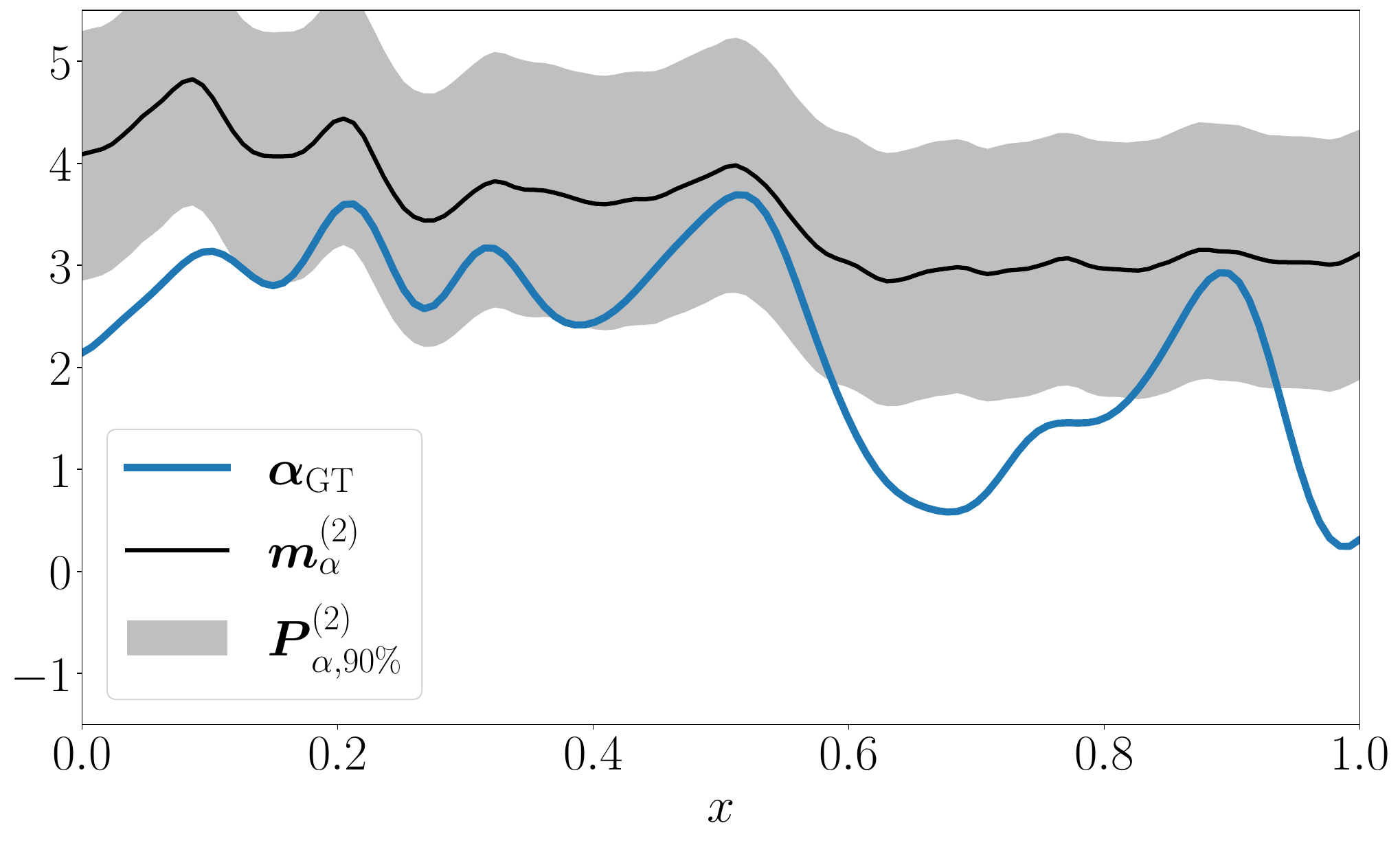}
    \includegraphics[width = 0.49\textwidth]{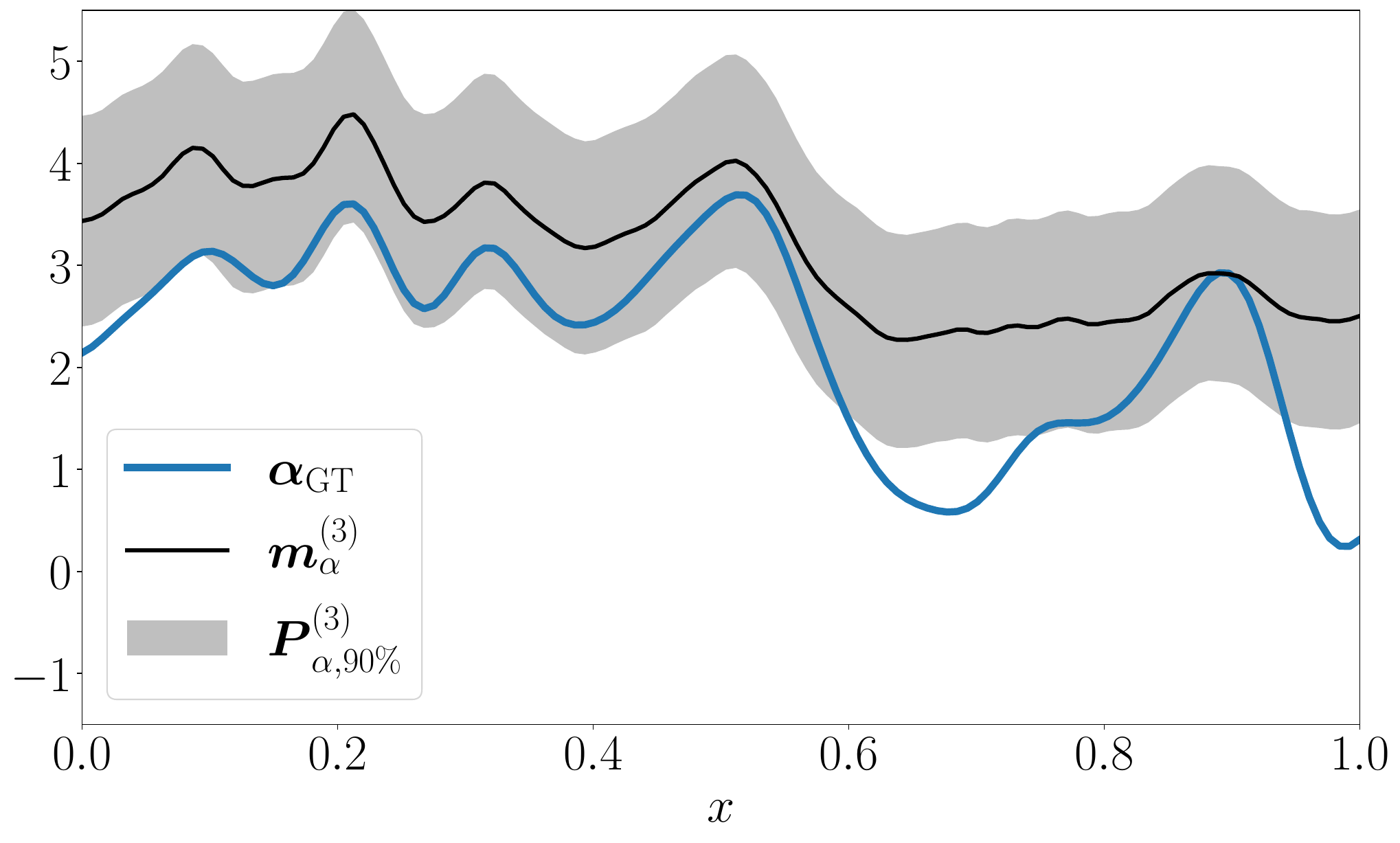}
    \includegraphics[width = 0.49\textwidth]{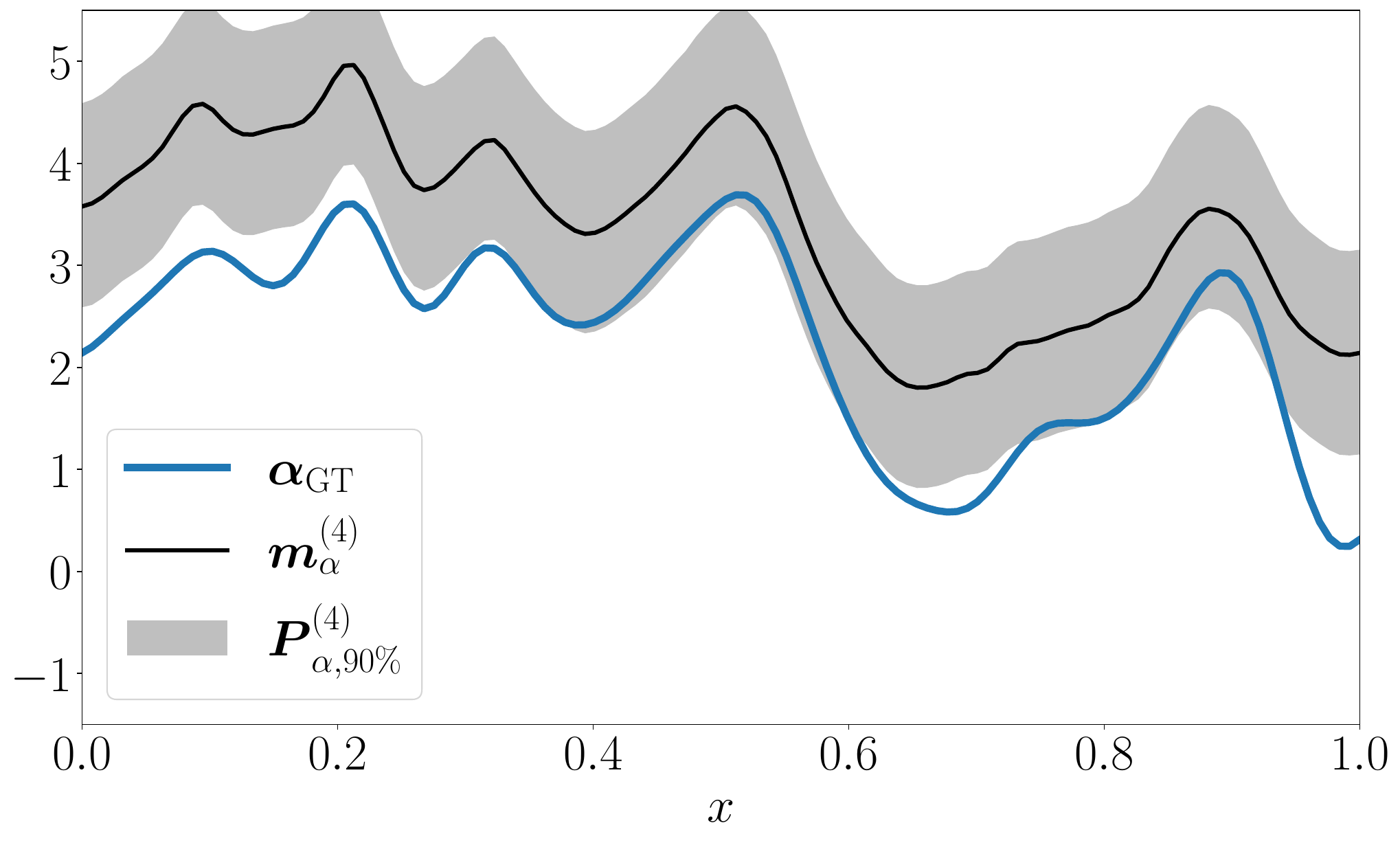}
    \includegraphics[width = 0.49\textwidth]{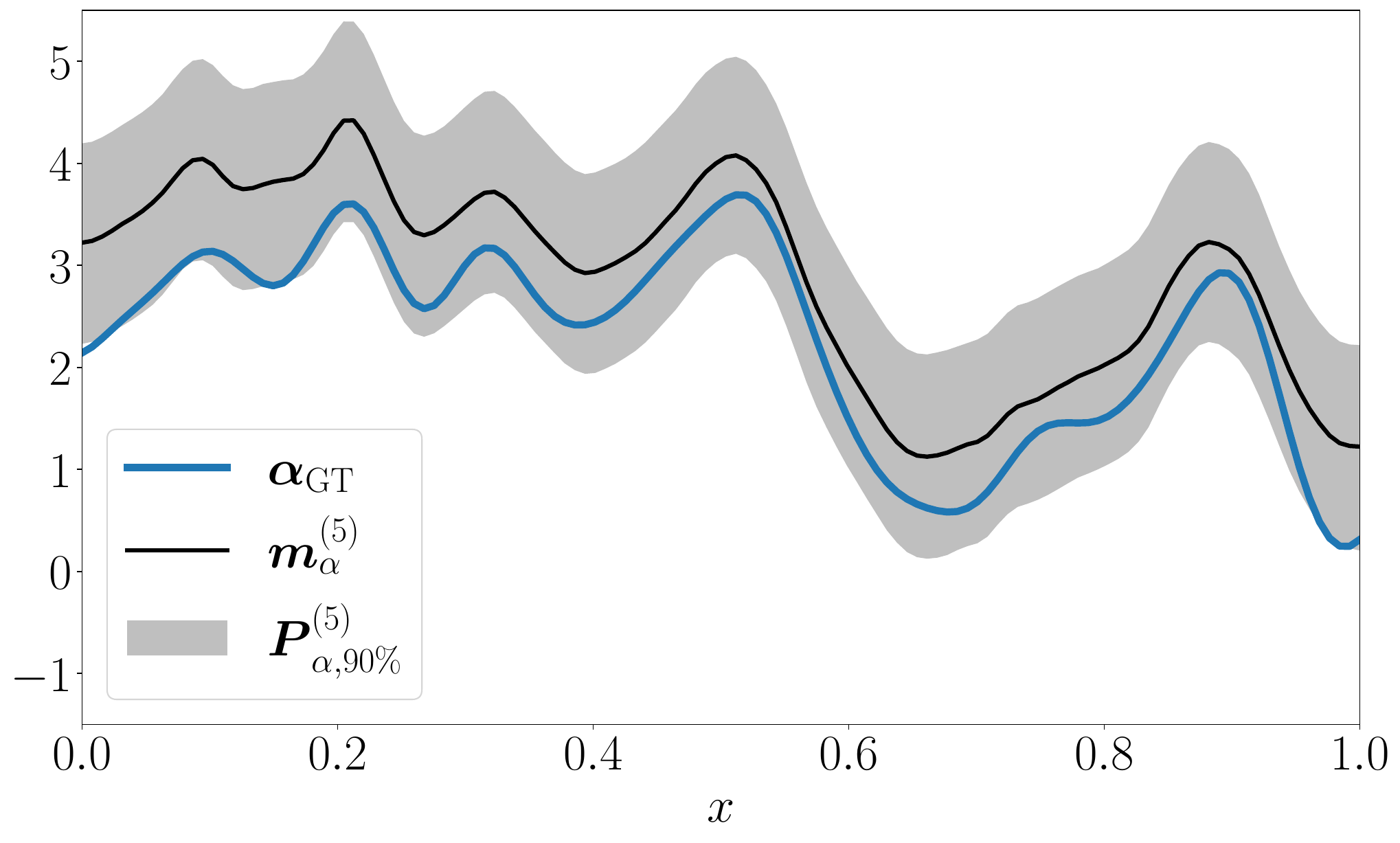}
    \caption{An example sequence of iterated posterior linearization for the log-shape process $\boldsymbol{\alpha}$ corresponding to the case shown in Figure \ref{im:lggpDiagram} with $T = 5$ iterations. Starting from the prior distribution at $t = 0$, the mean and covariance estimates are iteratively updated and refined until convergence. The 90\% marginal credible interval defined by the covariance $\boldsymbol{P}_{\alpha}^{(t)}$ is denoted by $\boldsymbol{P}_{\alpha, 90\%}^{(t)}$.} 
    \label{im:iteratedPosteriorLinearizationSequenceExample}
\end{figure}

For the prior covariance $\boldsymbol{P}^{(0)}$ estimator $ \hat{ \boldsymbol{P} }^{(0)}$, we generate ensembles $\left( \boldsymbol{\alpha}_1^{(0)}, \dots, \boldsymbol{\alpha}_J^{(0)} \right) \in \mathbb{R}^{K \times J}$ and $\left( \boldsymbol{\beta}_1^{(0)}, \dots, \boldsymbol{\beta}_J^{(0)} \right) \in \mathbb{R}^{K \times J}$  by first sampling the GP hyperparameter prior distribution $\pi_0( \mu_\alpha, \theta_\alpha, \mu_\beta, \theta_\beta)$ for $( \mu_\alpha, \theta_\alpha, \mu_\beta, \theta_\beta)_j^\intercal$ with $j \in ( 1, \dots, J)$.
For each $( \mu_\alpha, \theta_\alpha, \mu_\beta, \theta_\beta)_j^\intercal$, we draw realizations $\boldsymbol{\alpha}_j$ and $\boldsymbol{\beta}_j$ according to Equations~\eqref{eq:alphaPriorGP} and \eqref{eq:betaPriorGP}.
The prior covariance estimator $ \hat{ \boldsymbol{P} }^{(0)} $ can then be otained via Monte Carlo as
\begin{equation}
    \hat{ \boldsymbol{P} }^{(0)} = \frac{1}{J} \sum\limits_{j = 1}^J \left( \begin{bmatrix} \boldsymbol{\alpha}_1^{(0)}\\ \boldsymbol{\beta}_1^{(0)} \end{bmatrix} - \boldsymbol{m}^{(0)} \right) \left( \begin{bmatrix} \boldsymbol{\alpha}_1^{(0)}\\ \boldsymbol{\beta}_1^{(0)} \end{bmatrix} - \boldsymbol{m}^{(0)} \right)^\intercal.
    \label{eq:initialCovarianceMonteCarlo}
\end{equation}
For iterations $t > 0$, we draw ensembles $\left\{ \boldsymbol{\alpha}_1^{(t-1)}, \dots, \boldsymbol{\alpha}_J^{(t-1)} \right\}$ and $ \left\{ \boldsymbol{\beta}_1^{(t-1)}, \dots, \boldsymbol{\beta}_J^{(t-1)} \right\} $ from the approximate marginal posterior distribution $ \pi( \boldsymbol{\alpha}, \boldsymbol{\beta} \mid \boldsymbol{y} ) \approx \mathcal{N}\left( \begin{bmatrix}
        \boldsymbol{\alpha} \\ \boldsymbol{\beta}
    \end{bmatrix}; \boldsymbol{m}^{(t - 1)}, \boldsymbol{P}^{(t - 1)} \right) $ according to Equation~\eqref{eq:plMomentUpdate}. Given the log-shape and log-rate ensembles, we draw a data realization $ \boldsymbol{y}_j \sim \pi\left( \boldsymbol{y} \mid \boldsymbol{\alpha}_j^{(t - 1)}, \boldsymbol{\beta}_j^{(t - 1)}\right) $ to obtain an ensemble $ \left\{ \boldsymbol{y}_1^{(t-1)}, \dots, \boldsymbol{y}_J^{(t-1)} \right\} $.
With the above log-shape, log-rate, and data realization ensembles, the Monte Carlo estimators $\hat{ \boldsymbol{\mu} }^{+,(t - 1)}$, $\hat{ \boldsymbol{P} }^{yy,(t - 1)}$, and $\hat{ \boldsymbol{P} }^{ \alpha\beta y, (t - 1)}$  for the integrals in Equation~\eqref{eq:plLinearization} can be given as
\begin{equation}
\begin{split}
    \hat{ \boldsymbol{\mu} }^{+,(t - 1)} &= \frac{1}{J} \sum\limits_{j = 1}^J \boldsymbol{y}_j^{(t - 1)},\\
    \hat{ \boldsymbol{P} }^{yy,(t - 1)} &= \frac{1}{J} \sum\limits_{j = 1}^J \left( \boldsymbol{y}_j^{(t - 1)} - \hat{ \boldsymbol{\mu} }^{+,(t - 1)} \right) \left( \boldsymbol{y}_j^{(t - 1)} - \hat{ \boldsymbol{\mu} }^{+,(t - 1)} \right)^\intercal,\\
    \hat{ \boldsymbol{P} }^{ \alpha\beta y, (t - 1)} &= \frac{1}{J} \sum\limits_{j = 1}^J \left( \begin{bmatrix}
        \boldsymbol{\alpha}_j^{(t - 1)} \\\boldsymbol{\beta}_j^{(t - 1)}
    \end{bmatrix} - \boldsymbol{m}^{(t - 1)} \right) \left( \boldsymbol{y}_j - \hat{ \boldsymbol{\mu} }^{+,(t - 1)} \right)^\intercal,\\
\end{split}
\label{eq:plLinearizationMonteCarlo}
\end{equation}
where for $t = 1$, we can use the initial ensembles $\left\{ \boldsymbol{\alpha}_1^{(0)}, \dots, \boldsymbol{\alpha}_J^{(0)} \right\}$ and $ \left\{ \boldsymbol{\beta}_1^{(0)}, \dots, \boldsymbol{\beta}_J^{(0)} \right\} $.
With $T$ iterations of the above posterior linearization, we obtain approximate posterior means and covariances for the log-shape and log-rate processes as
\begin{equation}
\begin{split}
    \boldsymbol{m}^{(T)} =
    \begin{bmatrix}
        \boldsymbol{m}_\alpha^{(T)} \\ \boldsymbol{m}_\beta^{(T)} \end{bmatrix},
    \quad\boldsymbol{P}^{(T)} =
    \begin{bmatrix}
        \boldsymbol{P}^{(T)}_\alpha & \boldsymbol{P}^{(T)}_{\beta, \alpha} \\ \boldsymbol{P}^{(T)}_{\alpha,\beta} & \boldsymbol{P}^{(T)}_\beta
    \end{bmatrix},
\end{split}
\label{eq:alphaBetaIndependentMeanCovariance}
\end{equation}
where $\boldsymbol{P}^{(T)}_{\beta, \alpha} = \left[ \boldsymbol{P}^{(T)}_{\alpha, \beta} \right]^\intercal$ denote the cross-variance between the log-shape and log-rate processes.
An example of the above iterated posterior linearization sequence with $T = 5$ for $\boldsymbol{ \alpha }$ corresponding to the example shown in Figure \ref{im:lggpDiagram} is presented in Figure \ref{im:iteratedPosteriorLinearizationSequenceExample}.

Penultimately, we estimate the GP hyperparameters $\mu_\alpha, \theta_\alpha, \mu_\beta$, and $\theta_\beta$ using the approximate means $ \boldsymbol{m}^{(T)}_\alpha $ and $\boldsymbol{m}^{(T)}_\alpha$ as data for independent GPs with their respective covariances $\boldsymbol{P}^{(T)}_\alpha$ and $\boldsymbol{P}^{(T)}_\beta$ as synthetic error covariances.
This results in posterior distributions
\begin{equation}
\begin{split}
    \pi( \mu_\alpha, \theta_\alpha \mid \boldsymbol{y} ) &\propto \mathcal{L}( \boldsymbol{m}^{(T)}_\alpha, \boldsymbol{P}^{(T)}_\alpha \mid \mu_\alpha, \theta_\alpha) \pi_0( \mu_\alpha, \theta_\alpha),\\
    \pi( \mu_\beta, \theta_\beta \mid \boldsymbol{y} ) &\propto \mathcal{L}( \boldsymbol{m}^{(T)}_\beta, \boldsymbol{P}^{(T)}_\beta \mid \mu_\beta, \theta_\beta) \pi_0( \mu_\beta, \theta_\beta),
\end{split}
\label{eq:surrogatePosteriorAlphaBeta}
\end{equation}
where the prior distributions are defined in Table~\ref{table:hyperparameterPriorDistributions} and the likelihoods $ \mathcal{L}( \boldsymbol{m}^{(T)}_\alpha, \boldsymbol{P}^{(T)}_\alpha \mid \mu_\alpha, \theta_\alpha) $ and $\mathcal{L}( \boldsymbol{m}^{(T)}_\beta, \boldsymbol{P}^{(T)}_\beta \mid \mu_\beta, \theta_\beta)$ are given as
\begin{equation}
\begin{split}
    \mathcal{L}( \boldsymbol{m}^{(T)}_\alpha, \boldsymbol{P}^{(T)}_\alpha \mid \mu_\alpha, \theta_\alpha) = &\frac{1}{\sqrt{(2\pi)^K}} \left\vert \Sigma_\alpha( \boldsymbol{x}, \boldsymbol{x}; \theta_\alpha) + \boldsymbol{P}^{(T)}_\alpha \right\vert^{-1/2} \\ & \hspace{-1cm}\times\exp\left\{
     -\frac{1}{2} \left( \boldsymbol{m}^{(T)}_\alpha - \mu_\alpha \right)^T\left( \Sigma_\alpha( \boldsymbol{x}, \boldsymbol{x}; \theta_\alpha) + \boldsymbol{P}^{(T)}_\alpha \right)^{-1} \left( \boldsymbol{m}^{(T)}_\alpha - \mu_\alpha \right) \right\},\\
    \mathcal{L}( \boldsymbol{m}^{(T)}_\beta, \boldsymbol{P}^{(T)}_\beta \mid \mu_\beta, \theta_\beta) = &\frac{1}{\sqrt{(2\pi)^K}} \left\vert \Sigma_\beta( \boldsymbol{x}, \boldsymbol{x}; \theta_\beta) + \boldsymbol{P}^{(T)}_\beta \right\vert^{-1/2} \\ & \hspace{-1cm}\times\exp\left\{
     -\frac{1}{2} \left( \boldsymbol{m}^{(T)}_\beta - \mu_\beta \right)^T\left( \Sigma_\beta( \boldsymbol{x}, \boldsymbol{x}; \theta_\beta) + \boldsymbol{P}^{(T)}_\beta \right)^{-1} \left( \boldsymbol{m}^{(T)}_\beta - \mu_\beta \right) \right\}.
\end{split}
\label{eq:surrogatePosteriorAlphaBetaLikelihood}
\end{equation}

Lastly, to approximate the predictive distributions in Equations~\eqref{eq:alphaBetaPredictiveGP} and \eqref{eq:dataPosteriorPredictiveDistribution}, we can use the posterior linearization mean $\boldsymbol{m}^{(T)}$ and covariance $\boldsymbol{P}^{(T)}$ to draw an arbitrary number of posterior realizations $\boldsymbol{\alpha}_j$ and $\boldsymbol{\beta}_j$ as defined in Equation~\eqref{eq:alphaBetaApproximateMarginalPosteriorFinal}.
Given the realizations, we can compute the predictive means and covariance according to Equation~\eqref{eq:predictiveMeansCovariancesGP} with either a point estimate or samples of the GP hyperparameter posterior distributions $\pi( \mu_\alpha, \theta_\alpha \mid \boldsymbol{y} )$ and $\pi( \mu_\beta, \theta_\beta \mid \boldsymbol{y} )$.
We opt to use the posterior mean estimates, denoted by $\overline{\mu}_\alpha, \overline{\theta}_\alpha, \overline{\mu}_\beta$, and $\overline{\theta}_\beta$.
This allows us to precompute the predictive covariances and the matrix operations in the predictive means defined in Equations~\eqref{eq:predictiveMeansCovariancesGP}.
By sampling predictive samples $\boldsymbol{\alpha}_j^*$ and $\boldsymbol{\beta}_j^*$, we can by extension sample predictive data points $\boldsymbol{y}_j^*$ according to the data-generating function in Equation~\eqref{eq:dataDistribution}.
We present pseudocode for the above approximate posterior linearization scheme in Algorithm~\ref{alg:posteriorEstimationApproximate}.
While the approach described above is applicable as is, the result is an approximation for the desired posterior distribution in Eq.~\eqref{eq:lggpPosteriorDistribution}.
While approximate, the iterated posterior linearization provides an informed starting point for the log-shape and log-rate processes.
In addition, HMC sampling of the approximate posterior distribution in Eq.~\eqref{eq:surrogatePosteriorAlphaBeta} can be used to efficiently guide initial values for the hyperparameters of the latent log-Gaussian processes.
This is similar to sequential Monte Carlo methods, where it is common to construct a sequence of tempered posterior distributions that start by sampling the model parameters from a distribution, typically the prior distribution, that can be easily sampled from.
The initial sampling is followed by sampling of the subsequent distributions, which are most typically constructed by gradually including either the likelihood term or data points  \citep{Chopin:2020}.
In addition, tempering has been used in conjunction with HMC sampling \citep{Brooks:2011, Luo:2018, Sun:2019}.

We leverage the above information and construct a tempered sequence of posterior distributions
\begin{equation}
\begin{split}
    \pi_{t_\kappa}( \boldsymbol{\alpha}, \boldsymbol{\beta}, \mu_\alpha, \theta_\alpha, \mu_\beta, \theta_\beta \mid \boldsymbol{y}, \boldsymbol{m}^{(T)}, \boldsymbol{P}^{(T)} ) \propto \pi(& \boldsymbol{\alpha}, \boldsymbol{\beta}, \mu_\alpha, \theta_\alpha, \mu_\beta, \theta_\beta \mid \boldsymbol{y} )^{ \kappa( t_\kappa) } \\ &\times \mathcal{L}( \boldsymbol{m}^{(T)}_\alpha, \boldsymbol{P}^{(T)}_\alpha \mid \mu_\alpha, \theta_\alpha)^{ 1 - \kappa( t_\kappa) }\\ &\times \mathcal{L}( \boldsymbol{m}^{(T)}_\alpha, \boldsymbol{P}^{(T)}_\alpha \mid \mu_\alpha, \theta_\alpha)^{ 1 - \kappa( t_\kappa) }
\end{split}
\label{eq:temperedExactPosteriorDistribution}
\end{equation}
where $t_\kappa$ denotes the tempering iteration, or time step, and $\kappa{(t_\kappa)}$ is a strictly increasing sequence, $ 0 = \kappa(0) < \dots < \kappa( t_\kappa - 1) < \kappa(t_\kappa) < \dots < \kappa(T_\kappa) = 1 $, controlling the degree of tempering.
Importantly, $ \pi_{t_\kappa}( \boldsymbol{\alpha}, \boldsymbol{\beta}, \mu_\alpha, \theta_\alpha, \mu_\beta, \theta_\beta \mid \boldsymbol{y}, \boldsymbol{m}^{(T)}, \boldsymbol{P}^{(T)} ) = \pi( \boldsymbol{\alpha}, \boldsymbol{\beta}, \mu_\alpha, \theta_\alpha, \mu_\beta, \theta_\beta \mid \boldsymbol{y} )$ when $ t_\kappa = T_\kappa$, meaning that at the final tempering iteration, we are targeting the original posterior distribution in Eq.~\eqref{eq:lggpPosteriorDistribution}.

We target each tempered posterior distribution $ \pi_{t_\kappa}( \boldsymbol{\alpha}, \boldsymbol{\beta}, \mu_\alpha, \theta_\alpha, \mu_\beta, \theta_\beta \mid \boldsymbol{y}, \boldsymbol{m}^{(T)}, \boldsymbol{P}^{(T)} ) $ with NUTS while feeding the last element of the sample chain of the previous NUTS run as the initial state of the next tempering step. 
The computational cost of estimating Eq.~\eqref{eq:temperedExactPosteriorDistribution} is increased in comparison to Eq.~\eqref{eq:lggpPosteriorDistribution} due to the inclusion of additional Gaussian process likelihood terms.
Therefore, we propose running short chains for $ \kappa( t_\kappa ) \neq 1$ to obtain a reasonable starting state for the final iteration, and use most of the available computational budget at the last tempering step.
We present pseudocode for the tempering approach in Algorithm~\ref{alg:posteriorEstimationTempering}.
In contrast to Algorithm~\ref{alg:posteriorEstimationApproximate}, we present the tempering method using the obtained samples from the posterior distribution in Eq.~\eqref{eq:lggpPosteriorDistribution} to construct the predictive distribution.

We present the improved mixing achieved using the above tempering scheme in comparison to direct HMC sampling, following the results of the approximate iterated posterior linearization scheme in Section \ref{sec:results}.
\begin{algorithm}
\caption{Iterated posterior linearization for log-Gaussian gamma process log-shape and log-rate processes $\boldsymbol{\alpha}$ and $\boldsymbol{\beta}$.}
\label{alg:posteriorLinearization}
\begin{algorithmic}
\State \textbf{Initialize:}
\State\indt\textbf{for $j \in 1:J$}
\State\indt[2] Sample $( \mu_\alpha, \theta_\alpha, \mu_\beta, \theta_\beta)_j^\intercal$ from  $\pi_0( \mu_\alpha, \theta_\beta, \mu_\beta, \theta_\beta)$.
\State\indt[2] Sample $\boldsymbol{\alpha}_j^{(0)}$ and $\boldsymbol{\beta}_j^{(0)}$ according to Equations~\eqref{eq:alphaPriorGP} and \eqref{eq:betaPriorGP}.
\State\indt Set $\boldsymbol{m}^{(0)}$ and compute $\hat{ \boldsymbol{P} }^{(0)}$ according to Equation~\eqref{eq:initialCovarianceMonteCarlo}.
\State
\State \textbf{Posterior linearization for $ \pi\left( \boldsymbol{ \alpha }, \boldsymbol{\beta} \mid \boldsymbol{y} \right)$:}
\State\indt\textbf{for $t \in 1:T$}
\State\indt[2]\textbf{for $j \in 1:J$}
\State\indt[3] \textbf{if $t > 1$}
\State\indt[4] Sample $\boldsymbol{\alpha}_j^{(t-1)}$ and $\boldsymbol{\beta}_j^{(t-1)}$ according to Equation~\eqref{eq:alphaBetaApproximateMarginalPosterior}.
\State\indt[3] Sample $\boldsymbol{y}_j^{(t-1)}$ with Equation~\eqref{eq:dataDistribution} given $\boldsymbol{\alpha}_j^{(t-1)}$ and $\boldsymbol{\beta}_j^{(t-1)}$.
\State\indt[3] Compute estimates of $\hat{ \boldsymbol{\mu} }^{+,(t - 1)}$, $\hat{ \boldsymbol{P} }^{yy,(t - 1)}$, and $\hat{ \boldsymbol{P} }^{ \alpha\beta y, (t - 1)}$ according to \eqref{eq:plLinearizationMonteCarlo}.
\State\indt[3] Compute regression parameters $ \boldsymbol{A}^{(t)}$, $ \boldsymbol{b}^{(t)}$, and $ \boldsymbol{\Lambda}^{(t)} $ in Equation~\eqref{eq:slrParameters}.
\State\indt[3] Evaluate $ \boldsymbol{\mu}^{(t)} $, $\boldsymbol{S}^{(t)} $, and $\boldsymbol{K}^{(t)}$ given in Equation~\eqref{eq:plMomentUpdateParameters}.
\State\indt[3] Update the approximate mean $\boldsymbol{m}^{(t)}$ and covariance $ \boldsymbol{P}^{(t)}$ with Equation~\eqref{eq:plMomentUpdate}.
\end{algorithmic}
\end{algorithm}
\begin{algorithm}
\caption{Log-Gaussian gamma process approximate estimation scheme.}
\label{alg:posteriorEstimationApproximate}
\begin{algorithmic}
\State Perform iterated posterior linearization according to Algorithm \ref{alg:posteriorLinearization}.
\State
\State \textbf{Posterior estimation for $\mu_\alpha, \theta_\alpha, \mu_\beta,$ and $ \theta_\beta$:}
\State\indt[1] Extract $\boldsymbol{m}_\alpha^{(T)}$, $\boldsymbol{m}_\beta^{(T)}$, $\boldsymbol{P}_\alpha^{(T)}$, and $\boldsymbol{P}_\beta^{(T)}$ from $\boldsymbol{m}^{(T)}$ and $\boldsymbol{P}_\alpha^{(T)}$ by Equation~\eqref{eq:alphaBetaIndependentMeanCovariance}.
\State\indt[1] Target $\pi( \mu_\alpha, \theta_\alpha \mid \boldsymbol{y} )$ and $\pi( \mu_\beta, \theta_\beta \mid \boldsymbol{y} )$ in Equation~\eqref{eq:surrogatePosteriorAlphaBeta} with NUTS.
\State
\State \textbf{Sampling of predictive distributions $\pi( \boldsymbol{\alpha}^*, \boldsymbol{\beta}^* \mid \boldsymbol{y} )$ and $\pi( \boldsymbol{y}^* \mid \boldsymbol{y} )$}
\State\indt Compute the posterior means $\overline{ \mu }_\alpha, \overline{ \theta }_\alpha, \overline{ \mu }_\beta,$ and $ \overline{ \theta }_\beta$.
\State\indt Compute the predictive covariances $\Sigma_\alpha^*( \boldsymbol{x}^*, \boldsymbol{x}^*; \overline{ \theta }_\alpha)$ and $\Sigma_\beta^*( \boldsymbol{x}^*, \boldsymbol{x}^*; \overline{ \theta }_\beta)$ in Equation~\eqref{eq:predictiveMeansCovariancesGP}.
\State\indt\textbf{for $j \in 1:J$}
\State\indt[2] Sample $\boldsymbol{\alpha}_j$ and $\boldsymbol{\beta}_j$ according to Equation~\eqref{eq:alphaBetaApproximateMarginalPosteriorFinal}.
\State\indt[2] Compute predictive means $\mu_\alpha^*( \boldsymbol{x}^*; \boldsymbol{\alpha}_j, \overline{ \mu }_\alpha, \overline{ \theta }_\alpha)$ and $\mu_\beta^*( \boldsymbol{x}^*; \boldsymbol{\beta}_j, \overline{ \mu }_\beta, \overline{ \theta }_\beta)$ with \eqref{eq:predictiveMeansGP}.
\State\indt[2] Sample $\boldsymbol{\alpha}_j^* \sim \mathcal{N}\left( \mu_\alpha^*( \boldsymbol{x}^*; \boldsymbol{\alpha}_j, \overline{ \mu }_\alpha, \overline{ \theta }_\alpha), \Sigma_\alpha^*( \boldsymbol{x}^*, \boldsymbol{x}^*; \overline{ \theta }_\alpha) \right)$.
\State\indt[2] Sample $\boldsymbol{\beta}_j^* \sim \mathcal{N}\left( \mu_\beta^*( \boldsymbol{x}^*; \boldsymbol{\beta}_j, \overline{ \mu }_\beta, \overline{ \theta }_\beta), \Sigma_\beta^*( \boldsymbol{x}^*, \boldsymbol{x}^*; \overline{ \theta }_\beta) \right)$.
\State\indt[2] Sample $\boldsymbol{y}_j^* \sim \pi\left( \boldsymbol{y}^* \mid \boldsymbol{\alpha}_j^*, \boldsymbol{\beta}_j^* \right)$ with Equation~\eqref{eq:dataDistribution}.
\end{algorithmic}
\end{algorithm}
\begin{algorithm}
\caption{Log-Gaussian gamma process exact tempering estimation scheme.}
\label{alg:posteriorEstimationTempering}
\begin{algorithmic}
\State Perform iterated posterior linearization according to Algorithm \ref{alg:posteriorLinearization}.
\State
\State \textbf{Initialize:}
\State\indt[1] Set the tempering sequence $\boldsymbol{\kappa} = \left( \kappa( 0 ), \dots, \kappa(T_\kappa) \right)^\intercal$.
\State\indt[1] Extract $\boldsymbol{m}_\alpha^{(T)}$, $\boldsymbol{m}_\beta^{(T)}$, $\boldsymbol{P}_\alpha^{(T)}$, and $\boldsymbol{P}_\beta^{(T)}$ from $\boldsymbol{m}^{(T)}$ and $\boldsymbol{P}_\alpha^{(T)}$ by Equation~\eqref{eq:alphaBetaIndependentMeanCovariance}.
\State
\State \textbf{Sampling of the tempered posterior distributions}
\State\indt[1]\textbf{for $ t_\kappa \in 0 : T_\kappa$}
\State\indt[2]\textbf{if} $t_\kappa > 0$
\State\indt[3]Set the initial state of NUTS to the last sample obtained at the previous iteration, $t_\kappa - 1$.
\State\indt[2] Target $\pi_{t_\kappa}( \boldsymbol{\alpha}, \boldsymbol{\beta}, \mu_\alpha, \theta_\alpha, \mu_\beta, \theta_\beta \mid \boldsymbol{y}, \boldsymbol{m}^{(T)}, \boldsymbol{P}^{(T)} )$ in Eq.~\eqref{eq:temperedExactPosteriorDistribution} with NUTS.

\State\indt[2]\textbf{if} $t_\kappa = T_\kappa$
\State\indt[3]Save the posterior distribution samples.
\State
\State \textbf{Sampling of predictive distributions $\pi( \boldsymbol{\alpha}^*, \boldsymbol{\beta}^* \mid \boldsymbol{y} )$ and $\pi( \boldsymbol{y}^* \mid \boldsymbol{y} )$}
\State\indt\textbf{for $j \in 1:J$}
\State\indt[2] Sample $\boldsymbol{\alpha}_j$, $\boldsymbol{\beta}_j$, $\mu_{ \alpha, j}$, $\theta_{ \alpha, j}$, $\mu_{ \beta, j}$, and $\theta_{ \beta, j}$ from the posterior distribution samples.
\State\indt[2] Compute predictive means $\mu_\alpha^*( \boldsymbol{x}^*; \boldsymbol{\alpha}_j, \mu_{ \alpha, j}, \theta_{ \alpha, j} )$ and $\mu_\beta^*( \boldsymbol{x}^*; \boldsymbol{\beta}_j, \mu_{ \beta, j}, \theta_{ \beta, j} )$ with \eqref{eq:predictiveMeansGP}.
\State\indt[2] Compute the predictive covariances $\Sigma_\alpha^*( \boldsymbol{x}^*, \boldsymbol{x}^*; \theta_{\alpha, j})$ and $\Sigma_\beta^*( \boldsymbol{x}^*, \boldsymbol{x}^*; \theta_{ \beta, j} )$ in Equation~\eqref{eq:predictiveMeansCovariancesGP}.
\State\indt[2] Sample $\boldsymbol{\alpha}_j^* \sim \mathcal{N}\left( \mu_\alpha^*( \boldsymbol{x}^*; \boldsymbol{\alpha}_j, \mu_{ \alpha, j}, \theta_{ \alpha, j} ), \Sigma_\alpha^*( \boldsymbol{x}^*, \boldsymbol{x}^*; \theta_{\alpha, j}) \right)$.
\State\indt[2] Sample $\boldsymbol{\beta}_j^* \sim \mathcal{N}\left( \mu_\beta^*( \boldsymbol{x}^*; \boldsymbol{\beta}_j, \mu_{ \beta, j}, \theta_{ \beta, j} ), \Sigma_\beta^*( \boldsymbol{x}^*, \boldsymbol{x}^*; \theta_{ \beta, j} ) \right)$.
\State\indt[2] Sample $\boldsymbol{y}_j^* \sim \pi\left( \boldsymbol{y}^* \mid \boldsymbol{\alpha}_j^*, \boldsymbol{\beta}_j^* \right)$ with Equation~\eqref{eq:dataDistribution}.
\end{algorithmic}
\end{algorithm}
%
%
%
%
\section{Numerical examples}
\label{sec:results}
\begin{figure}
    \centering
    \begin{minipage}{0.49\textwidth}
        \subfloat[][Ground truth. ]{\includegraphics[width = \textwidth]{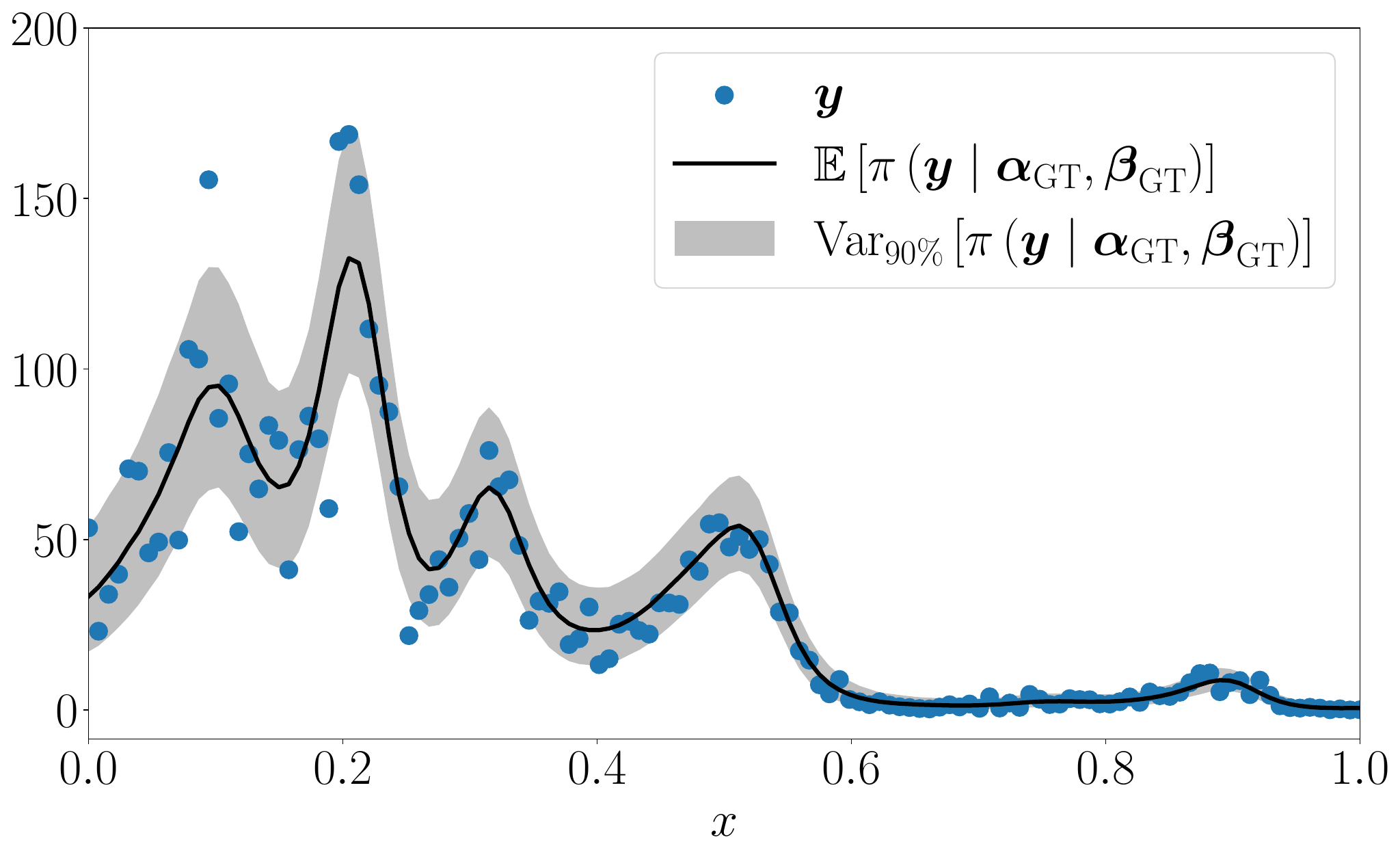}}
    \end{minipage}\\
    
    \begin{minipage}{0.49\textwidth}
        \subfloat[][HMC. ]{ \includegraphics[width = \textwidth]{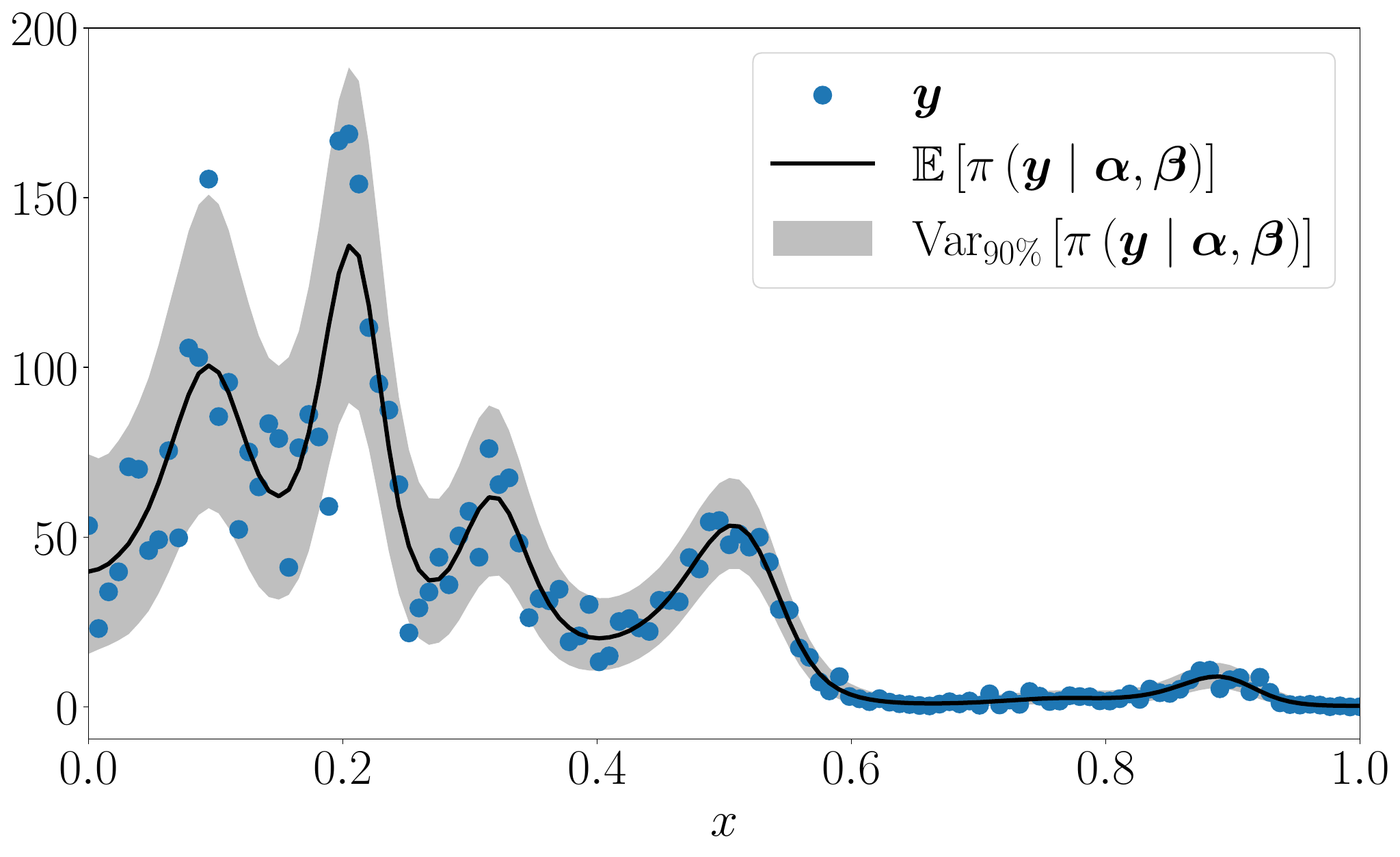} }
    \end{minipage}
    \begin{minipage}{0.49\textwidth}
        \subfloat[][HMC (short chain). ]{\includegraphics[width = \textwidth]{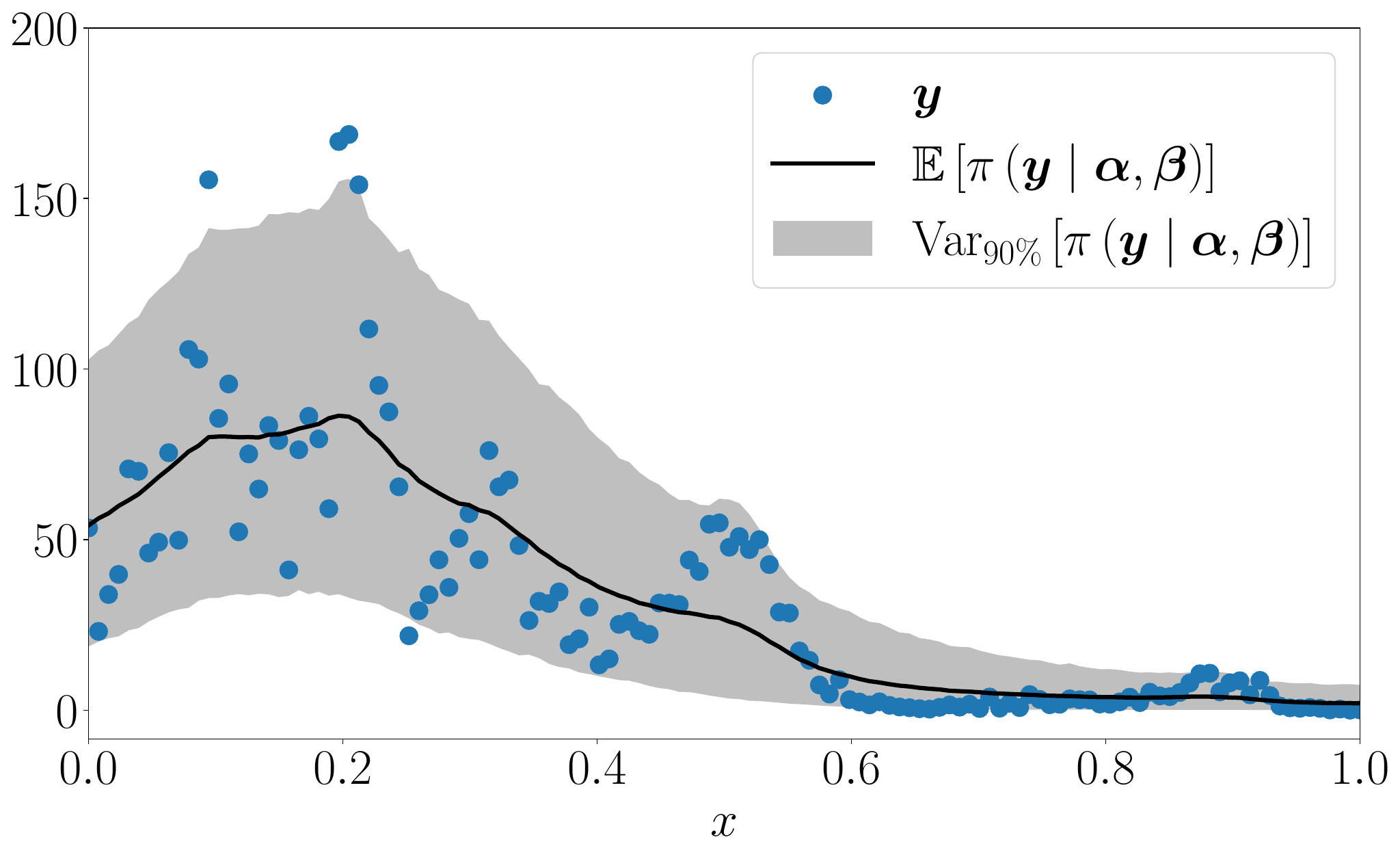}}
    \end{minipage}\\
    
    \begin{minipage}{0.49\textwidth}
        \subfloat[][PL + HMC. ]{ \includegraphics[width = \textwidth]{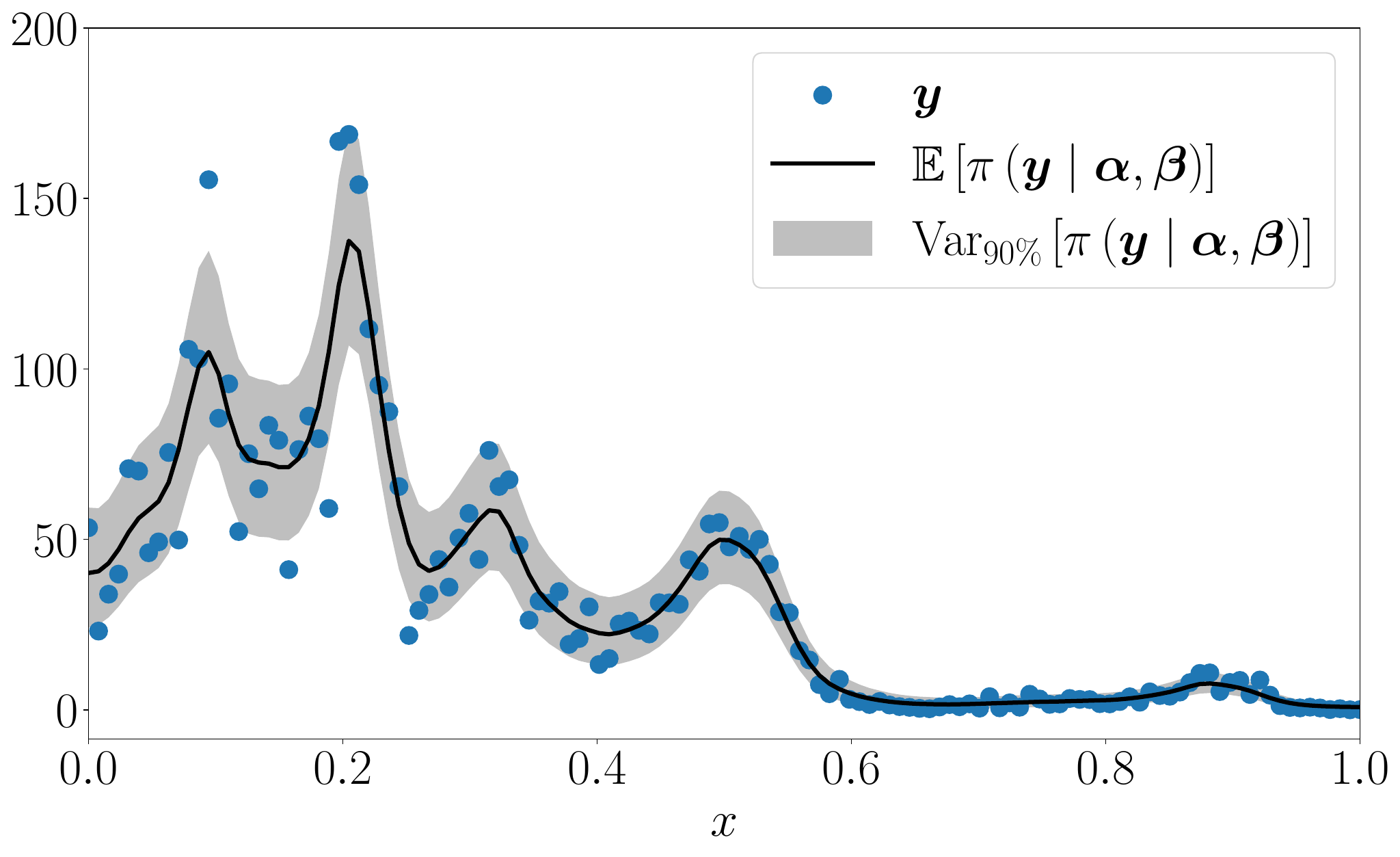} }
    \end{minipage}
    \begin{minipage}{0.49\textwidth}
        \subfloat[][PL + tempered HMC (short chain). ]{\includegraphics[width = \textwidth]{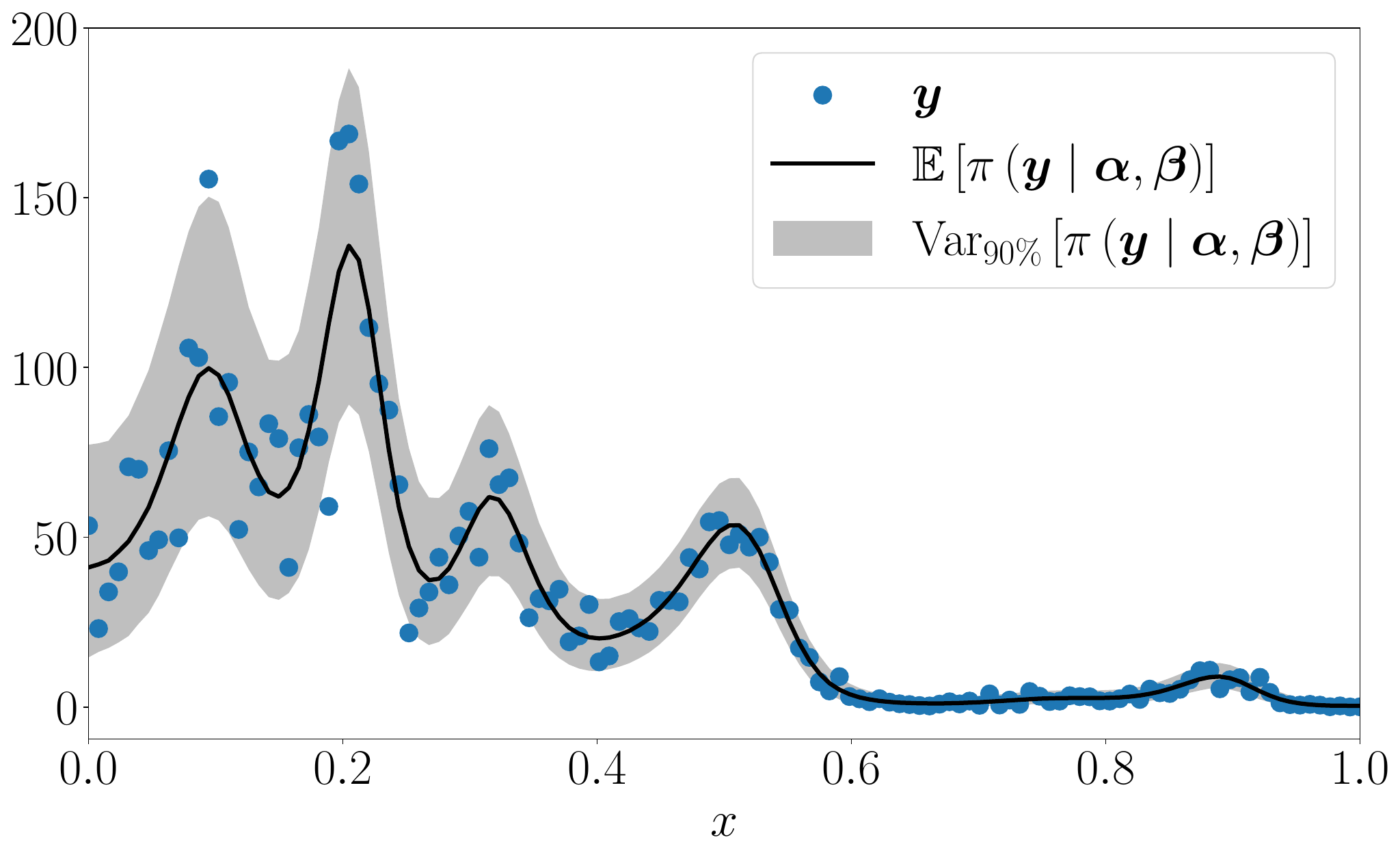}}
    \end{minipage}
    
    \caption{Synthetic dataset. In (a), data in blue, the ground truth expectation, and 90\% confidence interval of the data-generating function in black and gray, respectively, and $\boldsymbol{\alpha}_\textrm{GT} $ and $\boldsymbol{\beta}_\textrm{GT} $ denote the ground truth log-shape and log-rate processes. The mean estimate $\mathbb{E}\left[ \pi( \boldsymbol{y} \mid \boldsymbol{\alpha}, \boldsymbol{\beta} ) \right]$ and the corresponding 90\% credible interval $\textrm{Var}_{90\%}\left[ \pi( \boldsymbol{y} \mid \boldsymbol{\alpha}, \boldsymbol{\beta} ) \right]$ for the data-generating function obtained with (b) HMC, (c) short chain HMC, (d) iterated posterior linearization followed by HMC sampling, and (e) iterated posterior linearization and tempered short chain HMC sampling.}
    \label{im:dataMarginalPosterior_synthetic}
\end{figure}

\begin{figure}

    \centering
    \begin{minipage}{\textwidth}
        \subfloat[][HMC.]{ \includegraphics[width = 0.49\textwidth]{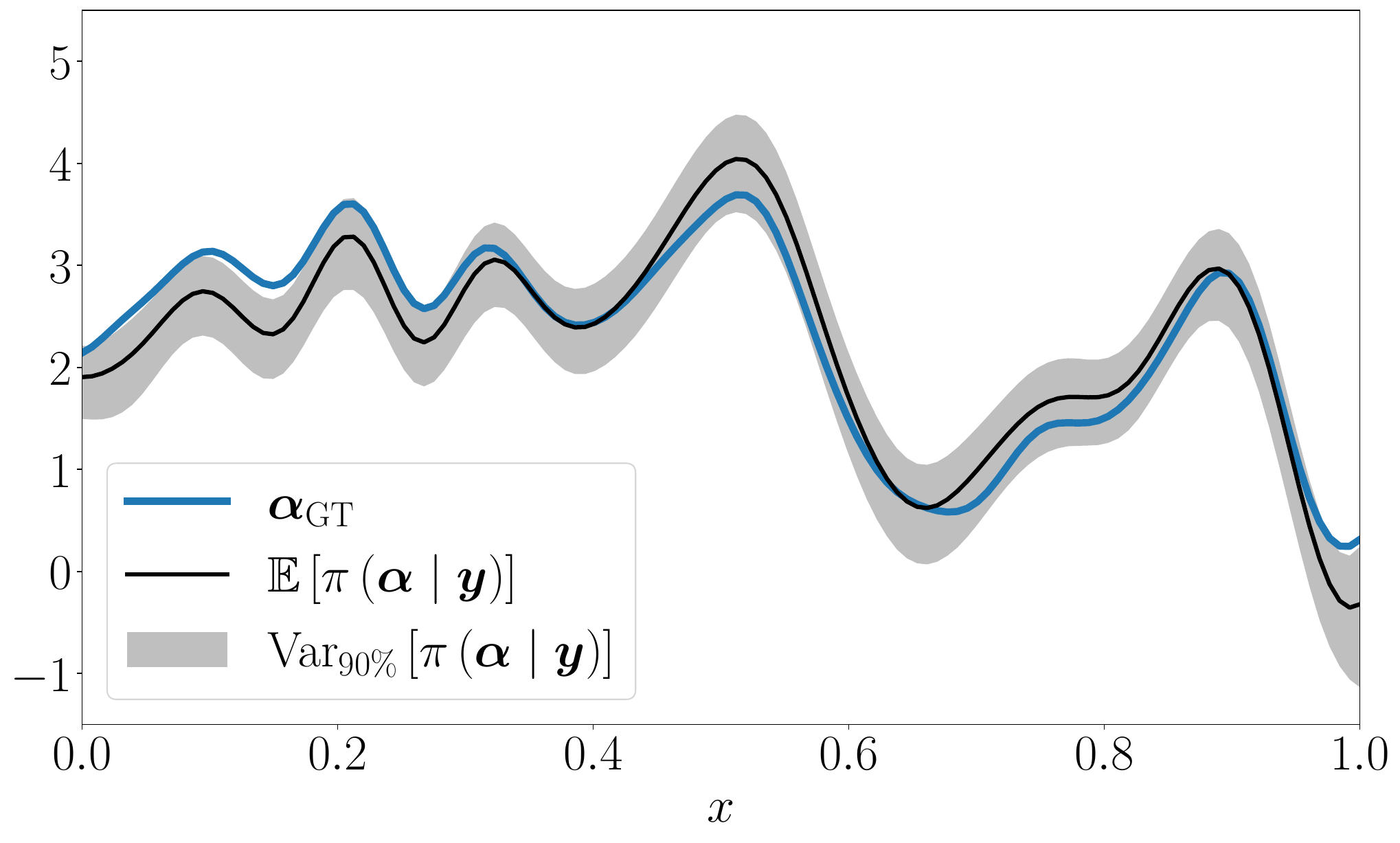} \includegraphics[width = 0.49\textwidth]{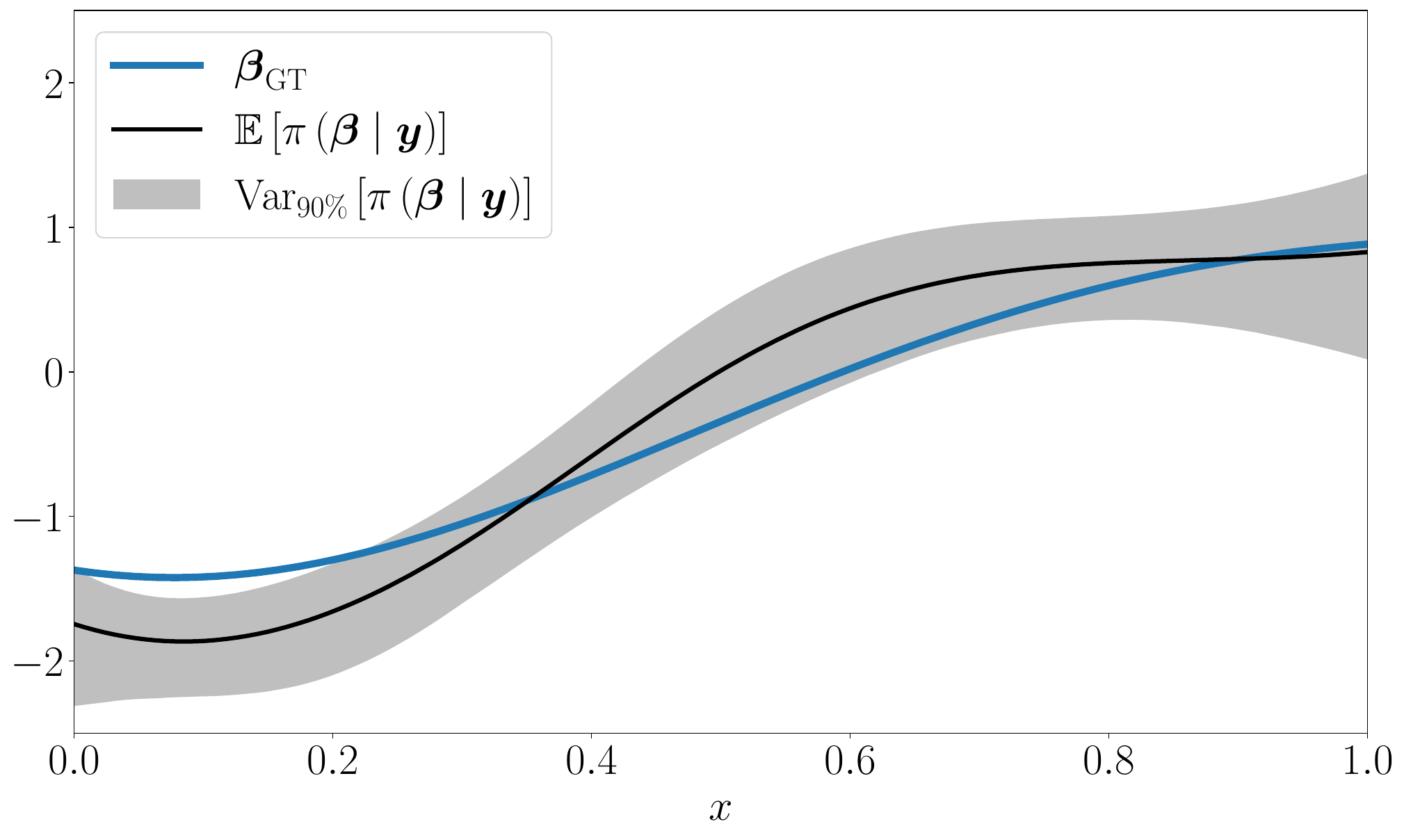} } 
    \end{minipage}\vspace{0.5cm}\\
    
    \begin{minipage}{\textwidth}
        \subfloat[][PL + HMC.]{ \includegraphics[width = 0.49\textwidth]{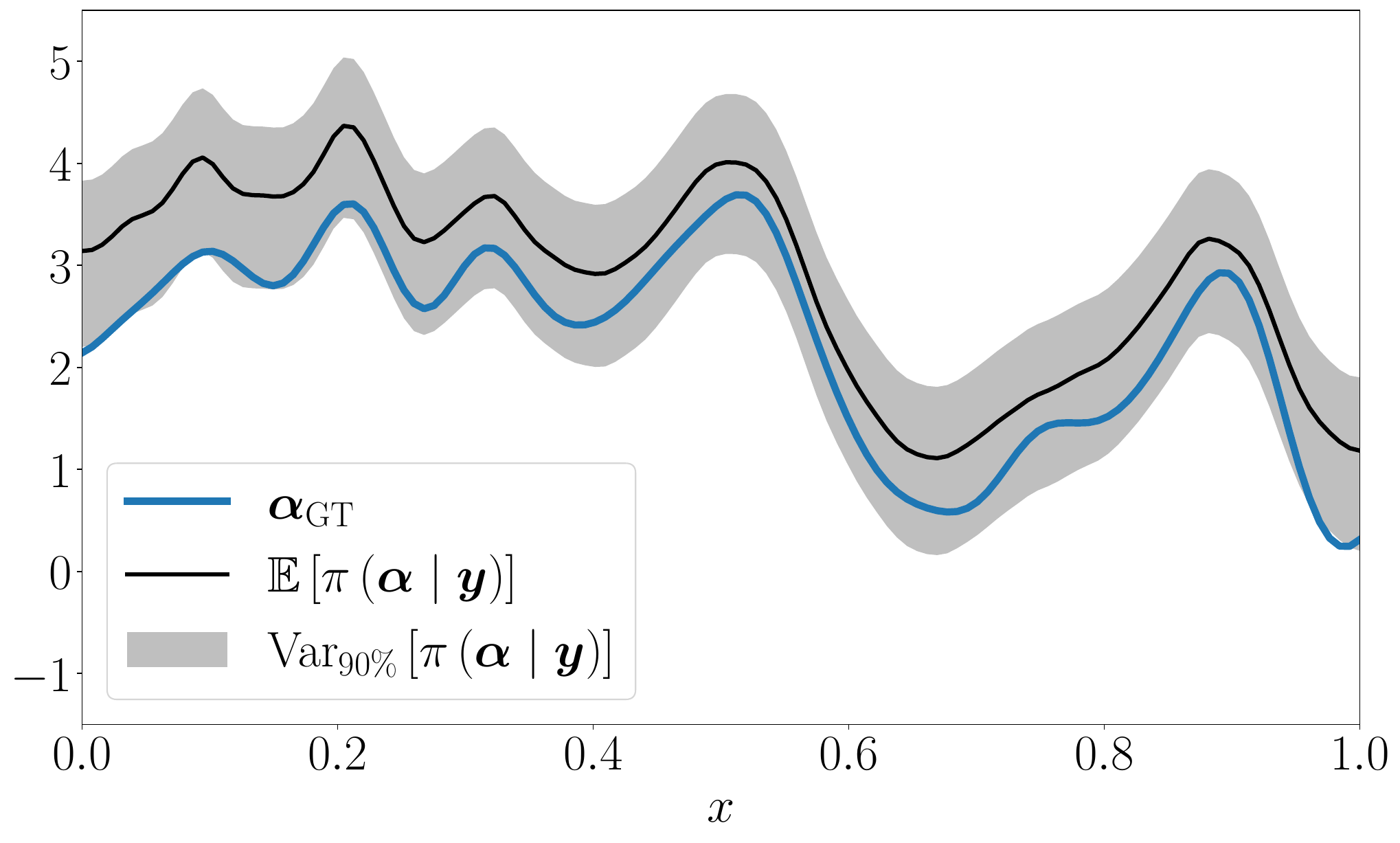} \includegraphics[width = 0.49\textwidth]{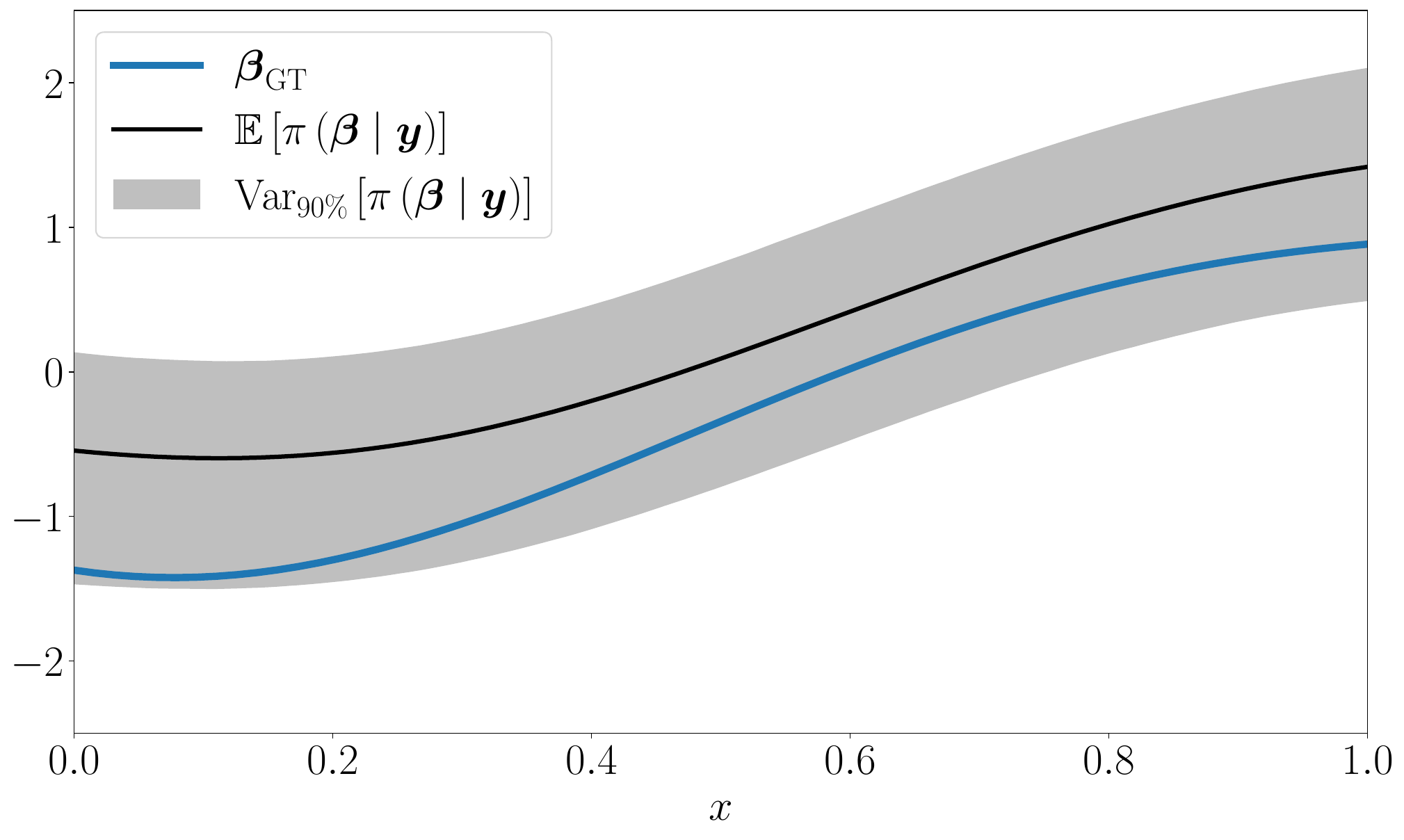} } 
    \end{minipage}\vspace{0.5cm}\\
    
    \caption{Synthetic dataset. The mean estimates $\mathbb{E}\left[ \pi( \boldsymbol{\alpha} \mid \boldsymbol{y} ) \right]$ and $\mathbb{E}\left[ \pi( \boldsymbol{\beta} \mid \boldsymbol{y} ) \right]$ with the corresponding 90\% credible intervals $\textrm{Var}_{90\%}\left[ \pi( \boldsymbol{\alpha} \mid \boldsymbol{y} ) \right]$ and $\textrm{Var}_{90\%}\left[ \pi( \boldsymbol{\beta} \mid \boldsymbol{y} ) \right]$ for the log-shape and log-rate processes obtained with (a) HMC sampling and (b) iterated posterior linearization followed by HMC sampling. The ground truth log-shape and log-rate processes are denoted by $\boldsymbol{\alpha}_\textrm{GT} $ and $\boldsymbol{\beta}_\textrm{GT} $.}
    \label{im:shapeRateMarginalPosterior_synthetic}
\end{figure}

\begin{figure}

    \centering
    \begin{minipage}{\textwidth}
        \subfloat[][HMC.]{ \includegraphics[width = 0.49\textwidth]{pics/vector/heteroscedastic/shape/shape_hmc.pdf} \includegraphics[width = 0.49\textwidth]{pics/vector/heteroscedastic/rate/rate_hmc.pdf} } 
    \end{minipage}\vspace{0.5cm}\\
    
    \begin{minipage}{\textwidth}
        \subfloat[][HMC (short chain).]{ \includegraphics[width = 0.49\textwidth]{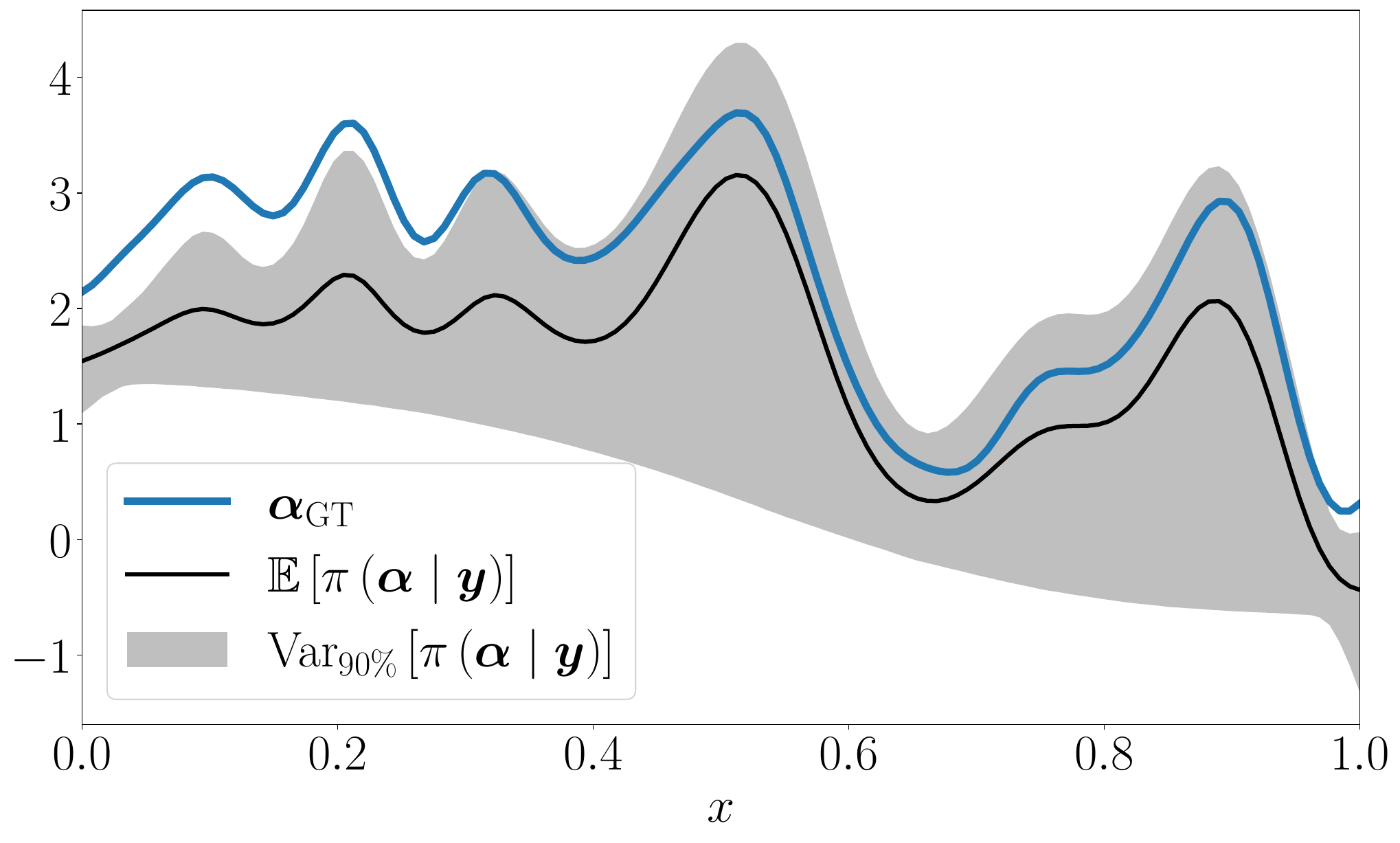} \includegraphics[width = 0.49\textwidth]{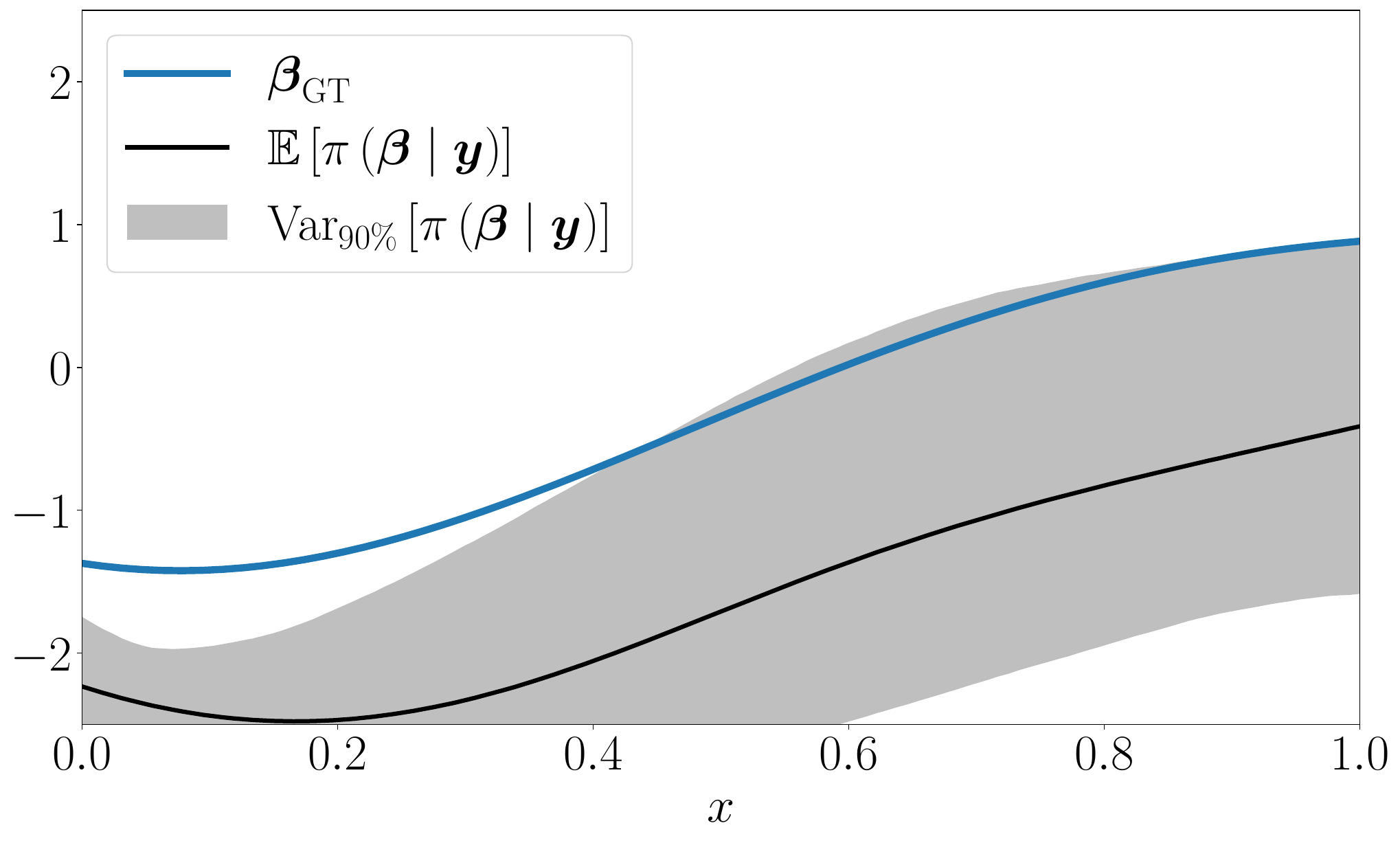} } 
    \end{minipage}\vspace{0.5cm}\\
    
    \begin{minipage}{\textwidth}
        \subfloat[][PL + tempered HMC (short chain).]{ \includegraphics[width = 0.49\textwidth]{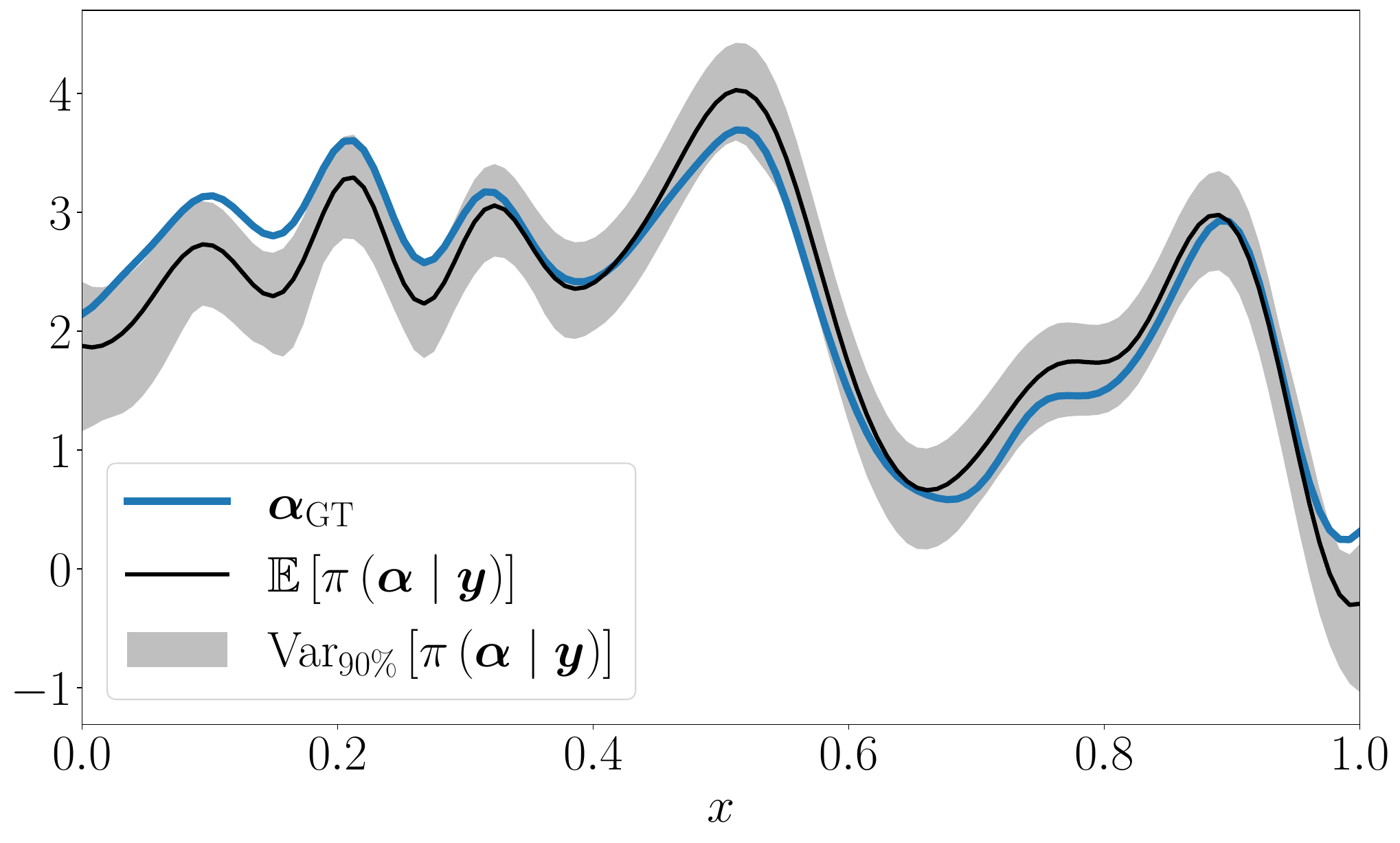} \includegraphics[width = 0.49\textwidth]{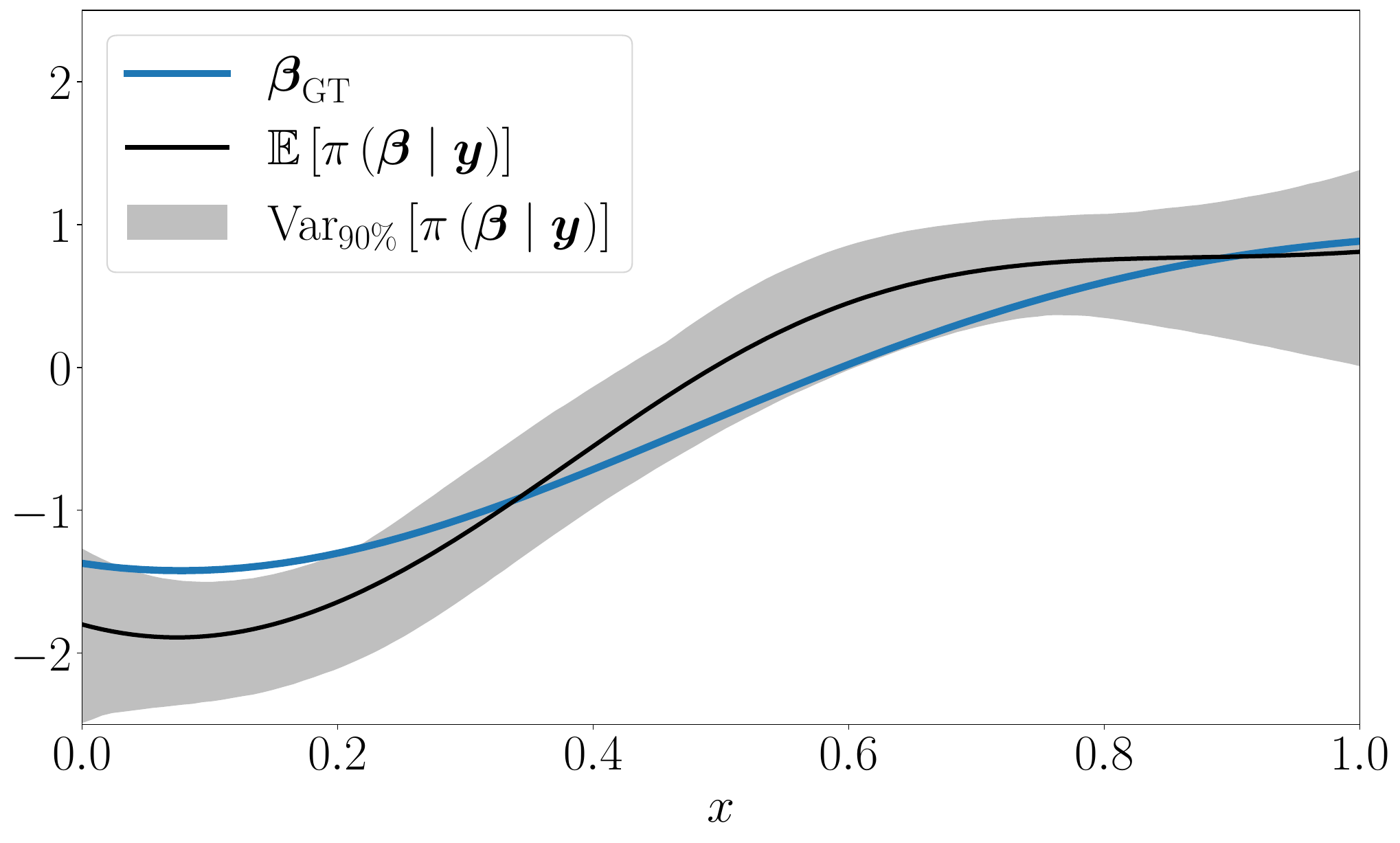} } 
    \end{minipage}
    
    \caption{Synthetic dataset. The mean estimates $\mathbb{E}\left[ \pi( \boldsymbol{\alpha} \mid \boldsymbol{y} ) \right]$ and $\mathbb{E}\left[ \pi( \boldsymbol{\beta} \mid \boldsymbol{y} ) \right]$ with the corresponding 90\% credible intervals $\textrm{Var}_{90\%}\left[ \pi( \boldsymbol{\alpha} \mid \boldsymbol{y} ) \right]$ and $\textrm{Var}_{90\%}\left[ \pi( \boldsymbol{\beta} \mid \boldsymbol{y} ) \right]$ for the log-shape and log-rate processes obtained with (a) HMC sampling, (b) short chain HMC sampling, and (c) iterated posterior linearization and tempered short chain HMC. The ground truth log-shape and log-rate processes are denoted by $\boldsymbol{\alpha}_\textrm{GT} $ and $\boldsymbol{\beta}_\textrm{GT} $.}
    \label{im:shapeRateMarginalPosteriorTempering_synthetic}
\end{figure}

We apply our proposed methods and direct HMC sampling with a shorter and longer Markov chain Monte Carlo chain to three datasets.
For the synthetic datasets, $ \boldsymbol{\alpha}_\textrm{GT} $ and $ \boldsymbol{\beta}_\textrm{GT} $ denote the ground truth log-shape and log-rate processes used to generate the datasets.
The first dataset is a synthetic dataset generated from the log-Gaussian gamma process in Eq.~\eqref{eq:lggpModel} with log-shape GP parameters $ \left( \mu_\alpha, l_\alpha, \sigma_{ \alpha, \textrm{s} } \right)^\intercal = ( 2, 0.05, 1)^\intercal$ and log-rate GP parameters $ \left( \mu_\beta, l_\beta,\sigma_{ \beta, \textrm{s} } \right)^\intercal = \left( 1, 0.5, 1 \right)^\intercal$ in a uniform grid with 128 data points.
Numerical results are presented in Figures \ref{im:dataMarginalPosterior_synthetic}-\ref{im:rateThetaMarginalPosterior_synthetic}.
Figure \ref{im:dataMarginalPosterior_synthetic} illustrates the synthetic data, true expectation, and 90\% confidence intervals of the data-generating function in conjunction with the estimated mean and marginal 90\% credible intervals obtained using the proposed methods and the two HMC sample chains.
Figure \ref{im:shapeRateMarginalPosterior_synthetic} shows the estimated log-shape and log-rate processes of the longer chain HMC and approximate iterated posterior linearization approach.
In Figure \ref{im:shapeRateMarginalPosteriorTempering_synthetic}, we present the estimated log-shape and log-rate processes of both the shorter and longer HMC chains and for the iterated posterior linearization approach followed by tempering.
One-dimensional marginal posterior distributions for the GP mean and covariance parameters obtained by jointly sampling the full log-Gaussian gamma process posterior in Eq.~\eqref{eq:lggpPosteriorDistribution}, by sampling Eq.~\eqref{eq:surrogatePosteriorAlphaBeta} after iterated posterior linearization, and by sampling the tempered sequence in Eq.~\eqref{eq:temperedExactPosteriorDistribution} are shown in Figures \ref{im:shapeThetaMarginalPosterior_synthetic} and \ref{im:rateThetaMarginalPosterior_synthetic} for $\boldsymbol{\alpha}$ and $\boldsymbol{\beta}$, respectively, together with the parameter values used for data generation.

\begin{figure}\centering

    \centering
    \begin{minipage}{\textwidth}
        \subfloat[][HMC.]{ \includegraphics[width = 0.49\textwidth]{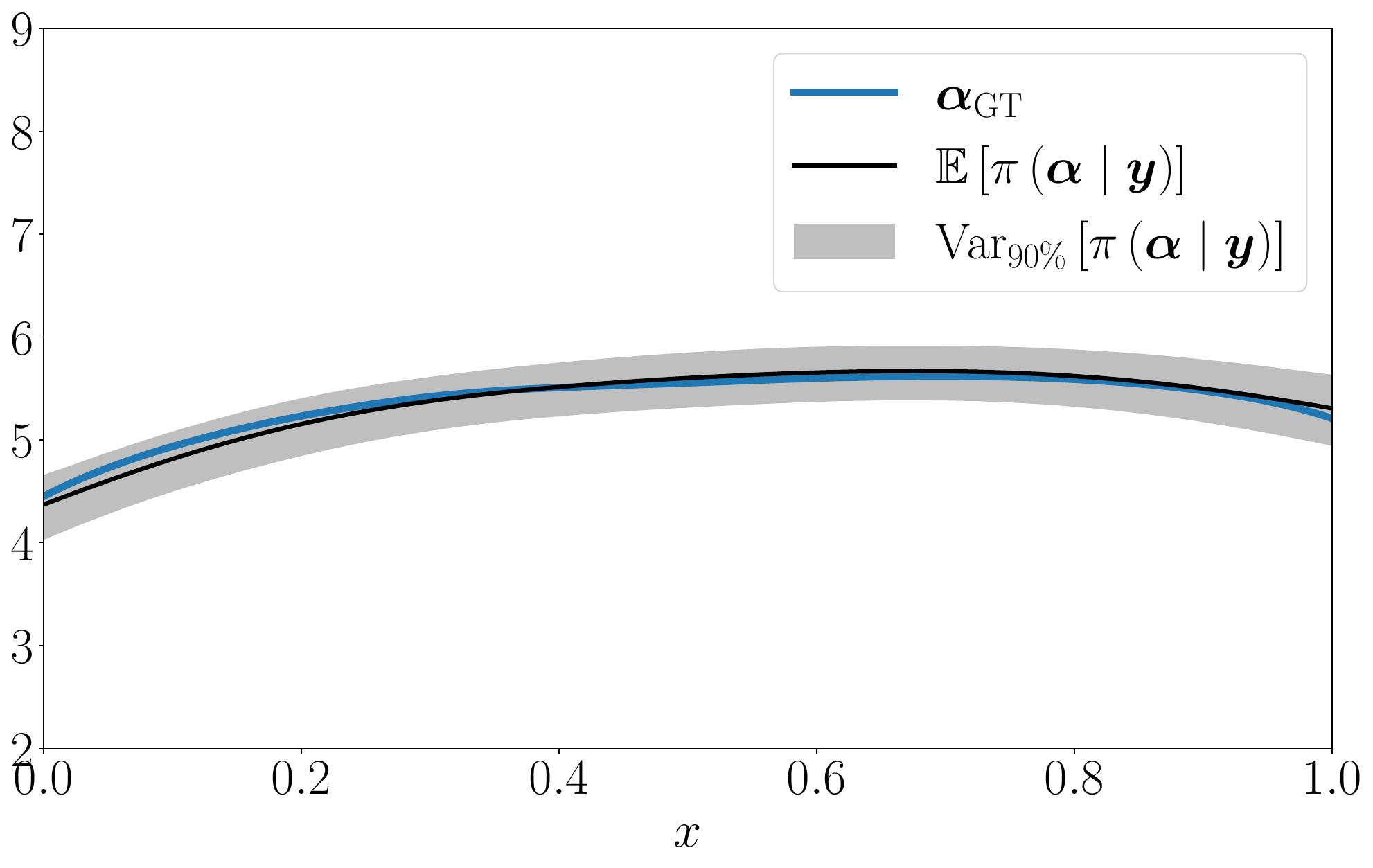} \includegraphics[width = 0.49\textwidth]{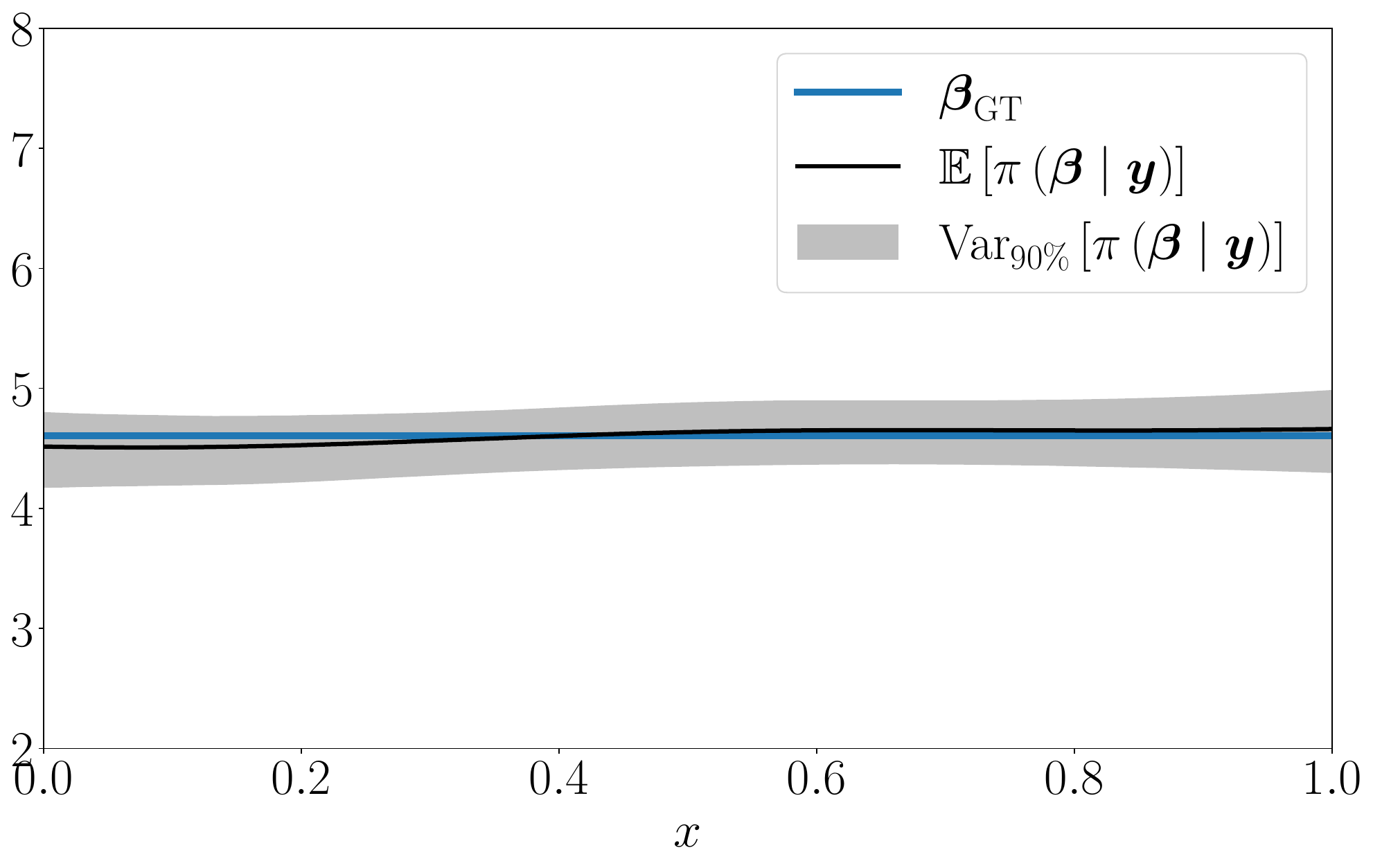} } 
    \end{minipage}\vspace{0.5cm}\\
    
    \begin{minipage}{\textwidth}
        \subfloat[][PL + HMC.]{ \includegraphics[width = 0.49\textwidth]{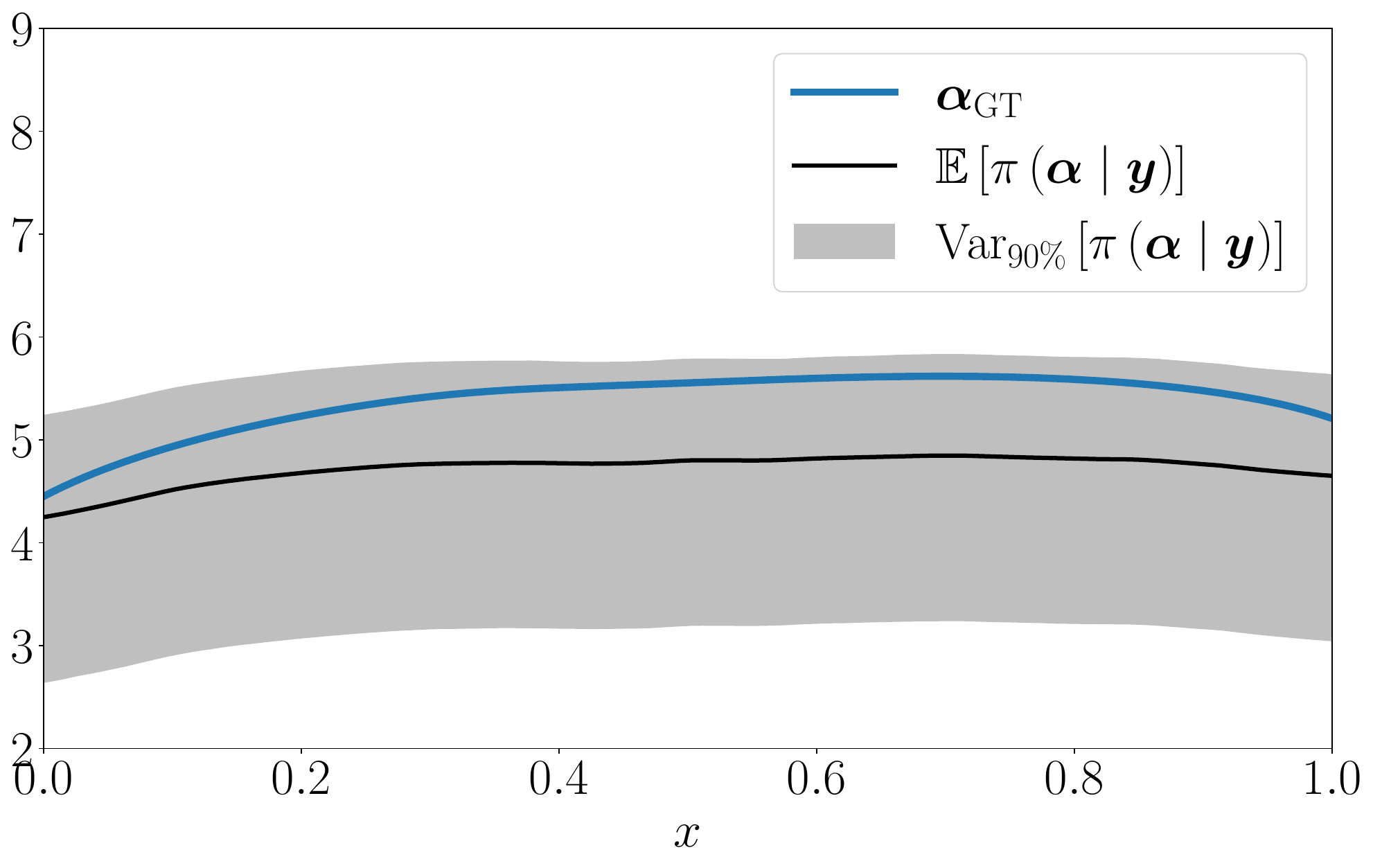} \includegraphics[width = 0.49\textwidth]{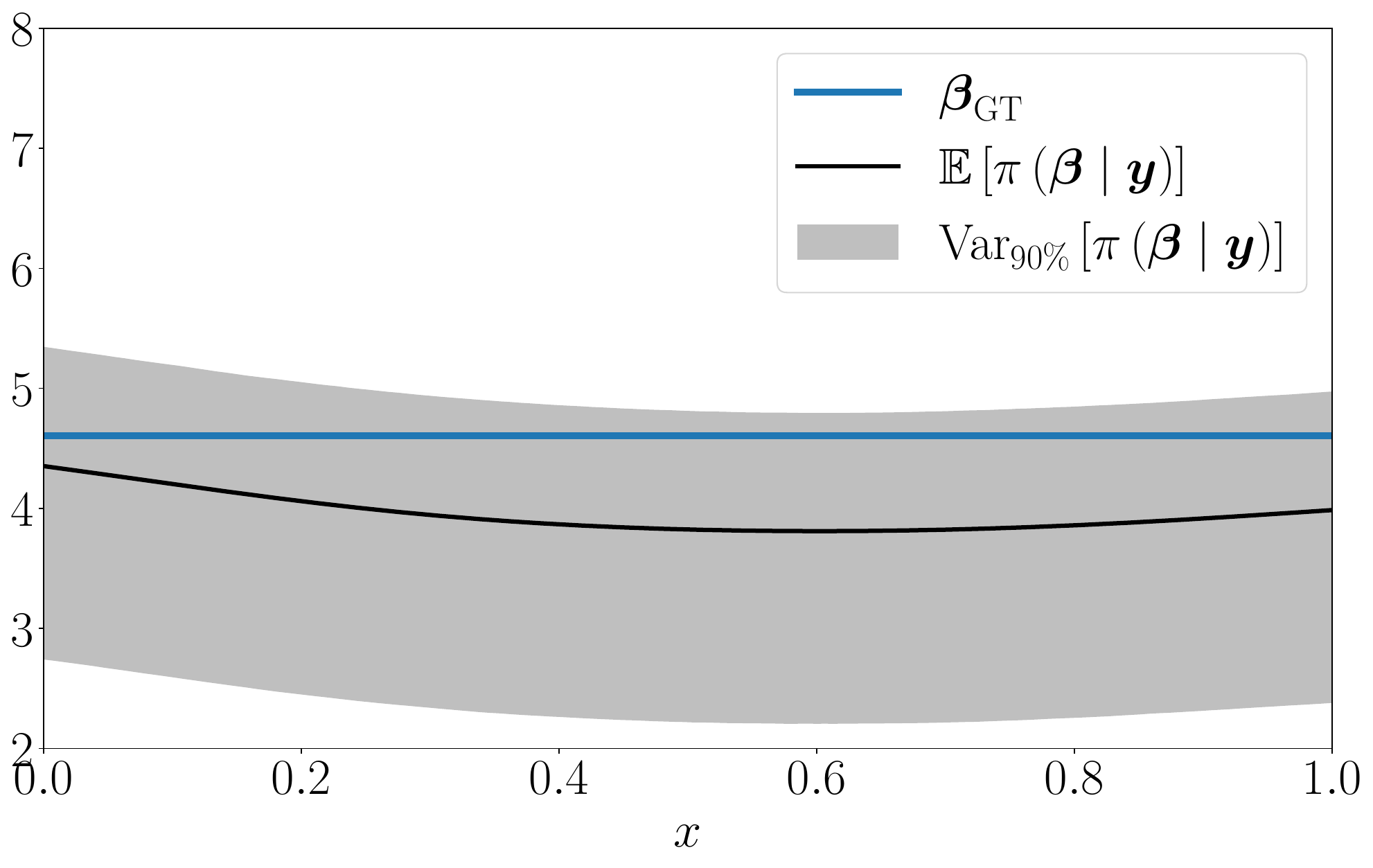} } 
    \end{minipage}
    
    \caption{Young's modulus dataset. The mean estimates $\mathbb{E}\left[ \pi( \boldsymbol{\alpha} \mid \boldsymbol{y} ) \right]$ and $\mathbb{E}\left[ \pi( \boldsymbol{\beta} \mid \boldsymbol{y} ) \right]$ with the corresponding 90\% credible intervals $\textrm{Var}_{90\%}\left[ \pi( \boldsymbol{\alpha} \mid \boldsymbol{y} ) \right]$ and $\textrm{Var}_{90\%}\left[ \pi( \boldsymbol{\beta} \mid \boldsymbol{y} ) \right]$ for the log-shape and log-rate processes obtained with (a) HMC sampling and (b) iterated posterior linearization followed by HMC sampling. The ground truth log-shape and log-rate processes are denoted by $\boldsymbol{\alpha}_\textrm{GT} $ and $\boldsymbol{\beta}_\textrm{GT} $.}
    \label{im:shapeRateMarginalPosterior_stiffness}
\end{figure}

\begin{figure}

    \centering
    \begin{minipage}{\textwidth}
        \subfloat[][HMC.]{ \includegraphics[width = 0.49\textwidth]{pics/vector/stiffness/shape/shape_hmc.pdf} \includegraphics[width = 0.49\textwidth]{pics/vector/stiffness/rate/rate_hmc.pdf} } 
    \end{minipage}\vspace{0.5cm}\\
    
    \begin{minipage}{\textwidth}
        \subfloat[][HMC (short chain).]{ \includegraphics[width = 0.49\textwidth]{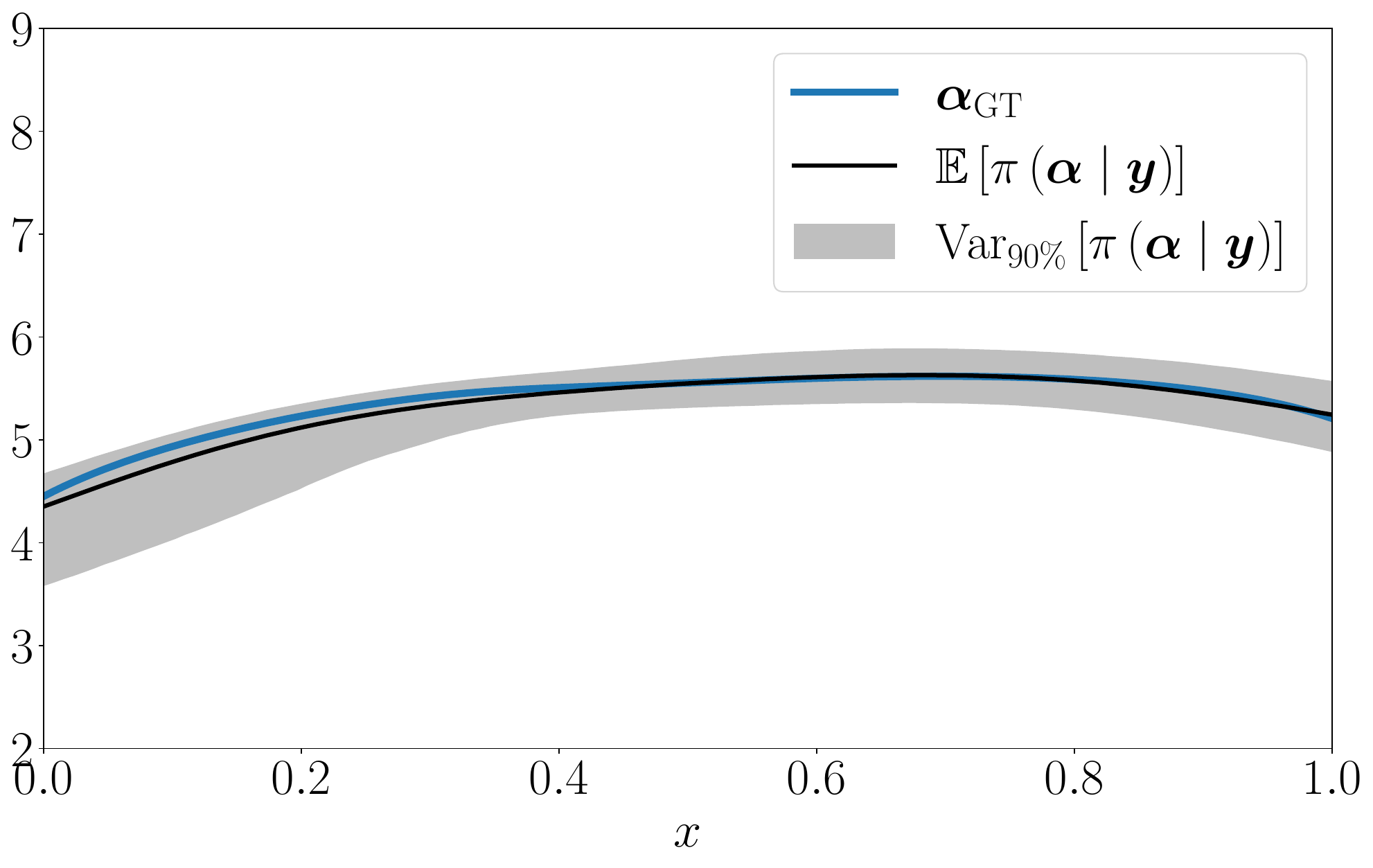} \includegraphics[width = 0.49\textwidth]{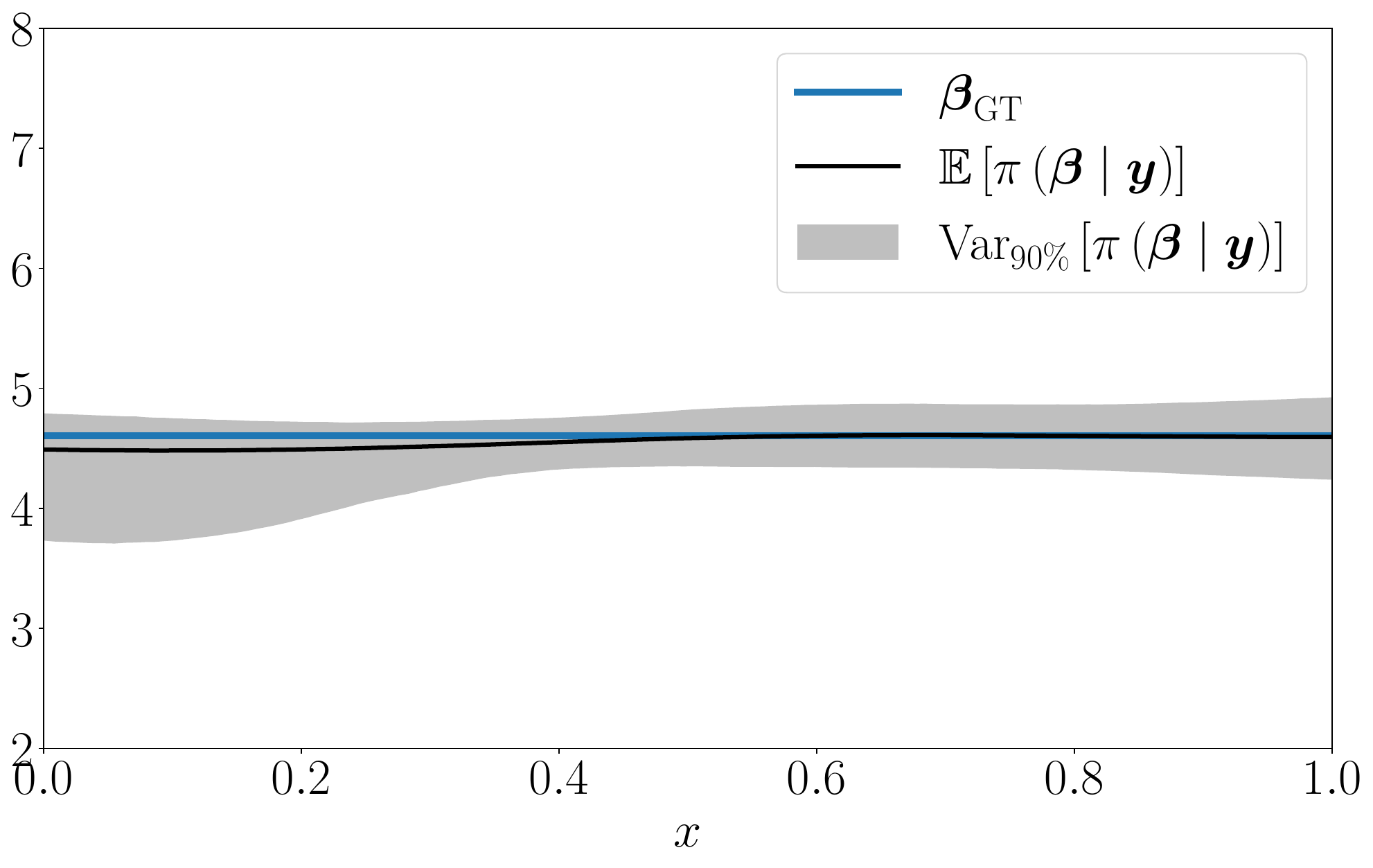} } 
    \end{minipage}\vspace{0.5cm}\\
    
    \begin{minipage}{\textwidth}
        \subfloat[][PL + tempered HMC (short chain).]{ \includegraphics[width = 0.49\textwidth]{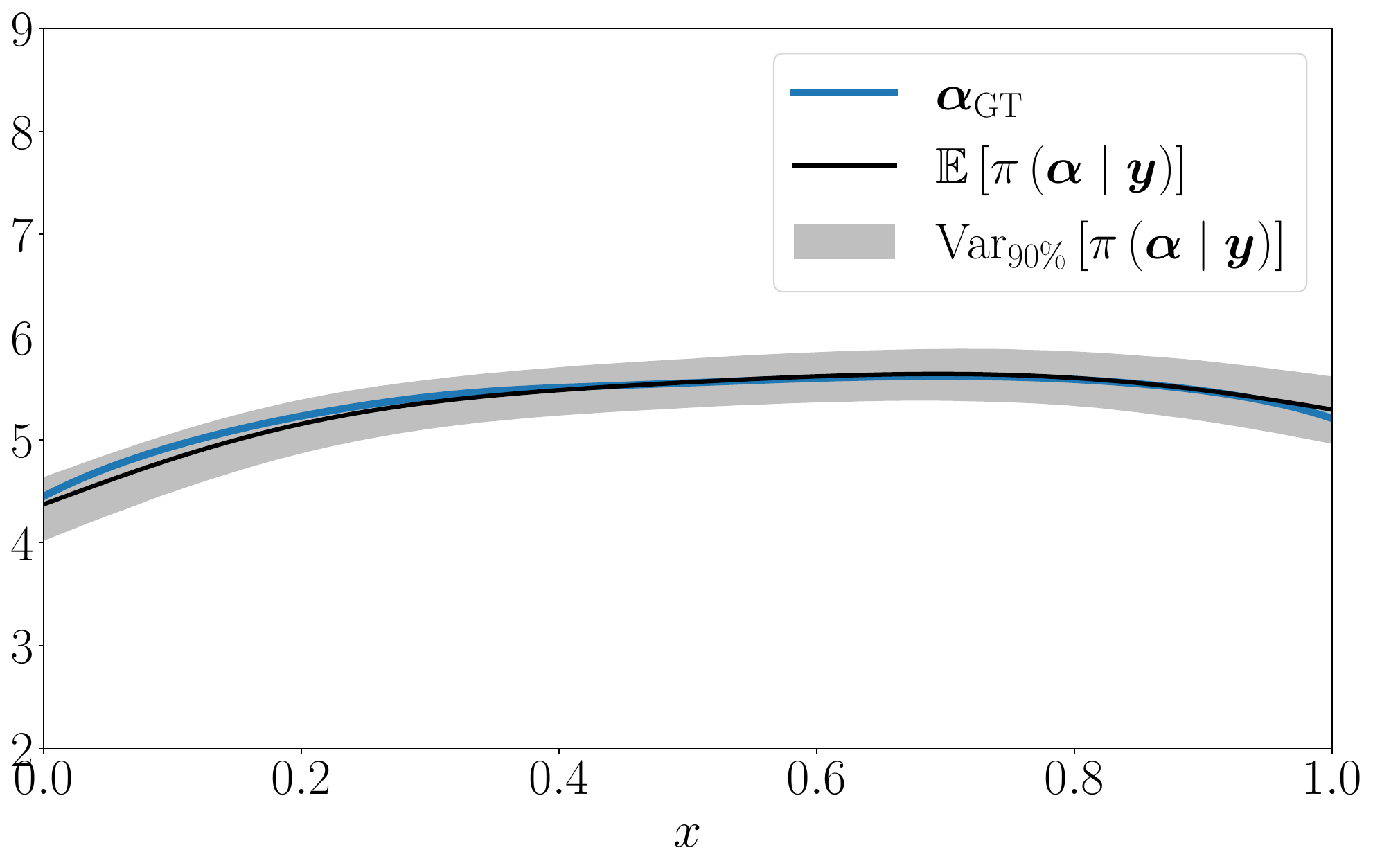} \includegraphics[width = 0.49\textwidth]{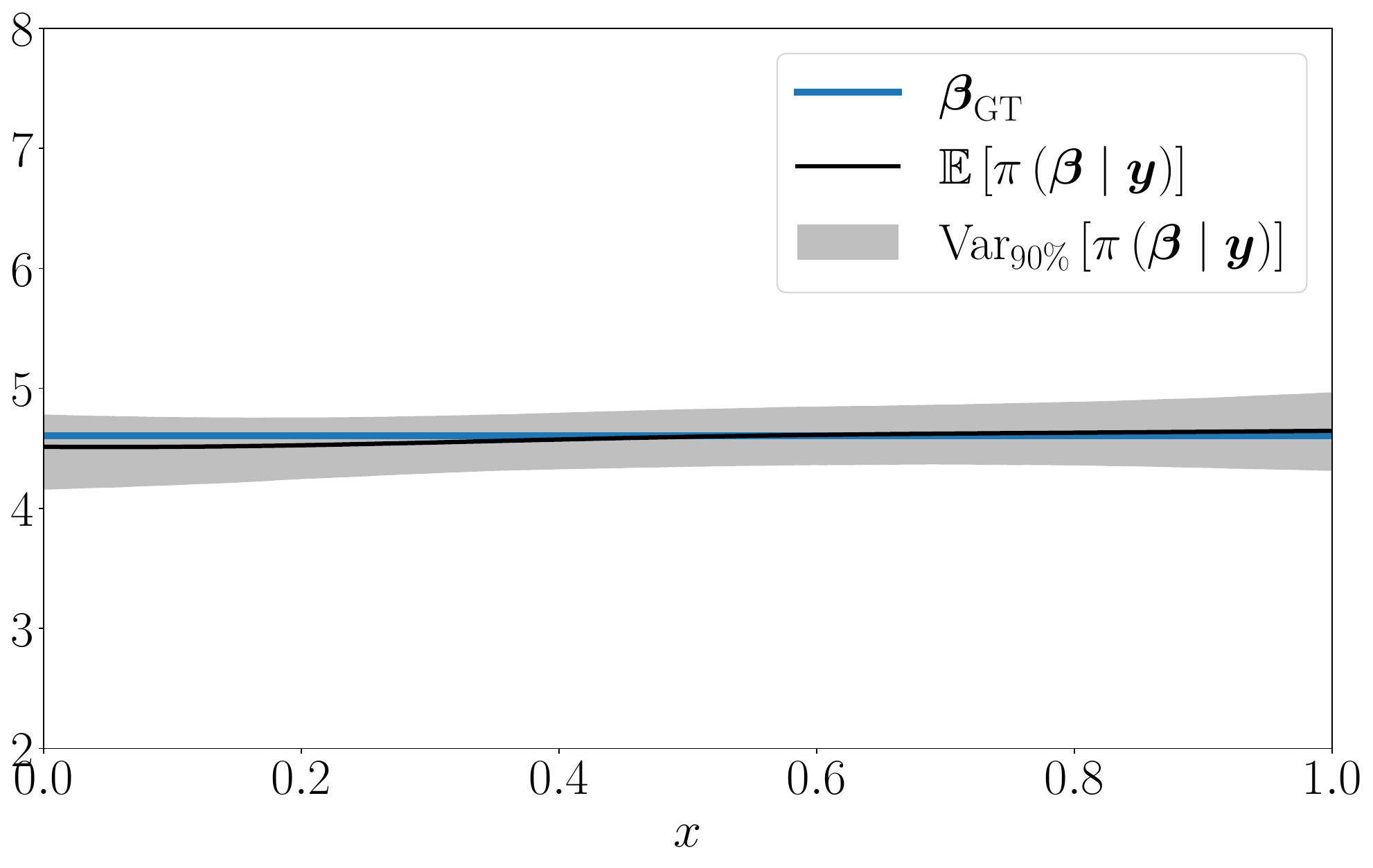} } 
    \end{minipage}
    
    \caption{Young's modulus dataset. The mean estimates $\mathbb{E}\left[ \pi( \boldsymbol{\alpha} \mid \boldsymbol{y} ) \right]$ and $\mathbb{E}\left[ \pi( \boldsymbol{\beta} \mid \boldsymbol{y} ) \right]$ with the corresponding 90\% credible intervals $\textrm{Var}_{90\%}\left[ \pi( \boldsymbol{\alpha} \mid \boldsymbol{y} ) \right]$ and $\textrm{Var}_{90\%}\left[ \pi( \boldsymbol{\beta} \mid \boldsymbol{y} ) \right]$ for the log-shape and log-rate processes obtained with (a) HMC sampling, (b) short chain HMC sampling, and (c) iterated posterior linearization and tempered short chain HMC. The ground truth log-shape and log-rate processes are denoted by $\boldsymbol{\alpha}_\textrm{GT} $ and $\boldsymbol{\beta}_\textrm{GT} $.}
    \label{im:shapeRateMarginalPosteriorTempering_stiffness}
\end{figure}

Similarly, the second dataset is generated as a synthetic dataset, albeit not directly from the log-Gaussian gamma process.
We generate a dataset of 128 Young's moduli using the biocomposite micromechanics model by  \citep{Konigsberger:2024}.
The vector of Young's moduli is used as the expected value of the log-Gaussian gamma process, $\mathbb{E}\left[ \boldsymbol{y} \right] = \exp\{ \boldsymbol{\alpha} \} / \exp\{ \boldsymbol{\beta} \}$ with the 
The synthetic measurement data is generated using a constant shape process $ \exp\{ \boldsymbol{\beta} \} = 1000$ combined with a shape process defined by the Young's modulus vector scaled by a factor of 1000, resulting in the above expectation relation.
We present the synthetic data, true expectation, and 90\% confidence intervals of the data-generating function in conjunction with the estimated mean and marginal 90\% credible intervals obtained using the proposed methods and the two HMC sample chains.
Figure \ref{im:shapeRateMarginalPosterior_stiffness} shows the estimated log-shape and log-rate processes of the longer chain HMC and approximate iterated posterior linearization approach.
Figure \ref{im:shapeRateMarginalPosteriorTempering_stiffness} shows the estimated log-shape and log-rate processes of the shorter and longer HMC chains and for the iterated posterior linearization approach followed by HMC sampling of the tempering sequence.
The obtained one-dimensional marginal posterior distributions for the GP mean and covariance parameters are shown in Figures \ref{im:shapeThetaMarginalPosterior_stiffness} and \ref{im:rateThetaMarginalPosterior_stiffness}.
Note that, even though the data is synthetic, it has not been directly generated from the log-Gaussian gamma process model in Eq.~\eqref{eq:lggpModel}.
Therefore, there are no ground truth values for the GP mean and covariance parameters as with the first synthetic dataset.

\begin{figure}
    \centering
    \begin{minipage}{0.49\textwidth}
        \subfloat[][Ground truth. ]{\includegraphics[width = \textwidth]{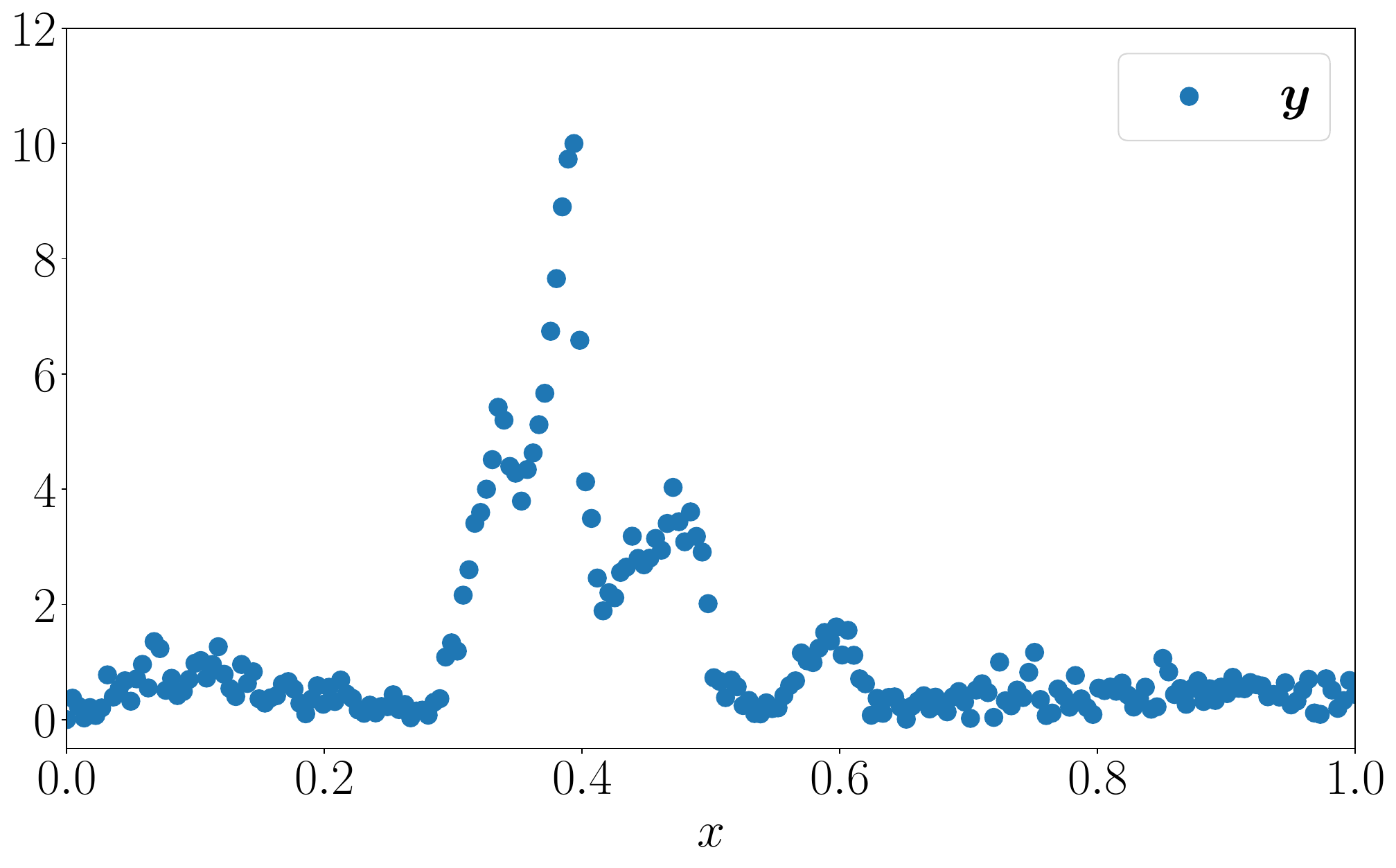}}
    \end{minipage}\\
    
    \begin{minipage}{0.49\textwidth}
        \subfloat[][HMC. ]{ \includegraphics[width = \textwidth]{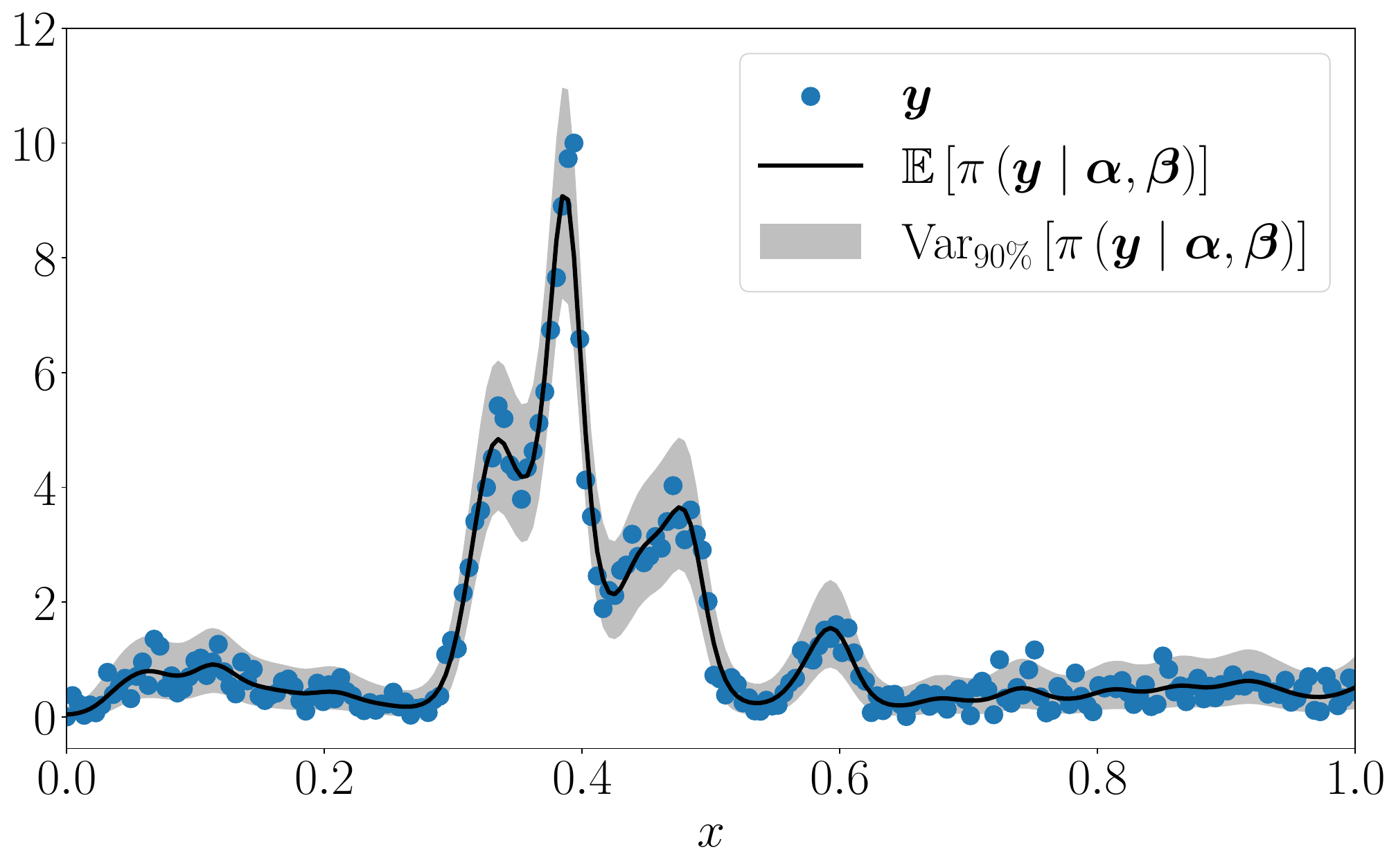} }
    \end{minipage}
    \begin{minipage}{0.49\textwidth}
        \subfloat[][HMC (short chain). ]{\includegraphics[width = \textwidth]{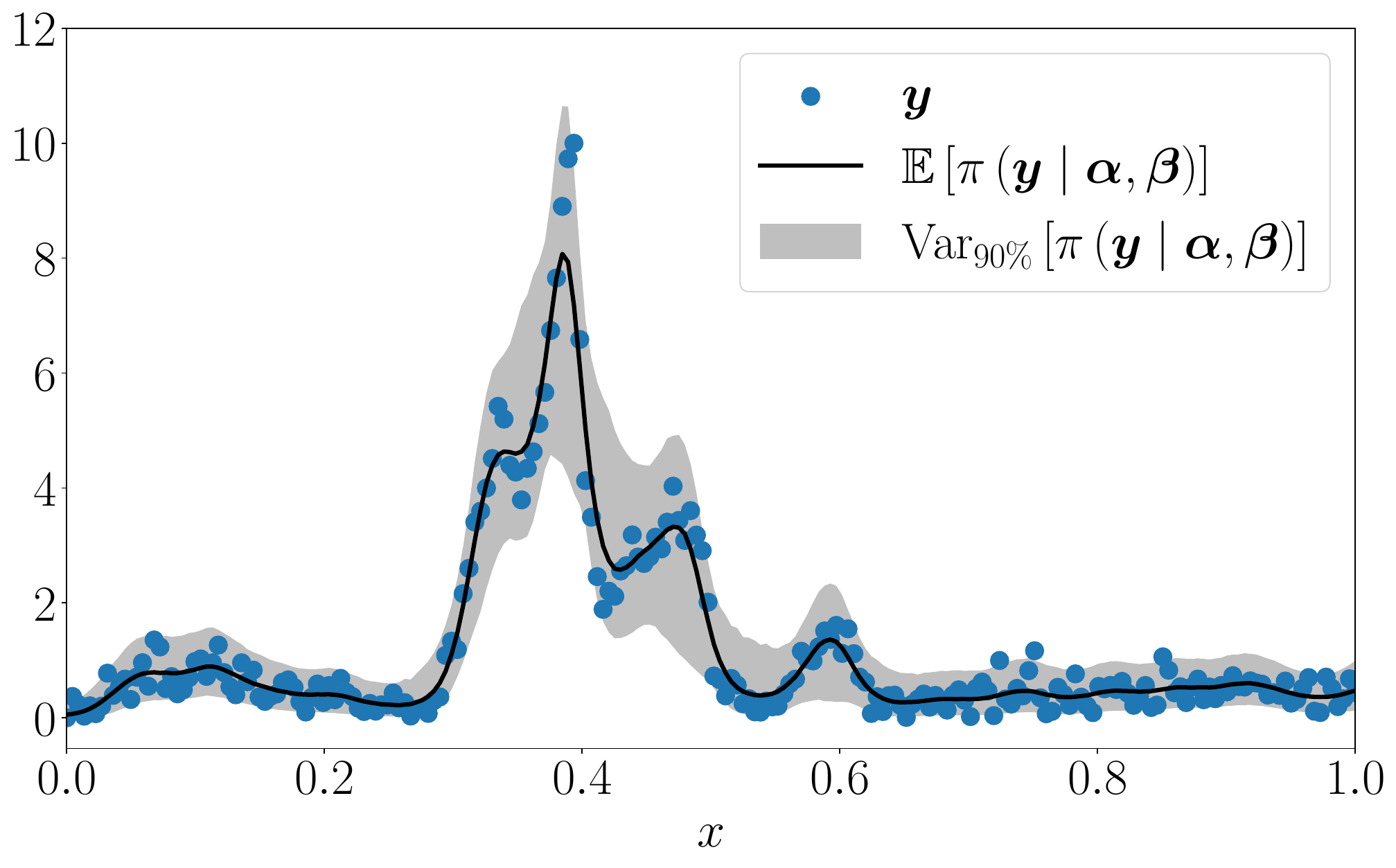}}
    \end{minipage}\\
    
    \begin{minipage}{0.49\textwidth}
        \subfloat[][PL + HMC. ]{ \includegraphics[width = \textwidth]{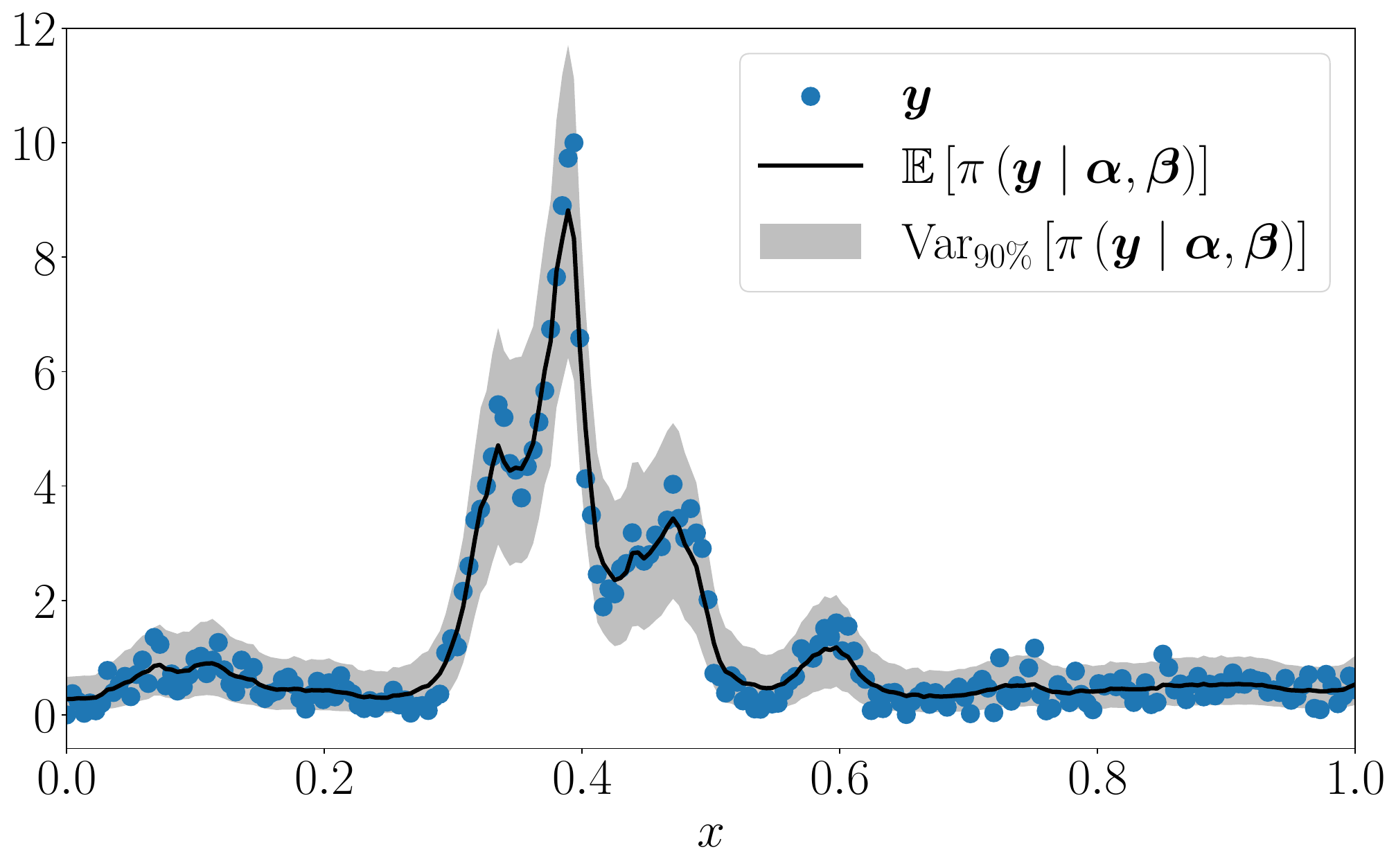} }
    \end{minipage}
    \begin{minipage}{0.49\textwidth}
        \subfloat[][PL + tempered HMC (short chain). ]{\includegraphics[width = \textwidth]{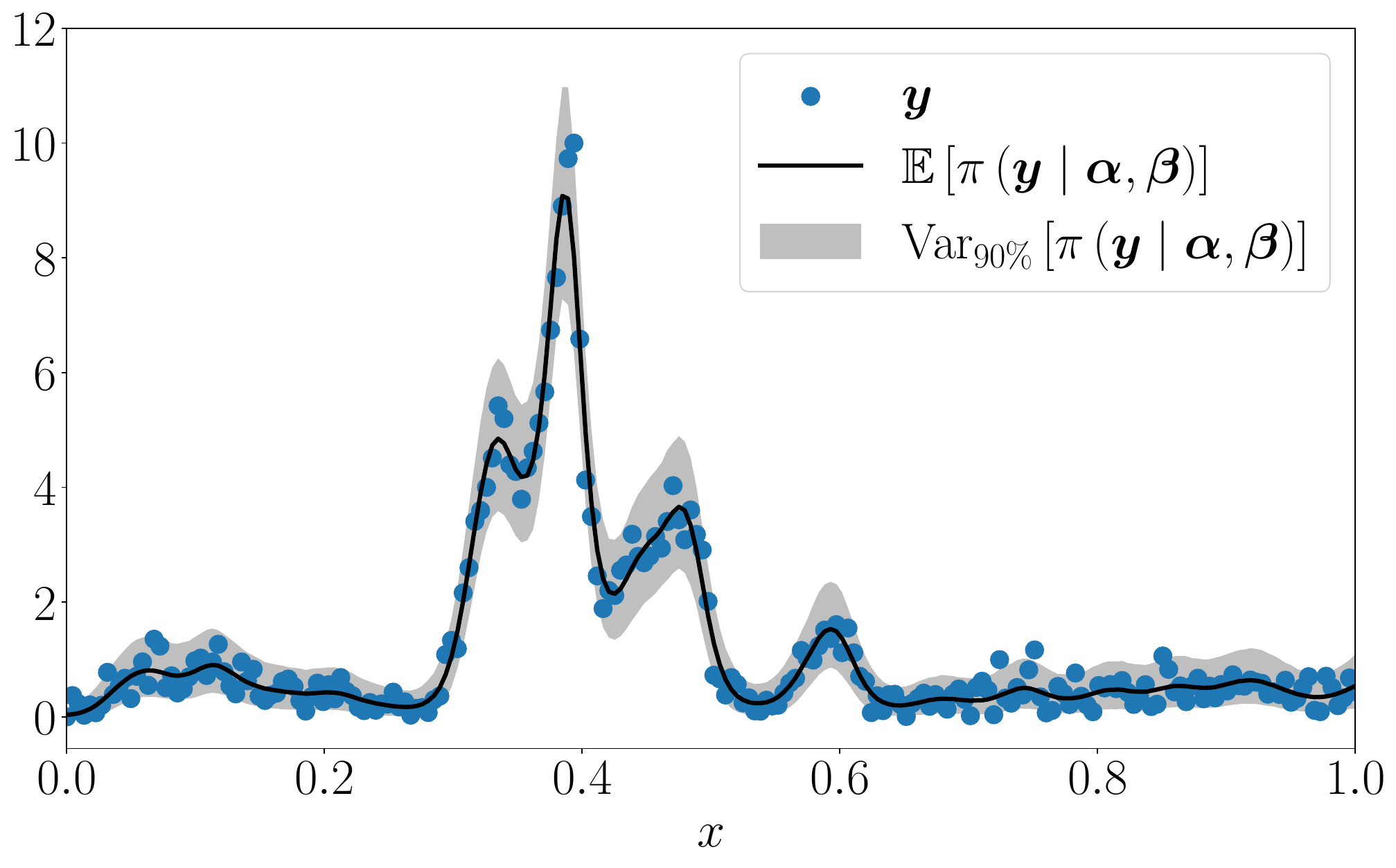}}
    \end{minipage}
    
    \caption{Argentopyrite Raman spectrum dataset. In (a), data in blue. The mean estimate $\mathbb{E}\left[ \pi( \boldsymbol{y} \mid \boldsymbol{\alpha}, \boldsymbol{\beta} ) \right]$ and the corresponding 90\% credible interval $\textrm{Var}_{90\%}\left[ \pi( \boldsymbol{y} \mid \boldsymbol{\alpha}, \boldsymbol{\beta} ) \right]$ for the data-generating function obtained with (b) HMC, (c) short-chain HMC, (d) iterated posterior linearization followed by HMC sampling, and (e) iterated posterior linearization and tempered short-chain HMC sampling.}
    \label{im:dataMarginalPosterior_spectrum}
\end{figure}

\begin{figure}

    \centering
    \begin{minipage}{\textwidth}
        \subfloat[][HMC.]{ \includegraphics[width = 0.49\textwidth]{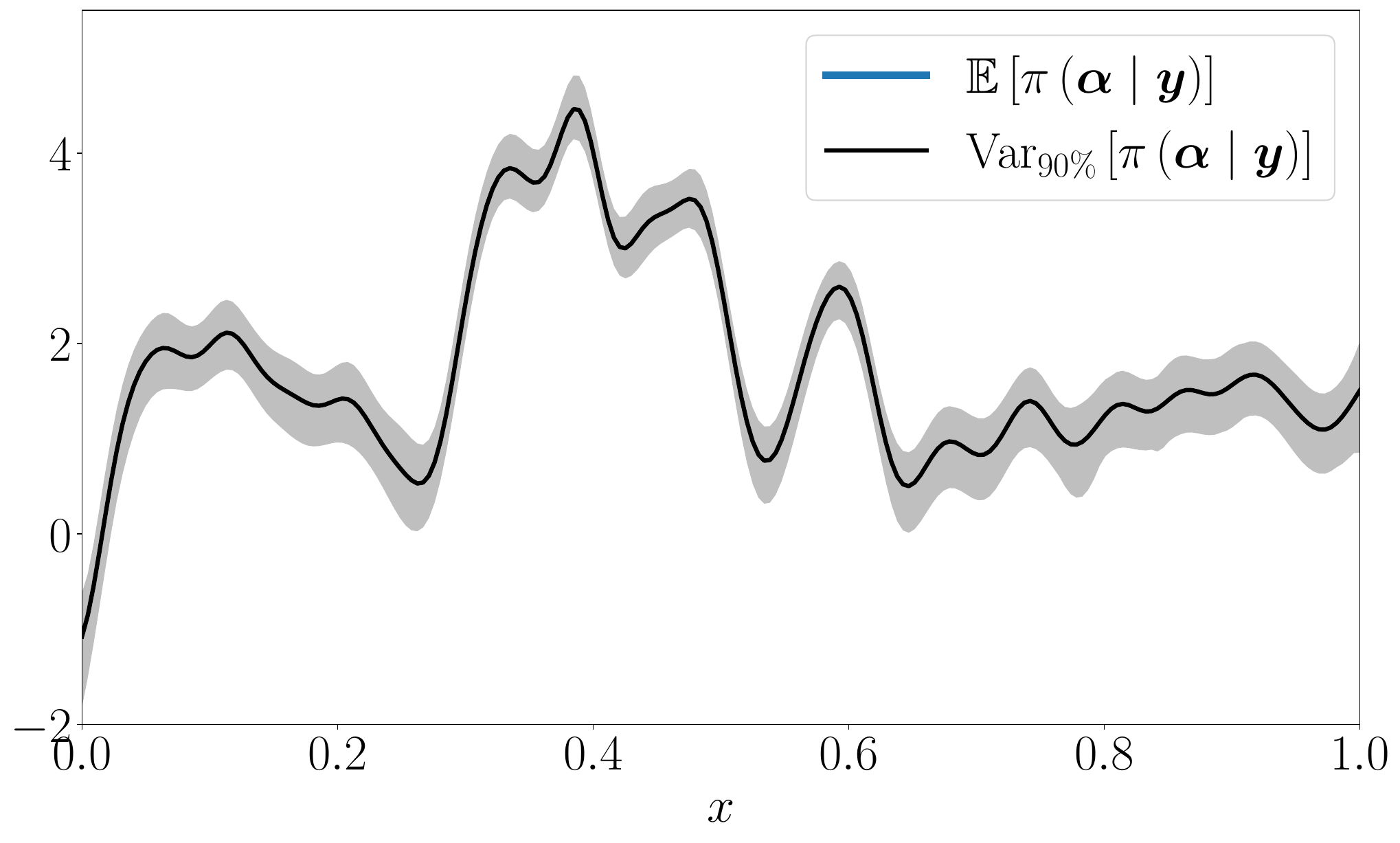} \includegraphics[width = 0.49\textwidth]{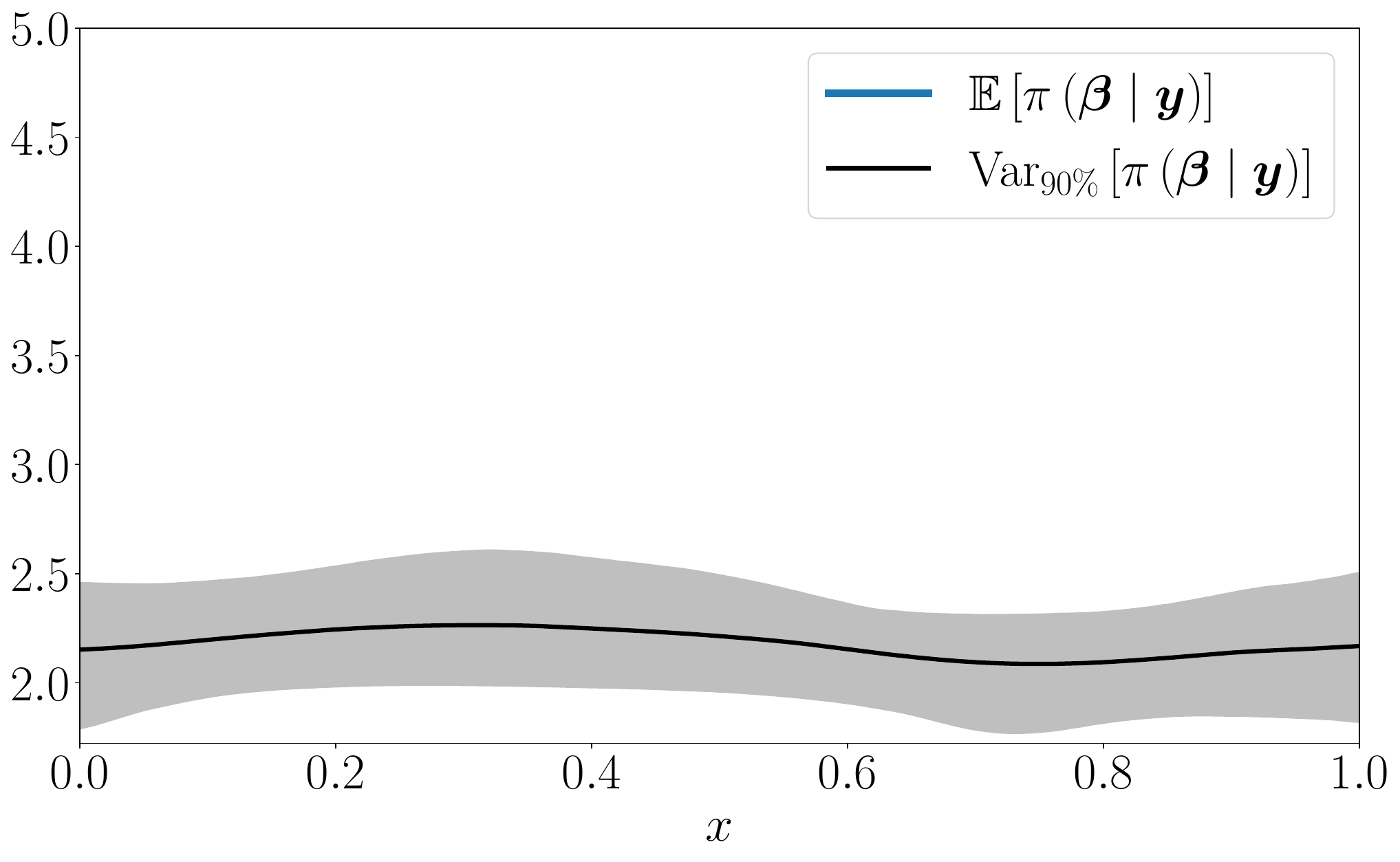} } 
    \end{minipage}\vspace{0.5cm}\\
    
    \begin{minipage}{\textwidth}
        \subfloat[][PL + HMC.]{ \includegraphics[width = 0.49\textwidth]{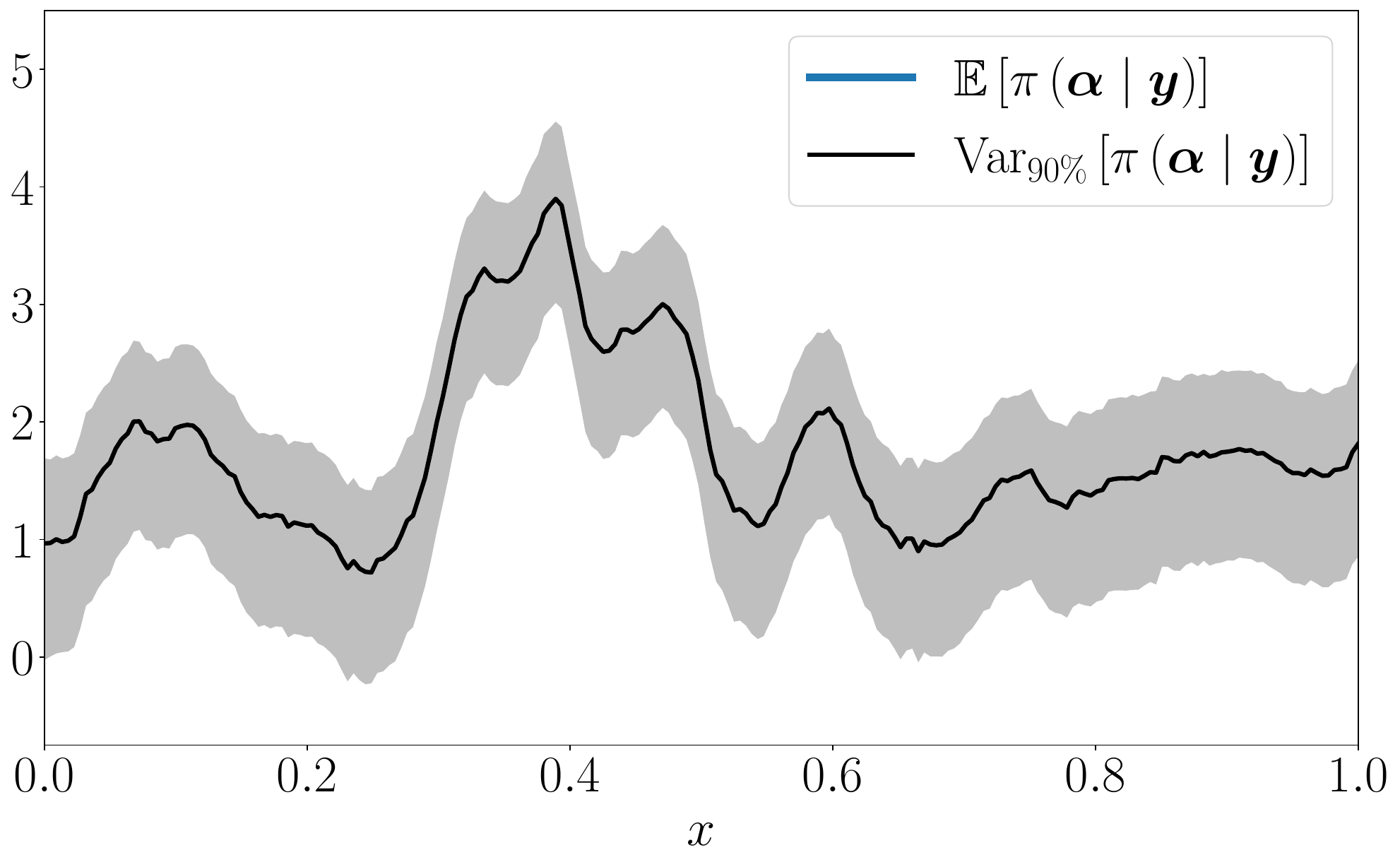} \includegraphics[width = 0.49\textwidth]{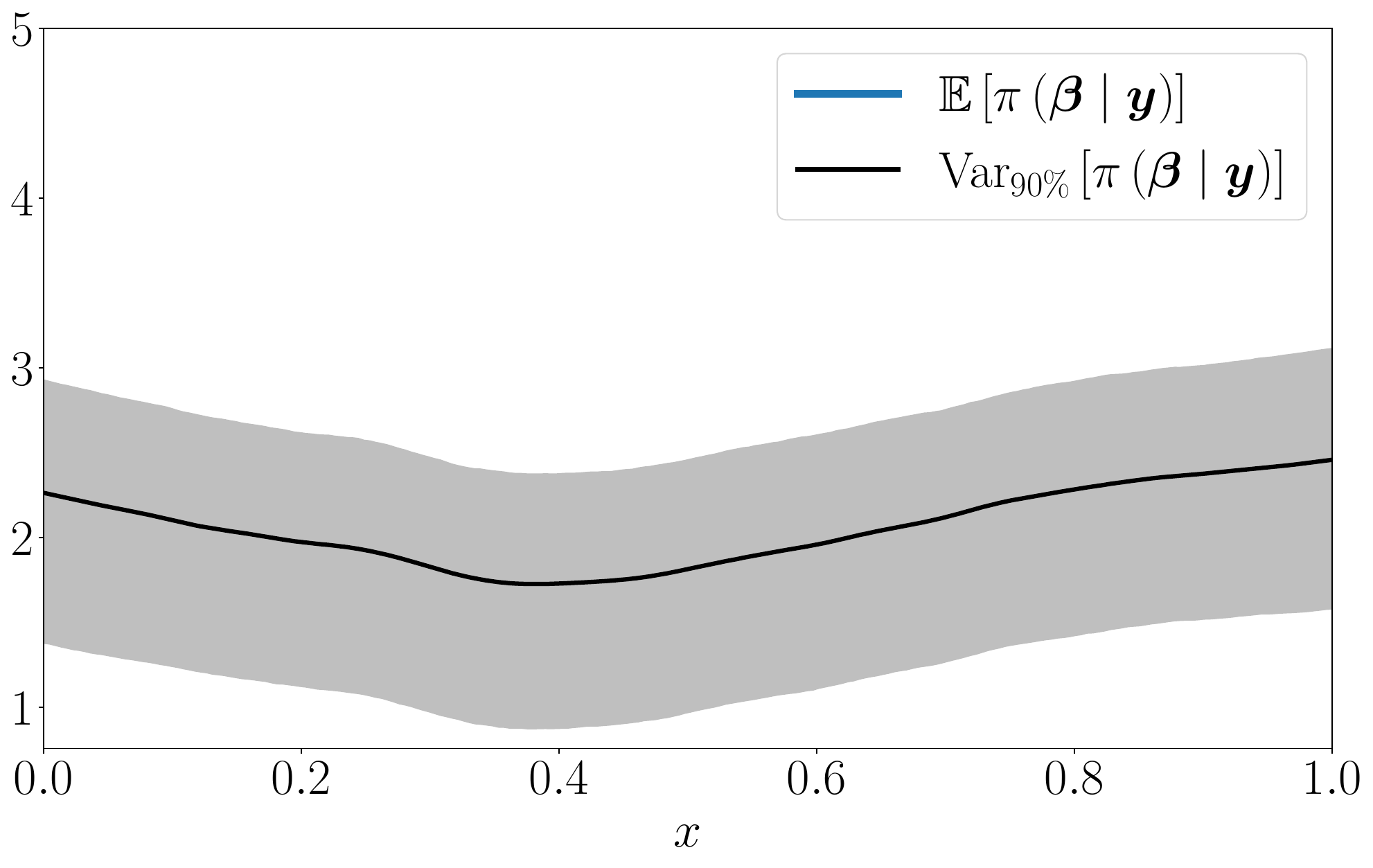} } 
    \end{minipage}
    
    \caption{Argentopyrite Raman spectrum dataset. The mean estimates $\mathbb{E}\left[ \pi( \boldsymbol{\alpha} \mid \boldsymbol{y} ) \right]$ and $\mathbb{E}\left[ \pi( \boldsymbol{\beta} \mid \boldsymbol{y} ) \right]$ with the corresponding 90\% credible intervals $\textrm{Var}_{90\%}\left[ \pi( \boldsymbol{\alpha} \mid \boldsymbol{y} ) \right]$ and $\textrm{Var}_{90\%}\left[ \pi( \boldsymbol{\beta} \mid \boldsymbol{y} ) \right]$ for the log-shape and log-rate processes obtained with (a) HMC sampling and (b) iterated posterior linearization followed by HMC sampling.}
    \label{im:shapeRateMarginalPosterior_spectrum}
\end{figure}

\begin{figure}

    \centering
    \begin{minipage}{\textwidth}
        \subfloat[][HMC.]{ \includegraphics[width = 0.49\textwidth]{pics/vector/argent/shape/shape_hmc.pdf} \includegraphics[width = 0.49\textwidth]{pics/vector/argent/rate/rate_hmc.pdf} } 
    \end{minipage}\vspace{0.5cm}\\
    
    \begin{minipage}{\textwidth}
        \subfloat[][HMC (short chain).]{ \includegraphics[width = 0.49\textwidth]{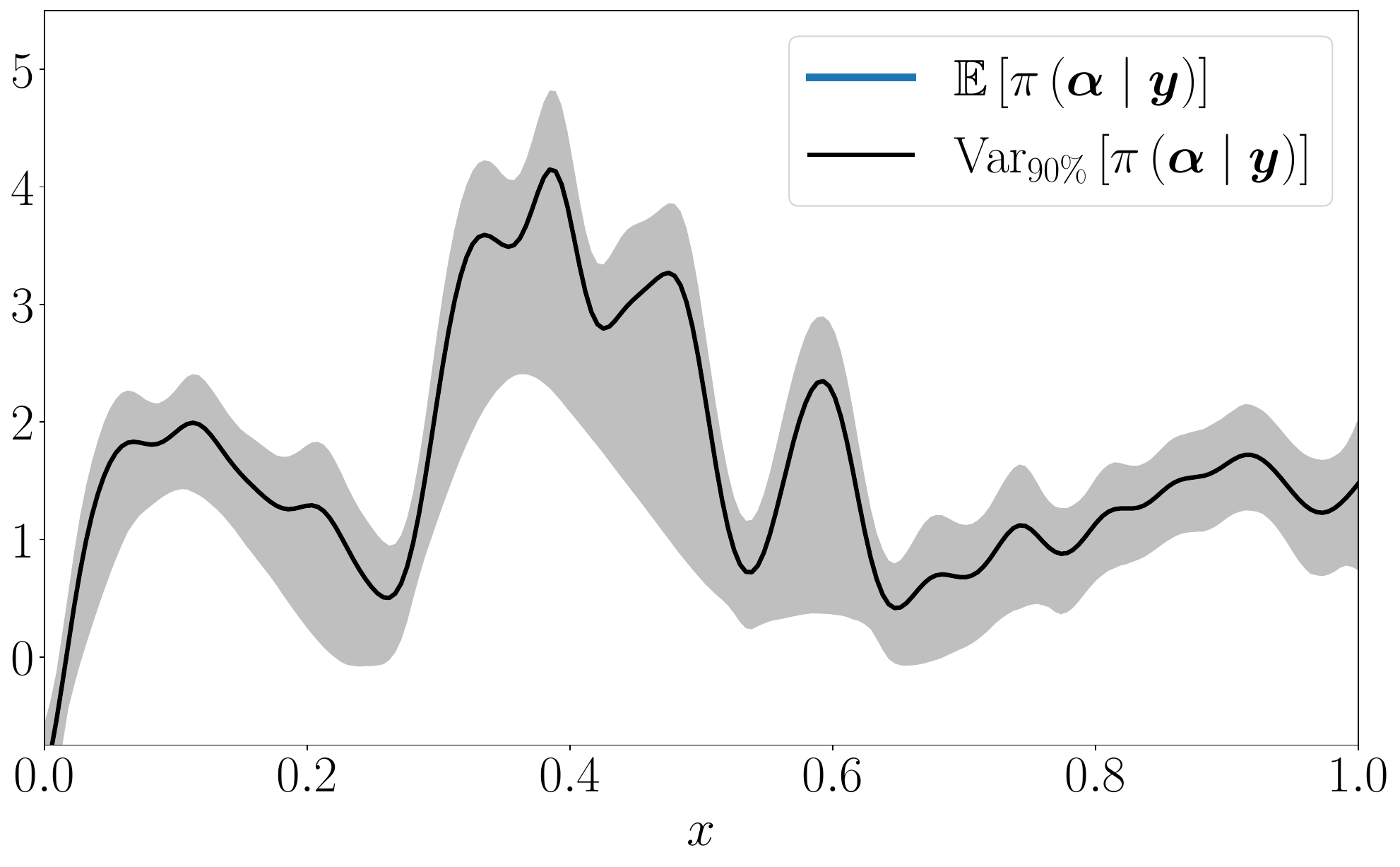} \includegraphics[width = 0.49\textwidth]{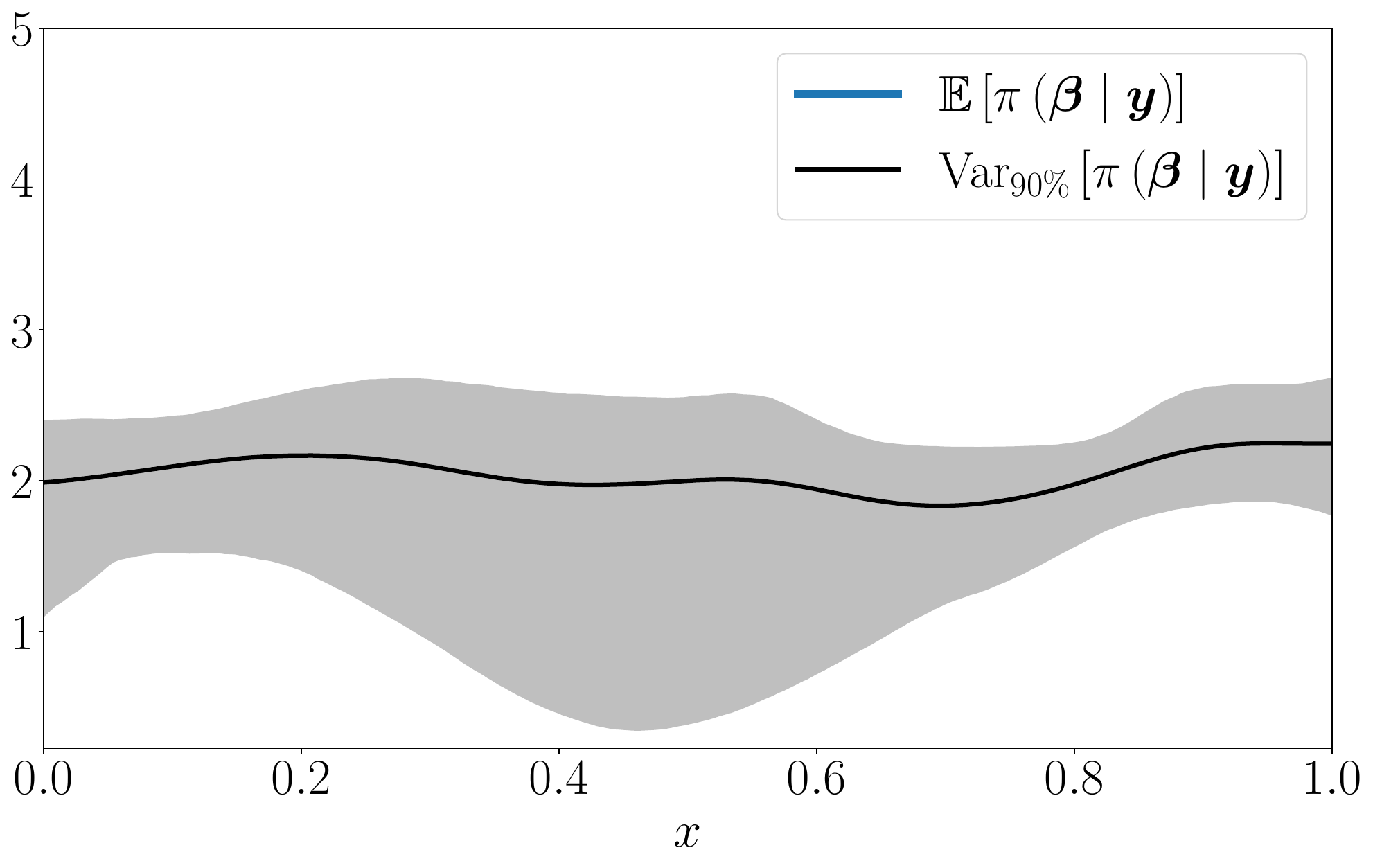} } 
    \end{minipage}\vspace{0.5cm}\\
    
    \begin{minipage}{\textwidth}
        \subfloat[][PL + tempered HMC (short chain).]{ \includegraphics[width = 0.49\textwidth]{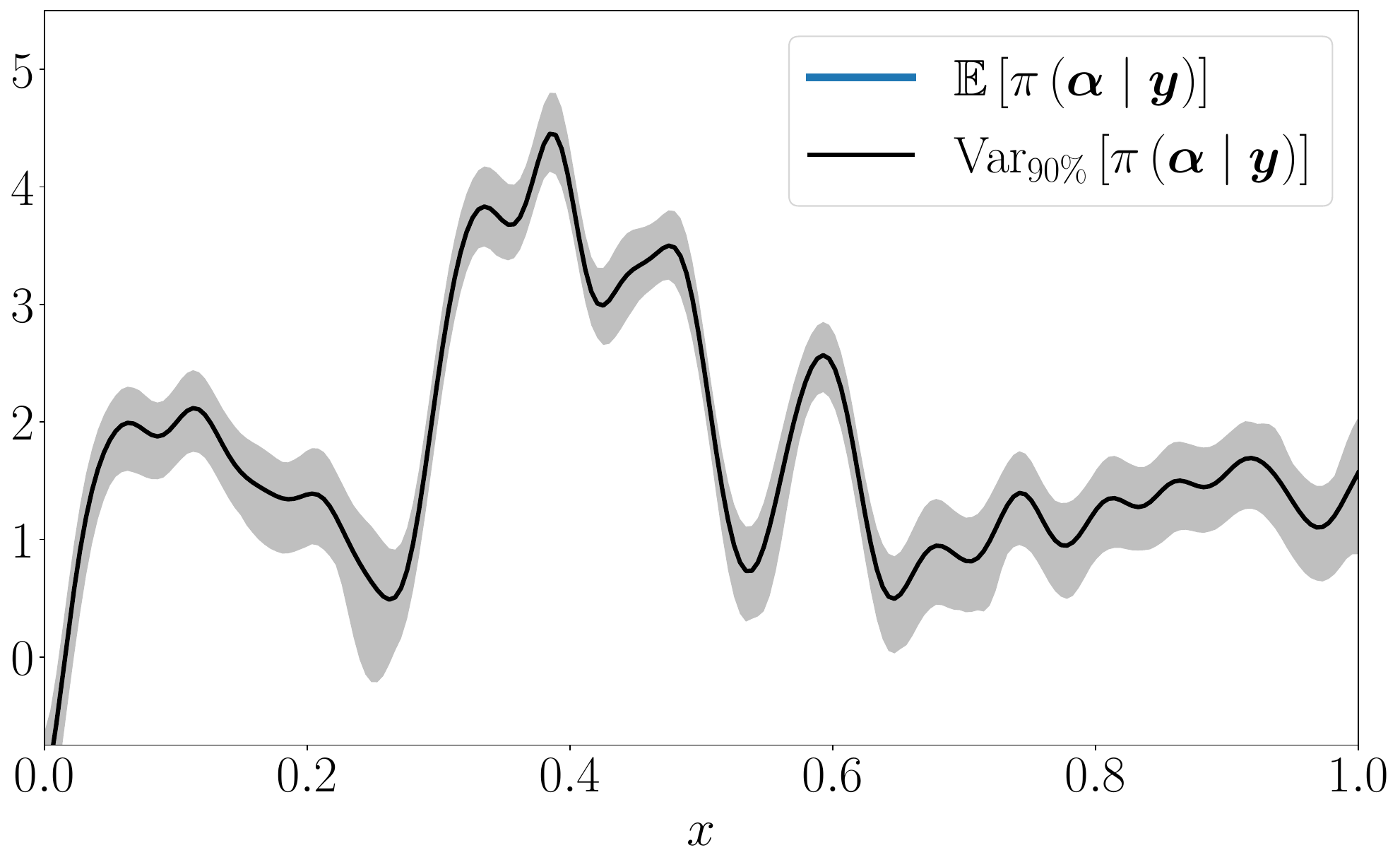} \includegraphics[width = 0.49\textwidth]{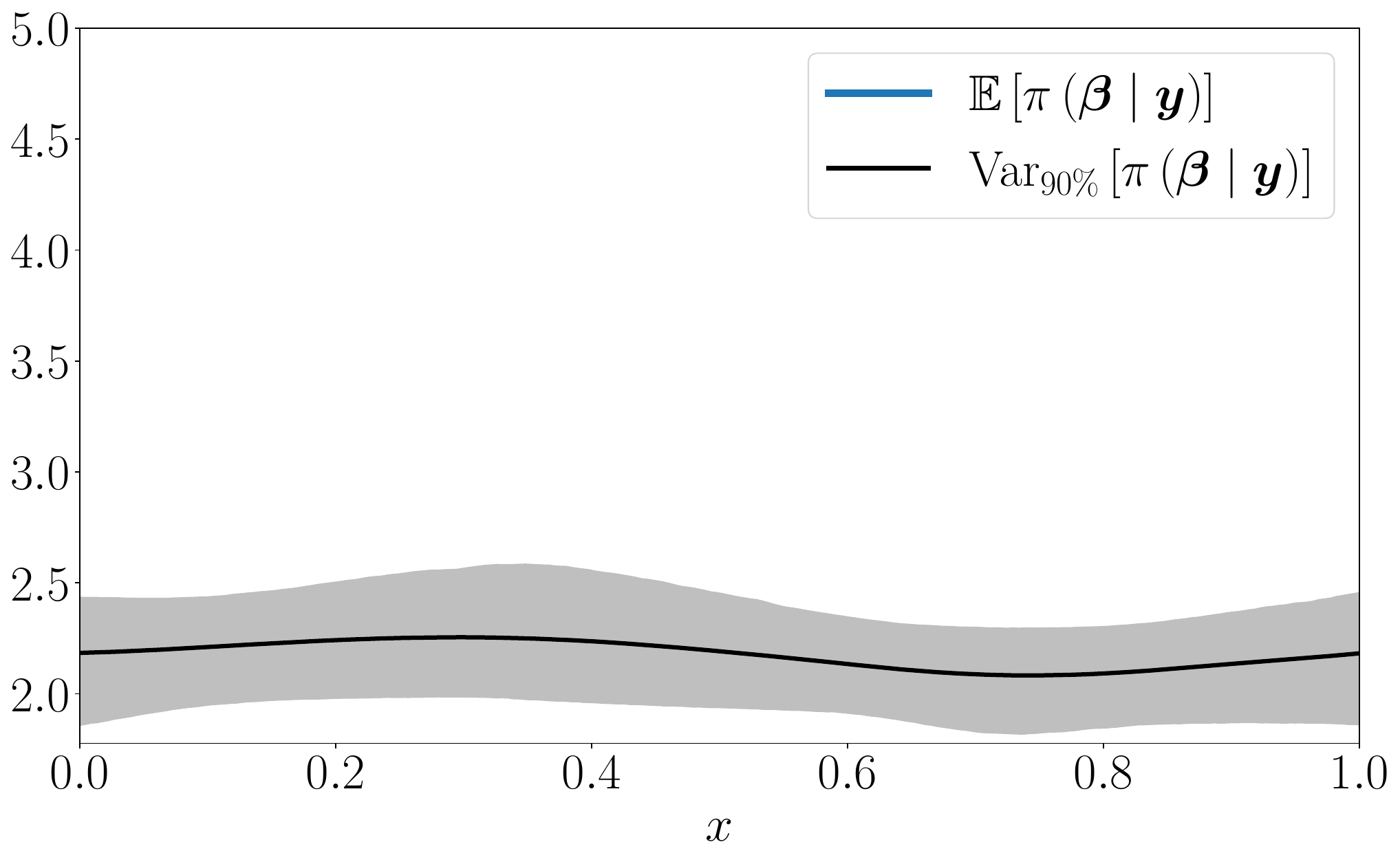} } 
    \end{minipage}
    
    \caption{Argentopyrite Raman spectrum dataset. The mean estimates $\mathbb{E}\left[ \pi( \boldsymbol{\alpha} \mid \boldsymbol{y} ) \right]$ and $\mathbb{E}\left[ \pi( \boldsymbol{\beta} \mid \boldsymbol{y} ) \right]$ with the corresponding 90\% credible intervals $\textrm{Var}_{90\%}\left[ \pi( \boldsymbol{\alpha} \mid \boldsymbol{y} ) \right]$ and $\textrm{Var}_{90\%}\left[ \pi( \boldsymbol{\beta} \mid \boldsymbol{y} ) \right]$ for the log-shape and log-rate processes obtained with (a) HMC sampling, (b) short chain HMC sampling, and (c) iterated posterior linearization and tempered short chain HMC.}
    \label{im:shapeRateMarginalPosteriorTempering_spectrum}
\end{figure}

Third, we apply the methods to an experimental Raman spectrum of argentopyrite.
The argentopyrite Raman spectrum is cutoff at 600nm and normalized such that $\max\{ \boldsymbol{y} \} = 10$ with $\min\{ \boldsymbol{x} \} = 0$ and $\max\{ \boldsymbol{x} \} = 1$.
The truncated spectrum consists of 222 data points.
We present the experimental data and estimated data-generating functions in Figure \ref{im:dataMarginalPosterior_spectrum}.
Similarly to the synthetic datasets, the mean and marginal 90\% credible intervals for the log-shape and log-rate processes of the longer chain HMC and approximate iterated posterior linearization approach are illustrated in Figure \ref{im:shapeRateMarginalPosterior_spectrum}.
In turn, Figure \ref{im:shapeRateMarginalPosteriorTempering_spectrum} shows the estimated log-shape and log-rate processes of the shorter and longer HMC chains and for the iterated posterior linearization approach followed by HMC sampling of the tempering sequence.
The respective one-dimensional marginal posterior distributions for the GP mean and covariance parameters are shown in Figures \ref{im:shapeThetaMarginalPosterior_spectrum} and \ref{im:rateThetaMarginalPosterior_spectrum}.

We present $L_1$ norms of differences of the proposed methods and short-chain HMC in comparison to the long-chain HMC sampling for the mean and marginal 90\% credible intervals for the log-shape and log-rate processes.
We use the long-chain HMC sampling results as the benchmark, as the posterior distribution is intractable, leaving the long HMC sampling run as the best available point of reference.
In addition, we present inference wall times for the three datasets and speed-ups provided by the proposed methods in Table \ref{tb:walltimes}.
The computed iterated posterior linearization was computed using $J = 10,000$.
For the longer and shorter HMC inferences, the NUTS sampler was run with 20,000 samples with 10,000 tuning steps and 1000 samples with 1200 tuning steps, respectively.
For estimating the sequence of tempered posterior distributions, we used three tempering steps with $ \boldsymbol{\kappa} = \left( \kappa( 0 ), \kappa( 1), \kappa( 2 ) \right)^\intercal = \left( 0, 1/2, 1 \right)^\intercal$ with 100 tuning samples for the first two tempered distributions and with 1000 samples and 1000 tunings steps for the final iteration.
This corresponds to the number of samples used for the shorter HMC chain.
In all cases, the HMC target acceptance rate was set to $0.99$.
The parameters of the used prior distributions are detailed in Table \ref{tb:priorDistributions}.
\begin{table}
    \centering
    \begin{tabular}{c|c|c|c}
         Parameter & Synthetic & Young's modulus & Argentopyrite \\
        \toprule
         $ \gamma_{ \mu, \alpha} $ & 2 & 4 & 1 \\
         $ \rho_{ \mu, \alpha} $ & 1 & 1 & 0.5 \\
         
         $ \gamma_{ \mu, \beta} $ & 1 & 4 & 3 \\
         $ \rho_{ \mu, \beta} $ & 0.5 & 1 & 0.5 \\

         \hline
         
         $ \rho_{ \sigma, \textrm{{e}}, \alpha} $ & 0.001 & 0.001 & 0.001 \\
         $ \rho_{ \sigma, \textrm{{e}}, \beta} $ & 0.001 & 0.001 & 0.001 \\
         
         $ \rho_{ \sigma, \textrm{{s}}, \alpha} $ & 0.5 & 0.5 & 0.5 \\
         $ \rho_{ \sigma, \textrm{{s}}, \beta} $ & 0.5 & 0.5 & 0.5 \\

         \hline
         
         $ \gamma_{ l, \alpha} $ & 0.1 & 0.5 & 0.1 \\
         $ \rho_{ l, \alpha} $ & 0.2 & 0.2 & 0.1 \\
         $ B_{\alpha} $ & 0.01 & 0.01 & 0.001 \\
         
         $ \gamma_{ l, \beta} $ & 0.5 & 0.5 & 0.5 \\
         $ \rho_{ l, \beta} $ & 0.2 & 0.2 & 0.2 \\
         $ B_{\beta} $ & 0.25 & 0.25 & 0.025 \\

    \end{tabular}
    \caption{Prior distribution parameter values for each dataset.}
    \label{tb:priorDistributions}
\end{table}

\begin{figure}
    \centering
    \begin{minipage}{0.49\textwidth}
        \subfloat[][Ground truth. ]{\includegraphics[width = \textwidth]{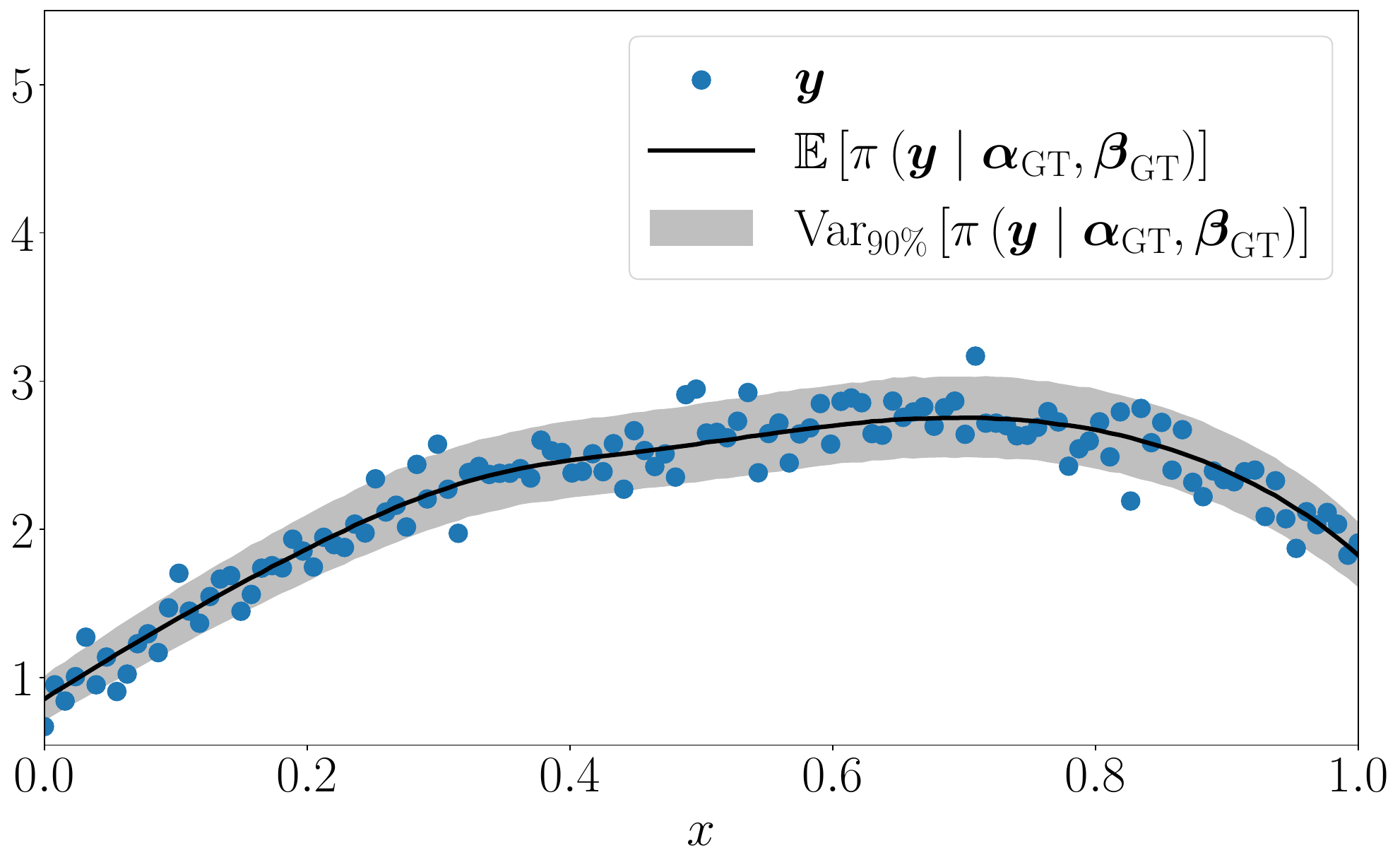}}
    \end{minipage}\\
    
    \begin{minipage}{0.49\textwidth}
        \subfloat[][HMC. ]{ \includegraphics[width = \textwidth]{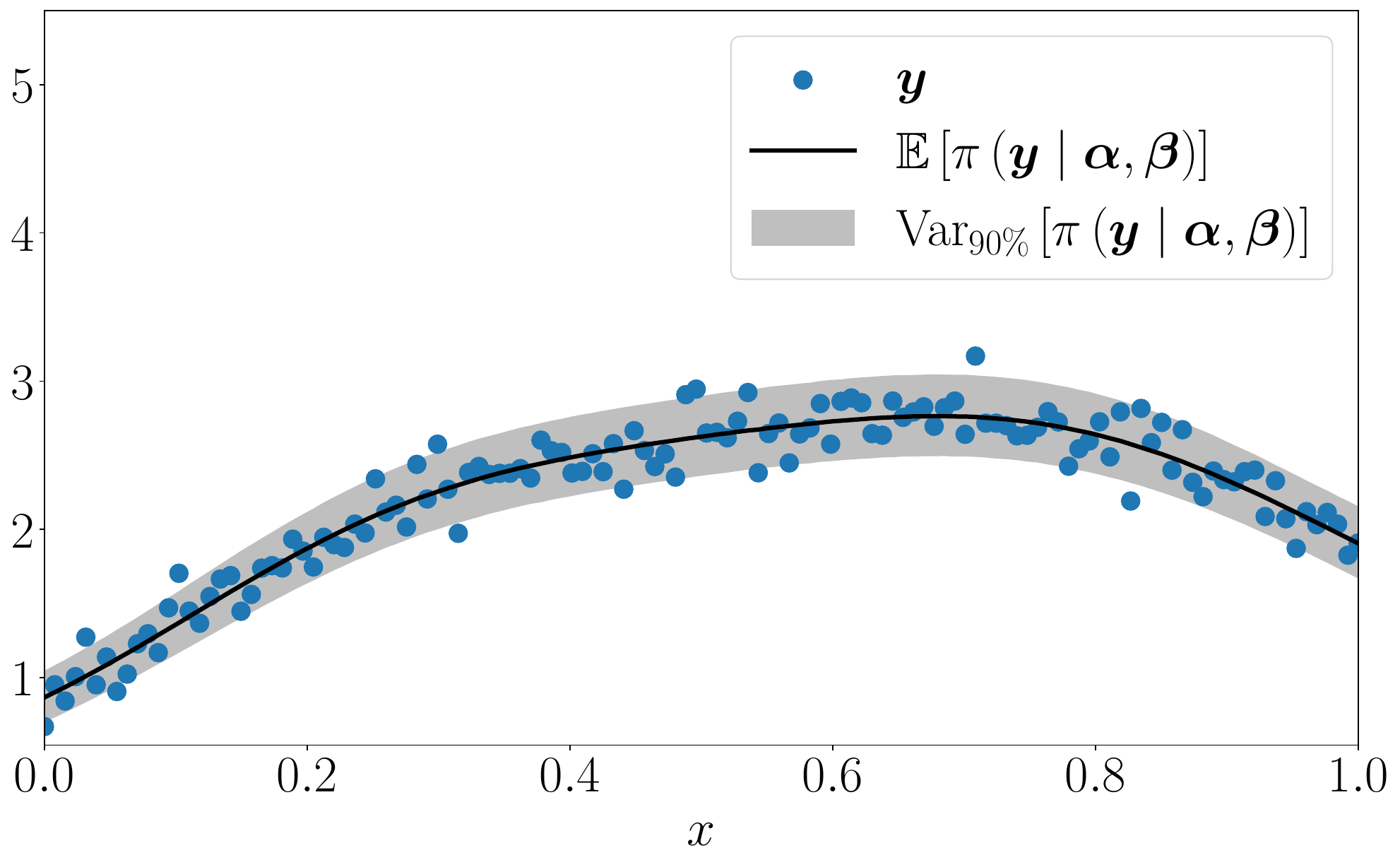} }
    \end{minipage}
    \begin{minipage}{0.49\textwidth}
        \subfloat[][HMC (short chain). ]{\includegraphics[width = \textwidth]{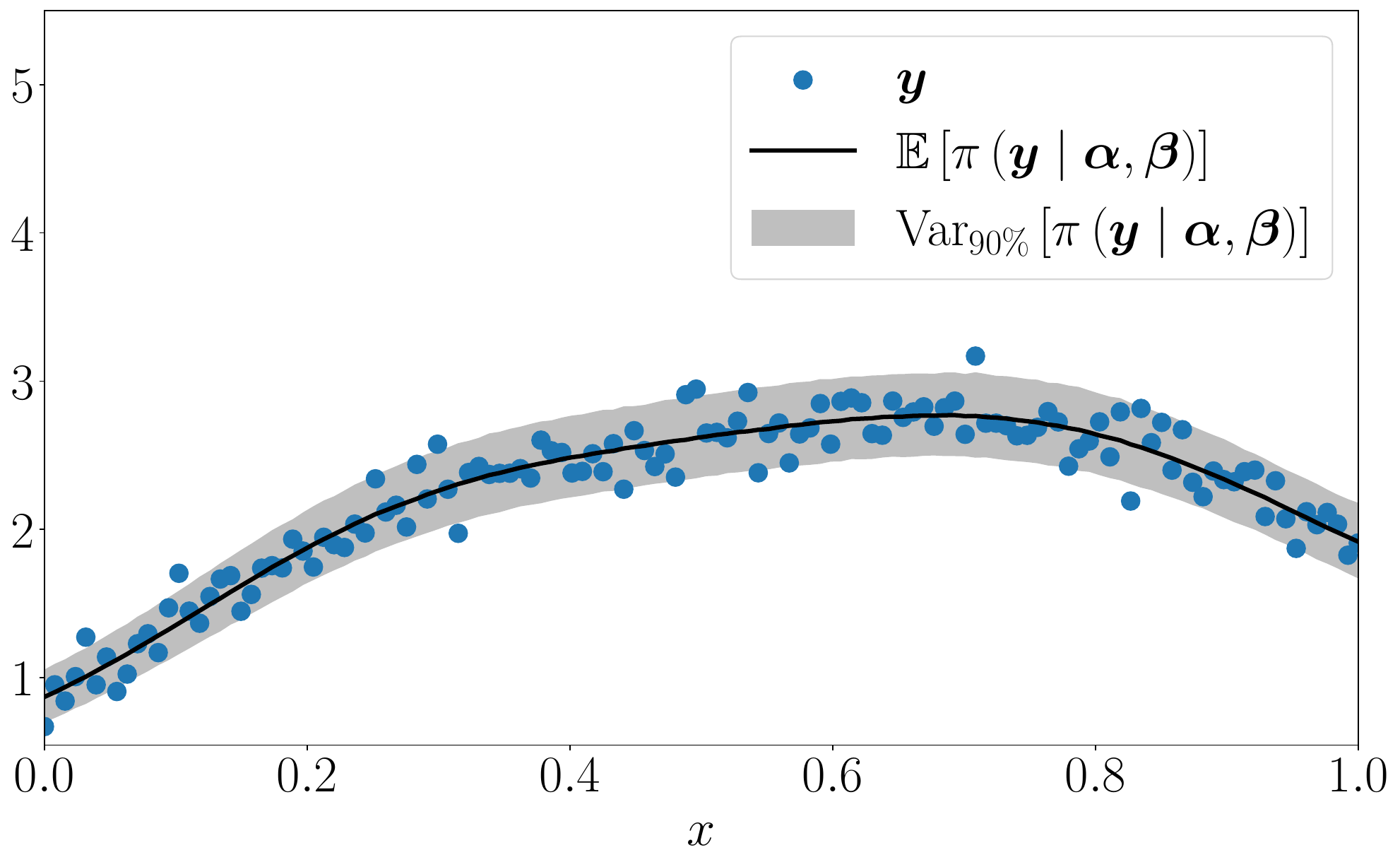}}
    \end{minipage}\\
    
    \begin{minipage}{0.49\textwidth}
        \subfloat[][PL + HMC. ]{ \includegraphics[width = \textwidth]{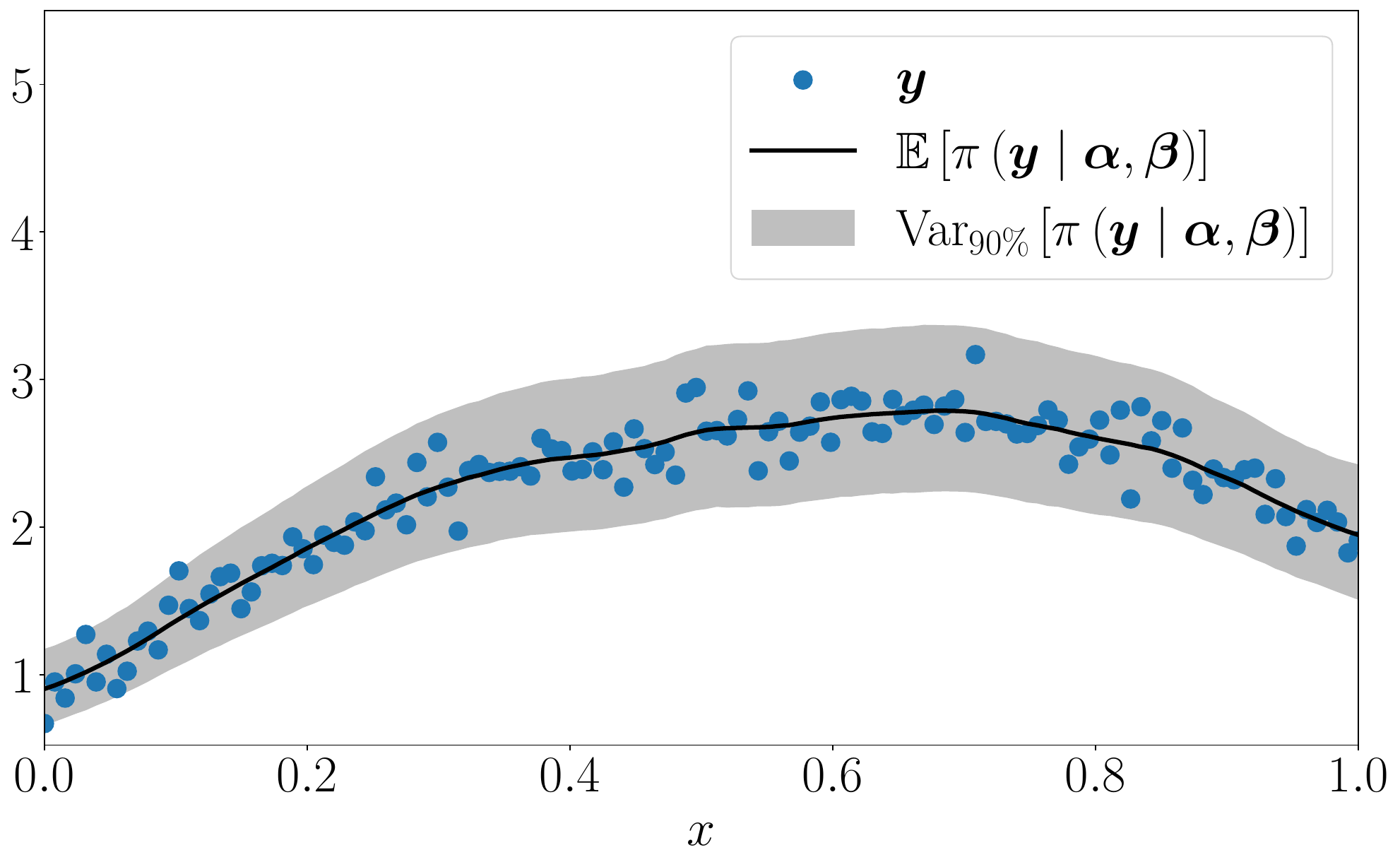} }
    \end{minipage}
    \begin{minipage}{0.49\textwidth}
        \subfloat[][PL + tempered HMC (short chain). ]{\includegraphics[width = \textwidth]{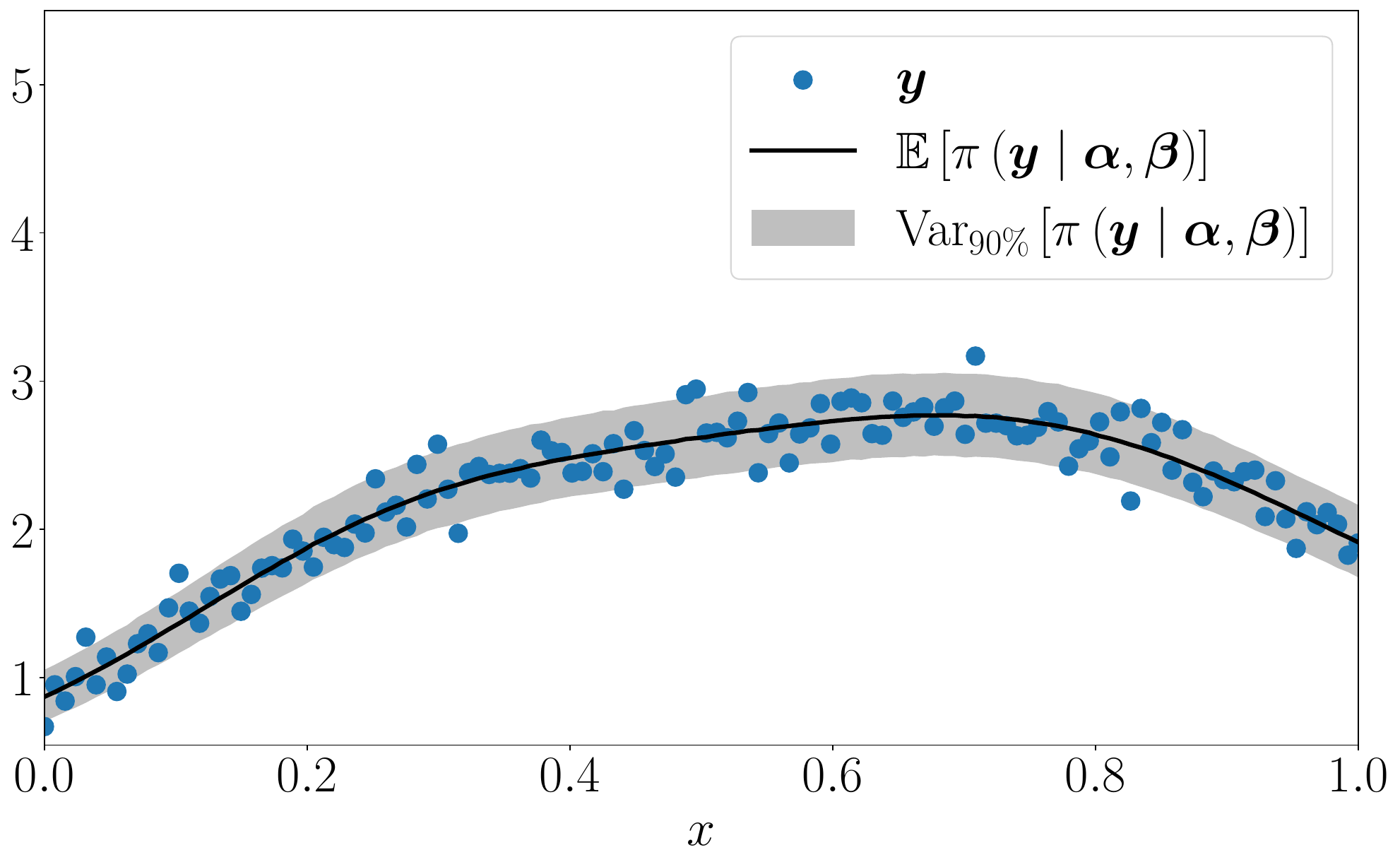}}
    \end{minipage}
    
    \caption{Young's modulus dataset. In (a), data in blue, the ground truth expectation, and 90\% confidence interval of the data-generating function in black and gray, respectively, and $\boldsymbol{\alpha}_\textrm{GT} $ and $\boldsymbol{\beta}_\textrm{GT} $ denote the ground truth log-shape and log-rate processes. The mean estimate $\mathbb{E}\left[ \pi( \boldsymbol{y} \mid \boldsymbol{\alpha}, \boldsymbol{\beta} ) \right]$ and the corresponding 90\% credible interval $\textrm{Var}_{90\%}\left[ \pi( \boldsymbol{y} \mid \boldsymbol{\alpha}, \boldsymbol{\beta} ) \right]$ for the data-generating function obtained with (b) HMC, (c) short-chain HMC, (d) iterated posterior linearization followed by HMC sampling, and (e) iterated posterior linearization and tempered short-chain HMC sampling.}
    \label{im:dataMarginalPosterior_stiffness}
\end{figure}

The resulting data-generating functions of the proposed methods in Figures \ref{im:dataMarginalPosterior_synthetic}, \ref{im:dataMarginalPosterior_stiffness}, and \ref{im:dataMarginalPosterior_spectrum} show good agreement with the true data-generating function, with some overestimation of the marginal credible intervals given by the approximate iterated posterior linearization compared to direct HMC sampling.
In particular, the estimates given by iterated posterior linearization and tempering are in close agreement with the results of the HMC sampling.
Most importantly, the estimates show superior mixing when compared to the shorter chain HMC sampling with the same number of samples.

The log-shape and log-rate estimates illustrated in Figures \ref{im:shapeRateMarginalPosterior_synthetic}, \ref{im:shapeRateMarginalPosteriorTempering_synthetic}, \ref{im:shapeRateMarginalPosterior_stiffness}, \ref{im:shapeRateMarginalPosteriorTempering_stiffness}, \ref{im:shapeRateMarginalPosterior_spectrum}, and \ref{im:shapeRateMarginalPosteriorTempering_spectrum} follow a similar pattern observed for the data-generating function.
With the approximate iterated posterior linearization approach, the overestimation of the marginal credible intervals for the log-shape and log-rate processes is systematic and more pronounced across all datasets in Figures \ref{im:shapeRateMarginalPosterior_synthetic}, \ref{im:shapeRateMarginalPosterior_stiffness}, and \ref{im:shapeRateMarginalPosterior_spectrum} while the tempering approach provides superior performance when compared to the shorter chain HMC sampling in Figures \ref{im:shapeRateMarginalPosteriorTempering_synthetic}, \ref{im:shapeRateMarginalPosteriorTempering_stiffness}, and \ref{im:shapeRateMarginalPosteriorTempering_spectrum}.
The larger uncertainty of the approximate iterated posterior linearization approach for the log-shape and log-rate estimates is reflected in the one-dimensional marginal posterior distributions yielded by the subsequent HMC sampling targeting Eq.~\eqref{eq:surrogatePosteriorAlphaBeta}, which exhibit higher variance, in comparison to direct HMC sampling of the full joint posterior distribution.
In turn, the tempering approach provides consistently better posterior distribution estimates in comparison to the shorter chain HMC without tempering.
This improved mixing is most prominent with the mean and length scale estimates.
\begin{figure}
    \foreach \x in {mu_alpha, l_shape, alphaSignalSTD, alphaNoiseSTD}{
        \centering
        \includegraphics[width = 0.2415\textwidth]{pics/vector/heteroscedastic/shape/\x_hmc.pdf}
    }\\
    \foreach \x in {mu_alpha, l_shape, alphaSignalSTD, alphaNoiseSTD}{
        \centering
        \includegraphics[width = 0.2415\textwidth]{pics/vector/heteroscedastic/shape/\x_hmc_low.pdf}
    }\\
    \foreach \x in {mu_alpha, l_shape, alphaSignalSTD, alphaNoiseSTD}{
        \centering
        \includegraphics[width = 0.2415\textwidth]{pics/vector/heteroscedastic/shape/\x_pl.pdf}
    }\\
    \foreach \x in {mu_alpha, l_shape, alphaSignalSTD, alphaNoiseSTD}{
        \centering
        \includegraphics[width = 0.2415\textwidth]{pics/vector/heteroscedastic/shape/\x_plex.pdf}
    }\\
    \caption{Synthetic dataset. From top to bottom, one-dimensional marginal posterior distribution for the log-shape Gaussian process mean and covariance parameters obtained with HMC, short chain HMC, iterated posterior linearization followed by HMC, and iterated posterior linearization and tempered short chain HMC, respectively. The ground truth parameter values are illustrated with vertical blue lines.}
    \label{im:shapeThetaMarginalPosterior_synthetic}
\end{figure}
\begin{figure}
    \foreach \x in {mu_beta, l_rate, betaSignalSTD, betaNoiseSTD}{
        \centering
        \includegraphics[width = 0.2415\textwidth]{pics/vector/heteroscedastic/rate/\x_hmc.pdf}
    }\\
    \foreach \x in {mu_beta, l_rate, betaSignalSTD, betaNoiseSTD}{
        \centering
        \includegraphics[width = 0.2415\textwidth]{pics/vector/heteroscedastic/rate/\x_hmc_low.pdf}
    }\\
    \foreach \x in {mu_beta, l_rate, betaSignalSTD, betaNoiseSTD}{
        \centering
        \includegraphics[width = 0.2415\textwidth]{pics/vector/heteroscedastic/rate/\x_pl.pdf}
    }\\
    \foreach \x in {mu_beta, l_rate, betaSignalSTD, betaNoiseSTD}{
        \centering
        \includegraphics[width = 0.2415\textwidth]{pics/vector/heteroscedastic/rate/\x_plex.pdf}
    }\\
    \caption{Synthetic dataset. One-dimensional marginal posterior distribution for the log-rate Gaussian process mean and covariance parameters obtained with HMC, short chain HMC, iterated posterior linearization followed by HMC, and iterated posterior linearization and tempered short chain HMC, respectively. The ground truth parameter values are illustrated with vertical blue lines.}
    \label{im:rateThetaMarginalPosterior_synthetic}
\end{figure}
\begin{figure}
    \foreach \x in {mu_alpha, l_shape, alphaSignalSTD, alphaNoiseSTD}{
        \centering
        \includegraphics[width = 0.2415\textwidth]{pics/vector/stiffness/shape/\x_hmc.pdf}
    }\\
    \foreach \x in {mu_alpha, l_shape, alphaSignalSTD, alphaNoiseSTD}{
        \centering
        \includegraphics[width = 0.2415\textwidth]{pics/vector/stiffness/shape/\x_hmc_low.pdf}
    }\\
    \foreach \x in {mu_alpha, l_shape, alphaSignalSTD, alphaNoiseSTD}{
        \centering
        \includegraphics[width = 0.2415\textwidth]{pics/vector/stiffness/shape/\x_pl.pdf}
    }\\
    \foreach \x in {mu_alpha, l_shape, alphaSignalSTD, alphaNoiseSTD}{
        \centering
        \includegraphics[width = 0.2415\textwidth]{pics/vector/stiffness/shape/\x_plex.pdf}
    }\\
    \caption{Young's modulus. From top to bottom, one-dimensional marginal posterior distribution for the log-shape Gaussian process mean and covariance parameters obtained with HMC, short chain HMC, iterated posterior linearization followed by HMC, and iterated posterior linearization and tempered short chain HMC, respectively.}
    \label{im:shapeThetaMarginalPosterior_stiffness}
\end{figure}
\begin{figure}
    \foreach \x in {mu_beta, l_rate, betaSignalSTD, betaNoiseSTD}{
        \centering
        \includegraphics[width = 0.2415\textwidth]{pics/vector/stiffness/rate/\x_hmc.pdf}
    }\\
    \foreach \x in {mu_beta, l_rate, betaSignalSTD, betaNoiseSTD}{
        \centering
        \includegraphics[width = 0.2415\textwidth]{pics/vector/stiffness/rate/\x_hmc_low.pdf}
    }\\
    \foreach \x in {mu_beta, l_rate, betaSignalSTD, betaNoiseSTD}{
        \centering
        \includegraphics[width = 0.2415\textwidth]{pics/vector/stiffness/rate/\x_pl.pdf}
    }\\
    \foreach \x in {mu_beta, l_rate, betaSignalSTD, betaNoiseSTD}{
        \centering
        \includegraphics[width = 0.2415\textwidth]{pics/vector/stiffness/rate/\x_plex.pdf}
    }\\
    \caption{Young's modulus. From top to bottom, one-dimensional marginal posterior distribution for the log-rate Gaussian process mean and covariance parameters obtained with HMC, short chain HMC, iterated posterior linearization followed by HMC, and iterated posterior linearization and tempered short chain HMC, respectively.}
    \label{im:rateThetaMarginalPosterior_stiffness}
\end{figure}
\begin{figure}
    \foreach \x in {mu_alpha, l_shape, alphaSignalSTD, alphaNoiseSTD}{
        \centering
        \includegraphics[width = 0.2415\textwidth]{pics/vector/argent/shape/\x_hmc.pdf}
    }\\
    \foreach \x in {mu_alpha, l_shape, alphaSignalSTD, alphaNoiseSTD}{
        \centering
        \includegraphics[width = 0.2415\textwidth]{pics/vector/argent/shape/\x_hmc_low.pdf}
    }\\
    \foreach \x in {mu_alpha, l_shape, alphaSignalSTD, alphaNoiseSTD}{
        \centering
        \includegraphics[width = 0.2415\textwidth]{pics/vector/argent/shape/\x_pl.pdf}
    }\\
    \foreach \x in {mu_alpha, l_shape, alphaSignalSTD, alphaNoiseSTD}{
        \centering
        \includegraphics[width = 0.2415\textwidth]{pics/vector/argent/shape/\x_plex.pdf}
    }\\
    \caption{Argentopyrite Raman spectrum dataset. From top to bottom, one-dimensional marginal posterior distribution for the log-shape Gaussian process mean and covariance parameters obtained with HMC, short chain HMC, iterated posterior linearization followed by HMC, and iterated posterior linearization and tempered short chain HMC, respectively.}
    \label{im:shapeThetaMarginalPosterior_spectrum}
\end{figure}
\begin{figure}
    \foreach \x in {mu_beta, l_rate, betaSignalSTD, betaNoiseSTD}{
        \centering
        \includegraphics[width = 0.2415\textwidth]{pics/vector/argent/rate/\x_hmc.pdf}
    }\\
    \foreach \x in {mu_beta, l_rate, betaSignalSTD, betaNoiseSTD}{
        \centering
        \includegraphics[width = 0.2415\textwidth]{pics/vector/argent/rate/\x_hmc_low.pdf}
    }\\
    \foreach \x in {mu_beta, l_rate, betaSignalSTD, betaNoiseSTD}{
        \centering
        \includegraphics[width = 0.2415\textwidth]{pics/vector/argent/rate/\x_pl.pdf}
    }\\
    \foreach \x in {mu_beta, l_rate, betaSignalSTD, betaNoiseSTD}{
        \centering
        \includegraphics[width = 0.2415\textwidth]{pics/vector/argent/rate/\x_plex.pdf}
    }\\
    \caption{Argentopyrite Raman spectrum dataset. From top to bottom, one-dimensional marginal posterior distribution for the log-rate Gaussian process mean and covariance parameters obtained with HMC, short-chain HMC, iterated posterior linearization followed by HMC, and iterated posterior linearization and tempered short-chain HMC, respectively.}
    \label{im:rateThetaMarginalPosterior_spectrum}
\end{figure}
\begin{table}
    \centering
    \begin{tabular}{l|c|c|c|c}
         Dataset & Long HMC & PL + tempered HMC & PL + HMC & Short HMC \\
        \toprule
        Synthetic & 22227.91 & 2494.80 & 898.12 & 2700.41\\
        Speed-up &  & 8.91 & 24.75 & \\
        \hline
        Young's modulus & 22076.55 & 2618.59 & 1074.10 & 2748.87\\
        Speed-up &  & 8.43 & 20.55 & \\
        \hline
        Argentopyrite & 73183.51 & 8179.13 & 3305.25 & 8737.67\\
        Speed-up &  & 8.95 & 22.14 &\\
    \end{tabular}
    \caption{Wall times in seconds for the numerical examples. From left to right, wall times of long-chain HMC, iterated posterior linearization and tempered short-chain HMC, iterated posterior linearization followed by HMC, and short chain HMC for each dataset. The speed-up factors of the proposed approaches are presented below their respective wall times.}
    \label{tb:walltimes}
\end{table}
\begin{table}
    \centering
    \footnotesize
    \begin{tabular}{l|c|l|c|c|c}
         Dataset & Variable & Estimate & HMC (short) & PL + HMC & PL + tempered HMC \\
        \toprule
            Synthetic & $\boldsymbol{\alpha}$ & Median & 1.483 & 0.5518 & 0.01994\\
            Synthetic & $\boldsymbol{\alpha}$ & Lower bound & 1.512 & 0.3805 & 0.05298\\
            Synthetic & $\boldsymbol{\alpha}$ & Upper bound & 0.4683 & 0.9221 & 0.03768\\
            Synthetic & $\boldsymbol{\beta}$ & Median & 1.752 & 0.4690 & 0.01926\\
            Synthetic & $\boldsymbol{\beta}$ & Lower bound & 1.848 & 0.3720 & 0.05342\\
            Synthetic & $\boldsymbol{\beta}$ & Upper bound & 0.4809 & 0.8332 & 0.03596\\
            \hline
            Young's modulus & $\boldsymbol{\alpha}$ & Median & 0.04613 & 0.9556 & 0.02577\\
            Young's modulus & $\boldsymbol{\alpha}$ & Lower bound & 0.1285 & 1.958 & 0.01005\\
            Young's modulus & $\boldsymbol{\alpha}$ & Upper bound & 0.04729 & 0.1502 & 0.02927\\
            Young's modulus & $\boldsymbol{\beta}$ & Median & 0.04650 & 0.9560 & 0.02592\\
            Young's modulus & $\boldsymbol{\beta}$ & Lower bound & 0.1271 & 1.958 & 0.009469\\
            Young's modulus & $\boldsymbol{\beta}$ & Upper bound & 0.04888 & 0.1493 & 0.02985\\
            \hline
            Argentopyrite & $\boldsymbol{\alpha}$ & Median & 0.1261 & 0.3565 & 0.02749\\
            Argentopyrite & $\boldsymbol{\alpha}$ & Lower bound & 0.5985 & 0.6521 & 0.03417\\
            Argentopyrite & $\boldsymbol{\alpha}$ & Upper bound & 0.06694 & 0.3494 & 0.03122\\
            Argentopyrite & $\boldsymbol{\beta}$ & Median & 0.1230 & 0.3223 & 0.02048\\
            Argentopyrite & $\boldsymbol{\beta}$ & Lower bound & 0.7667 & 0.8197 & 0.01335\\
            Argentopyrite & $\boldsymbol{\beta}$ & Upper bound & 0.08223 & 0.2462 & 0.02694\\
            \bottomrule
    \end{tabular}
    \caption{$L_1$ norms of the differences between the median, $90\%$ marginal predictive interval lower bound, and $90\%$ marginal predictive interval upper bound estimates given by long HMC sampling and short-chain HMC, and the proposed methods for the marginal posterior distributions $\pi( \boldsymbol{\alpha} \mid \boldsymbol{y} )$ and $\pi( \boldsymbol{\beta} \mid \boldsymbol{y} )$.}
    \label{tb:estimateErrorNorms}
\end{table}
\section{Conclusions}
\label{sec:conclusions}
We propose two inference schemes for estimating latent log-shape and log-rate processes together with their hyperparameters for log-Gaussian gamma processes.
Initially, both inference methods use iterated posterior linearization to estimate the latent log-shape and log-shape processes.

In our first approach, we approximate the posterior distributions of the latent processes as normal distributions with their means and covariances given by the iterated posterior linearization.
The approximation is followed by Hamiltonian Monte Carlo sampling of the mean and covariance parameters for both of the latent processes.
The combined samples of the latent processes and their associated hyperparameters are used to construct approximate samples from the log-Gaussian gamma process posterior distribution.

The second approach uses identical iterated posterior linearization as in the first approach.
However, in contrast to our first approach, we target the true posterior distribution using a tempered sequence of probability distributions.
The sequence is initialized using the results of the iterated posterior linearization, with the last element of the sequence corresponding to the original posterior distribution of the log-Gaussian gamma process.
Each of the tempered distributions is sampled with Hamiltonian Monte Carlo, with the last estimate of the current tempered distribution used as the starting point for the sampling of the next tempered distribution.

For both methods, the resulting posterior distributions are comparable to those obtained from direct Hamiltonian Monte Carlo sampling of the full joint posterior distribution of the log-Gaussian gamma process parameters, but at a reduced computational cost.
Our first approach had speed-up factors of 17 to 26 for the considered datasets.
The second approach shows superior mixing when compared with direct Hamiltonian Monte Carlo sampling, thereby providing a computational benefit.
Both sampling methods were applied to two synthetic datasets for validation.
Additionally, the methods were applied to an experimental Raman spectrum of argentopyrite.
Furthermore, the method is readily applicable to similar models with latent Gaussian process structures, such as the log-Gaussian Cox process.
\section*{Author contributions}
Both authors -- conceptualization, methodology, and manuscript editing.
Teemu Härkönen -- formal analysis, investigation, software, validation, and visualization.
Simo Särkkä -- funding acquisition, supervision. 

\section*{Acknowledgments}
The authors thank the HORIZON AI-TRANSPWOOD (AI-Driven Multiscale Methodology to Develop Transparent Wood as Sustainable Functional Material) project, Grant no. 101138191, co-funded by the European Union. Views and opinions expressed are however those of the author(s) only and do not necessarily reflect those of the European Union or HaDEA. Neither the European Union nor the granting authority can be held responsible for them.

\section*{Financial disclosure}

None reported.

\section*{Conflict of interest}

The authors declare no potential conflict of interests.

\bibliography{ref.bib}


\end{document}